\documentstyle[sprocl]{article}      

\def\Journal#1#2#3#4{{#1} {\bf #2}, #3 (#4)}

\def\NP{{\em Nucl. Phys.}}
\def\AP{{\em Ann. Phys.} (N.Y.)}
\def\YF{\em Yad. Fyz.}
\def\NPB{{\em Nucl. Phys.} B}
\def\NPA{{\em Nucl. Phys.} A}
\def\NPBPS{{\em Nucl. Phys.} B(Proc.~Suppl.)}
\def\JHEP{{\em J. High Energy Phys.}}
\def\PLB{{\em Phys. Lett.} B}
\def\PLA{{\em Phys. Lett.} A}
\def\RMP{\em Rev. Mod. Phys.}
\def\JDG{\em J. Diff. Geom.}
\def\PR{\em Phys. Rev.}
\def\PRL{\em Phys. Rev. Lett.}
\def\PRD{{\em Phys. Rev.} D}
\def\PRP{\em Phys. Rep.}
\def\JMP{\em J. Math. Phys.}
\def\JPA{{\em J. Phys.} A}
\def\EPJ{{\em Eur. Phys. J.} C}
\def\CMP{\em Comm. Math. Phys.}

\newcommand{\beq}{\begin{equation}}
\newcommand{\eeq}{\end{equation}}
\newcommand{\bea}{\begin{eqnarray}}
\newcommand{\eea}{\end{eqnarray}}
\newcommand{\half}{{\scriptstyle{{1\over 2}}}}
\newcommand{\twoth}{{\scriptstyle{{2\over 3}}}}
\newcommand{\quart}{{\scriptstyle{{1\over 4}}}}
\newcommand{\real}{\relax{\rm I\kern-.18em R}}
\newcommand{\zahlen}{{\rm Z \!\! Z}}
\newcommand{\quat}{{\rm I \! H}}
\newcommand{\Tr}{\mbox{\,Tr\,}}
\newcommand{\tr}{\mbox{\,tr\,}}
\newcommand{\zN}{{\rm z}_{{}_N}}
\newcommand{\ad}{{\rm ad}}
\newcommand{\Ad}{{\rm Ad}}

\newcommand{\eff}{{\rm eff}}
\newcommand{\diag}{{\rm diag}}
\newcommand{\norm}[1]{\left\| #1 \right\|}
\newcommand{\sgbar}{\bar{\sg}}
\newcommand{\al}{\alpha}
\newcommand{\Gm}{\Gamma}
\newcommand{\dl}{\delta}
\newcommand{\eps}{\epsilon}
\newcommand{\veps}{\varepsilon}
\newcommand{\lm}{\lambda}
\newcommand{\Lm}{\Lambda}
\newcommand{\Om}{\Omega}
\newcommand{\sg}{\sigma}
\newcommand{\Ss}[1]{\mbox{$\cal #1$}}
\newcommand{\pr}{\partial}
\newcommand{\Lkw}{\vek{L}_1^2}
\newcommand{\ip}[1]{\left\langle #1 \right \rangle }
\newcommand{\Order}[1]{\Ss{O}\left(#1\right)}

\newcommand{\phd}{{\vphantom{\dagger}}}
\newcommand{\coker}{\rm coker}
\newcommand{\ch}{\rm ch}

\newcommand{\Cite}[1]{$\,$\cite{#1}}
\newcommand{\Ref}[1]{(\ref{#1})}

\def\vek#1{{\bf #1}}
\def\overbar{\overline}
\def\vv{{\bf V}}
\def\rd{{\rm d}}
\def\mco{\multicolumn}

\begin{document}
\title{
\hskip3.25cm                                            
QCD IN A FINITE VOLUME
\footnote{To appear in the Boris Ioffe Festschrift,     
 edited by M. Shifman (World Scientific).}              
\hfill{\tiny INLO-PUB-08/00}\vskip-6.5mm                
}
\author{PIERRE VAN BAAL}
\address{Instituut-Lorentz for Theoretical Physics, University of Leiden,\\
P.O. Box 9506, Nl-2300 RA Leiden, The Netherlands} 
\vskip-5mm
\maketitle
\abstracts{We will review our understanding of non-abelian gauge theories in 
finite physical volumes. It allows one in a reliable way to trace some of
the non-perturbative dynamics. The role of gauge fixing ambiguities related 
to large field fluctuations is an important lesson that can be learned. The 
hamiltonian formalism is the main tool, partly because semiclassical 
techniques are simply inadequate once the coupling becomes strong. Using 
periodic boundary conditions, continuum results can be compared to those 
on the lattice. Results in a spherical finite volume will be discussed
as well.}

\section{Introduction}

We have decided to take this opportunity of contributing to a handbook of QCD, 
in honor of Boris Lazarevich Ioffe, to bring together results and methods that 
were developed in solving for the low-lying spectrum in a finite volume. The 
emphasis will be on the dynamical aspects of the classically scale invariant 
theory for non-abelian gauge theories in 3+1 dimensions.\cite{YaMi} The 
challenge is to understand how the mass gap is generated. Due to the need 
for regularization, at the quantum level breaking the scale invariance, a 
running coupling appears. The difficulty lies in the fact that this coupling 
increases at low-energies and long-distance scales, beyond the point where one 
has control over the field fluctuations, which become so large that they probe 
the essential non-linearities of the theory. A finite volume explicitly breaks 
the scale invariance. However, it does so in a rather mild way. Classically one
can use the scale transformation to go from one physical volume to another, and
only the running of the coupling constant prevents us from taking the infinite 
volume limit. As the longest distance scale, i.e. the lowest energy scale, is 
set by the volume, one can keep in check the growth of the running coupling 
constant. 

We will describe the results mainly in the context of a hamiltonian 
picture\Cite{Chle} with wave functionals on configuration space. Although 
rather cumbersome from a perturbative point of view, where the covariant 
path integral approach of Feynman is vastly superior, it provides more 
intuition on how to deal with non-perturbative contributions in situations
where semi-classical techniques can no longer be used, like for observables 
that do not vanish in perturbation theory. The high energy modes can be 
well-approximated by a harmonic oscillator contribution to the wave functional.
In the direction of these field modes the potential energy rises steeply. 
Their contribution, including regulating the ultraviolet behavior, is 
treated perturbatively, giving in particular rise to the running of the 
coupling. The finite volume allows us to have a well-defined mode expansion. 
Due to the classical scale invariance, the hamiltonian can be formulated
in terms of dimensionless fields. This can be extended to the quantum 
theory, as Ward identities allow for a field definition without anomalous
scaling. Apart from the overall scaling dimension of the hamiltonian, only 
the running coupling introduces a non-trivial volume dependence. 

Due to the non-abelian nature of the theory the physical configuration space,
formed by the set of gauge orbits $\Ss{A}/\Ss{G}$ ($\Ss{A}$ is the collection 
of connections, $\Ss{G}$ the group of local gauge transformations) is 
non-trivial.\cite{Bavi} Most frequently, coordinates of this orbit space 
are chosen by picking a representative gauge field on the orbit in a smooth 
and preferably unique way. Linear gauge conditions like the Landau or Coulomb 
gauge suffer from what is known as Gribov ambiguities.\cite{Grib} The region 
of field space that contains no further gauge copies is called a fundamental 
domain for non-abelian gauge theories.\cite{Sefr}

Having arranged the low-energy field modes (and all those modes not affected 
by the cutoff) to be scale invariant, the spreading of the wave functional is 
completely caused by an increasing coupling. This is what leads to 
non-perturbative effects. Asymptotic freedom on the other hand guarantees 
that in small volumes the running coupling is small and it thus keeps the 
wave functional localized near the classical vacuum manifold. In a periodic 
geometry, this perturbative analysis was pioneered by Bjorken\Cite{Bjor} and 
further developed by L\"uscher\Cite{Lue1}. The essential ingredient we have 
added to address non-perturbative effects is boundary conditions in {\em field 
space}, at the boundary of the fundamental domain, with gauge invariance 
implemented properly at all stages. 

The non-perturbative features due to spreading of the wave functional can  
be followed out to a physical volume of about one cubic fermi (setting the 
scale by the string tension or lowest glueball state). This is, on the 
basis of a comparison with lattice Monte Carlo data, the point where in 
the pure gauge theory the confining string is being formed, with no 
significant finite volume dependence beyond this volume. What has become
clear is that the transition from the finite to the infinite volume is 
driven by field fluctuations that cross the barrier which is associated 
with tunneling between different classical vacua. This is natural, since this 
barrier (the finite volume sphaleron), will be the direction beyond which 
the wave functional can first spread most significantly, as it provides the 
lowest mountain pass in the energy landscape.

In Sec.~\ref{sec:copies} we describe the process of complete gauge fixing, and 
the fact that the boundary of the fundamental domain, unlike its interior, 
has gauge copies that implement the non-trivial topology of configuration 
space. This is applied in Sec.~\ref{sec:torus} to formulating non-abelian 
gauge theories on a torus, both for SU(2) and in Sec.~\ref{subsec:SUN} for 
SU(3). The influence of massless quarks on the small volume vacuum structure 
is discussed in Sec.~\ref{subsec:quarks}, whereas Sec.~\ref{subsec:running} 
gives a short review of the non-perturbative evaluation of the running 
renormalized coupling, making explicit use of the finite volume 
geometry.\cite{LSWW} In Sec.~\ref{subsec:twist} we consider the case of 
twisted boundary conditions,\cite{Tho1} and in Sec.~\ref{subsec:susy} we 
discuss supersymmetric Yang-Mills theory in a finite volume in the light 
of some recent developments concerning the Witten index.\cite{WiIn} 
Sec.~\ref{sec:inst} outlines what is known about instantons on the torus, 
reviewing the Nahm transformation\Cite{Nahm} in Sec.~\ref{subsec:Nahm}.
Sec.~\ref{sec:sphere} analyzes the situation for a spherical geometry, 
particularly well adapted to study the role of instantons in the low-lying 
glueball spectrum. Finally, we shall consider the behavior in large finite 
volumes. We discuss how the previous results fit together and what this 
implies. In Secs.~\ref{subsec:stable} and \ref{subsec:scatt} we review the 
volume dependence whenever the polarization clouds are well-contained in 
the finite volume.\cite{Lue0} In Sec.~\ref{subsec:chiral} we briefly mention 
the situation in the absence of a mass gap, particularly relevant for QCD 
with its light up and down quarks, for which chiral perturbation theory 
applies.\cite{Leut} We conclude with a review of 't~Hooft's electric-magnetic 
duality on the torus\Cite{Tho1} in Sec.~\ref{subsec:EMdual}.

\section{Complete Gauge Fixing}\label{sec:copies}

An (almost) unique representative of the gauge orbit is found by minimizing 
the $L^2$ norm of the vector potential along the gauge orbit\Cite{Sefr,Del1}
\beq
G_A(h)\equiv\norm{[h]A}^2=-\int_M\tr\left(\left( h^{-1}(\vek x)A_i(\vek x) 
h(\vek x)+h^{-1}(\vek x)\pr_i h(\vek x)\right)^2\right),\label{eq:gAnorm}
\eeq
where the vector potential is taken anti-hermitian, the integral over the 
finite spatial volume $M$ is with the appropriate canonical volume form and
$h(\vek x)$ is a Lie-group element, with $[h]A$ a short-hand notation for 
the associated gauge transformation. Note that in these conventions the 
field strength is given by 
\beq
F_{\mu\nu}(\vek x,t)=\pr_\mu A_\nu(\vek x,t)-\pr_\nu A_\mu(\vek x,t)+
[A_\mu(\vek x,t),A_\nu(\vek x,t)] \label{eq:F}
\eeq 
and the action by 
\beq
L(t)=\half\int_M g^{-2}\tr(F_{\mu\nu}^{\,2}(\vek x,t)).\label{eq:L}
\eeq
Expanding around the minimum of Eq.~\Ref{eq:gAnorm}, writing $h(\vek x)=
\exp(X(\vek x))$ ($X(\vek x)$ is, like the gauge field $A_i(\vek x)$, an 
element of the Lie-algebra) one easily finds
\bea 
\norm{\,[h]A}^2 &=& \norm{A}^2+2\int_M \tr(X \partial_i A_i)+\int_M \tr 
(X^\dagger FP (A) X) \\ &&\hskip-1.3cm+\frac{1}{3}\int_M\tr\left(X\left[[A_i,X],
\partial_i X\right]\right)+\frac{1}{12}\int_M\tr\left([\Ss{D}_iX,X][\partial_i 
X,X]\right)+\Order{X^5},\nonumber\label{eq:Xexp}
\eea
where $FP(A)$ is the Faddeev-Popov operator $(\ad(A)X\equiv[A,X])$
\beq
FP (A)=-\partial_i \Ss{D}_i (A)\equiv-\partial_i(\partial_i+\ad(A_i)).
\label{eq:FP}
\eeq

At a local minimum the vector potential is therefore transverse, $\partial_i 
A_i=0$, and $FP(A)$ is a positive operator. The set of all these vector 
potentials is by definition the Gribov region $\Omega$. Using the fact that 
$FP(A)$ is linear in $A$, $\Omega$ is seen to be a convex subspace of the set 
of transverse connections $\Gamma$. Its boundary $\partial \Omega$ is called 
the Gribov horizon. At the Gribov horizon, the {\em lowest} non-trivial 
eigenvalue of the Faddeev-Popov operator vanishes, and points on $\partial
\Omega$ are associated with coordinate singularities. Any point on $\partial
\Omega$ can be seen to have a finite distance to the origin of field space 
and in some cases even uniform bounds can be derived.\cite{Dezw,Zwan}

The Gribov region is the set of {\em local} minima of the norm functional, 
Eq.~\Ref{eq:gAnorm}, and needs to be further restricted to the {\em absolute} 
minima to form a fundamental domain,\Cite{Sefr} which will be denoted by 
$\Lambda$. The fundamental domain is clearly contained within the Gribov 
region. To show that also $\Lambda$ is convex, note that
\bea
&&\norm{\,[h]A}^2-\norm{A}^2=\int\tr\left(A_i^2\right)-\int\tr
  \left(\left(h^{-1}A_ih+h^{-1}\pr_ih\right)^2\right)\label{eq:FPf}\\
&&\quad=\int\tr\left(h^\dagger FP_f(A)\,h\right)\equiv\ip{h,FP_f(A)\,h},
  \quad FP_f(A)\equiv-\pr_iD_i(A),\nonumber
\eea
where $FP_f(A)$ acts on the fundamental representation and is similar to the 
Faddeev-Popov operator. Both $FP(A)$ and $FP_f(A)$ are hermitian operators 
when $A$ is a critical points of the norm functional, i.e. for $A$ transverse.
We can define $\Lm$ in terms of the absolute minima over $h\in\Ss{G}$ of 
$\ip{h,FP_f(A)\,h}$
\beq
\Lm =\{A\in\Gm|\min_{h\in\Ss{G}}\ip{h,FP_f(A)\,h}=0\}.\label{eq:Lm}
\eeq
Using that $FP_f(A)$ is linear in $A$, the convexity of $\Lm$ is automatic: A 
line connecting two points in $\Lm$ lies within $\Lm$. 

If we would not specify anything further, as a convex space is contractible,
the fundamental region could never reproduce the non-trivial topology of the
configuration space. This means that $\Lm$ should have a boundary.\cite{Vba1}
Indeed, as $\Lambda$ is contained in $\Omega$, this means $\Lm$ is also 
bounded in each direction. Clearly $A=0$ is in the interior of $\Lm$, which
allows us to consider a ray extending out from the origin into a given 
direction, where it will have to cross the boundary of $\Lm$ and $\Om$.
For any point along this ray in $\Lm$, the norm functional is at its 
absolute minimum as a function of the gauge orbit. However, for points in
$\Om$ that are not also in $\Lm$, the norm functional is necessarily at
a relative minimum. The absolute minimum for this orbit is an element
of $\Lm$, but in general not along the ray. Continuity therefore tells
us that at some point along the ray, this absolute minimum has to pass
the local minimum. At the point they are exactly degenerate, there are
two gauge equivalent vector potentials with the same norm, both at the 
absolute minimum. As in the interior the norm functional has a unique
minimum, again by continuity, these two degenerate configurations have
to both lie on the boundary of $\Lm$. 

When $L$ denotes the linear size of $M$, we may express the gauge fields in 
the dimensionless combination of $L A$ (in our conventions the fields have no 
anomalous scale dependence), and {\em the shape and geometry of the Gribov and 
fundamental regions are scale independent}. We should note that the norm 
functional is degenerate along the constant gauge transformations and indeed 
the Coulomb gauge does not fix these gauge degrees of freedom. We simply demand
that the wave functional is in the singlet representation under the constant 
gauge transformations and we have $\Lm/G=\Ss{A}/\Ss{G}$. Here $\Lm$ is assumed 
to include the non-trivial boundary identifications that restore the 
non-trivial topology of $\Ss{A}/\Ss{G}$. 

\begin{figure}[htb]
\vspace{5.7cm}
\includegraphics{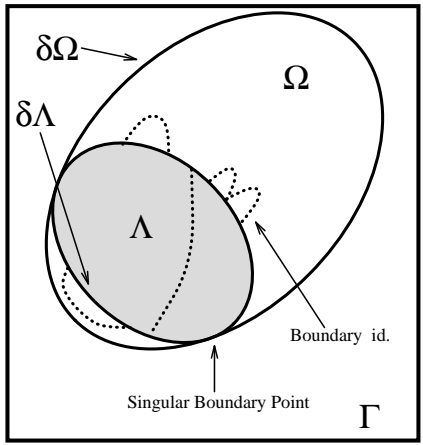}
\caption{Sketch of the fundamental (shaded) and Gribov regions. The dotted 
lines indicate the boundary identifications.}
\label{fig:Grib}
\end{figure}

If a degeneracy at the boundary is continuous, other than by constant gauge 
transformations, one necessarily has at least one non-trivial zero eigenvalue 
for $FP(A)$ and the Gribov horizon will touch the boundary of the fundamental 
domain at these so-called singular boundary points. We sketch the general 
situation in Fig.~\ref{fig:Grib}. In principle, by choosing a different gauge 
fixing in the neighborhood of these points one could resolve the singularity. 
If singular boundary points would not exist, all that would have been required 
is to complement the hamiltonian in the Coulomb gauge with the appropriate 
boundary conditions in field space. Since the boundary identifications are 
by gauge transformations the boundary condition on the wave functionals is 
simply that they are identical under the boundary identifications, possibly 
up to a phase in case the gauge transformation is homotopically non-trivial.

\begin{figure}[htb]
\vspace{3.5cm}
\includegraphics{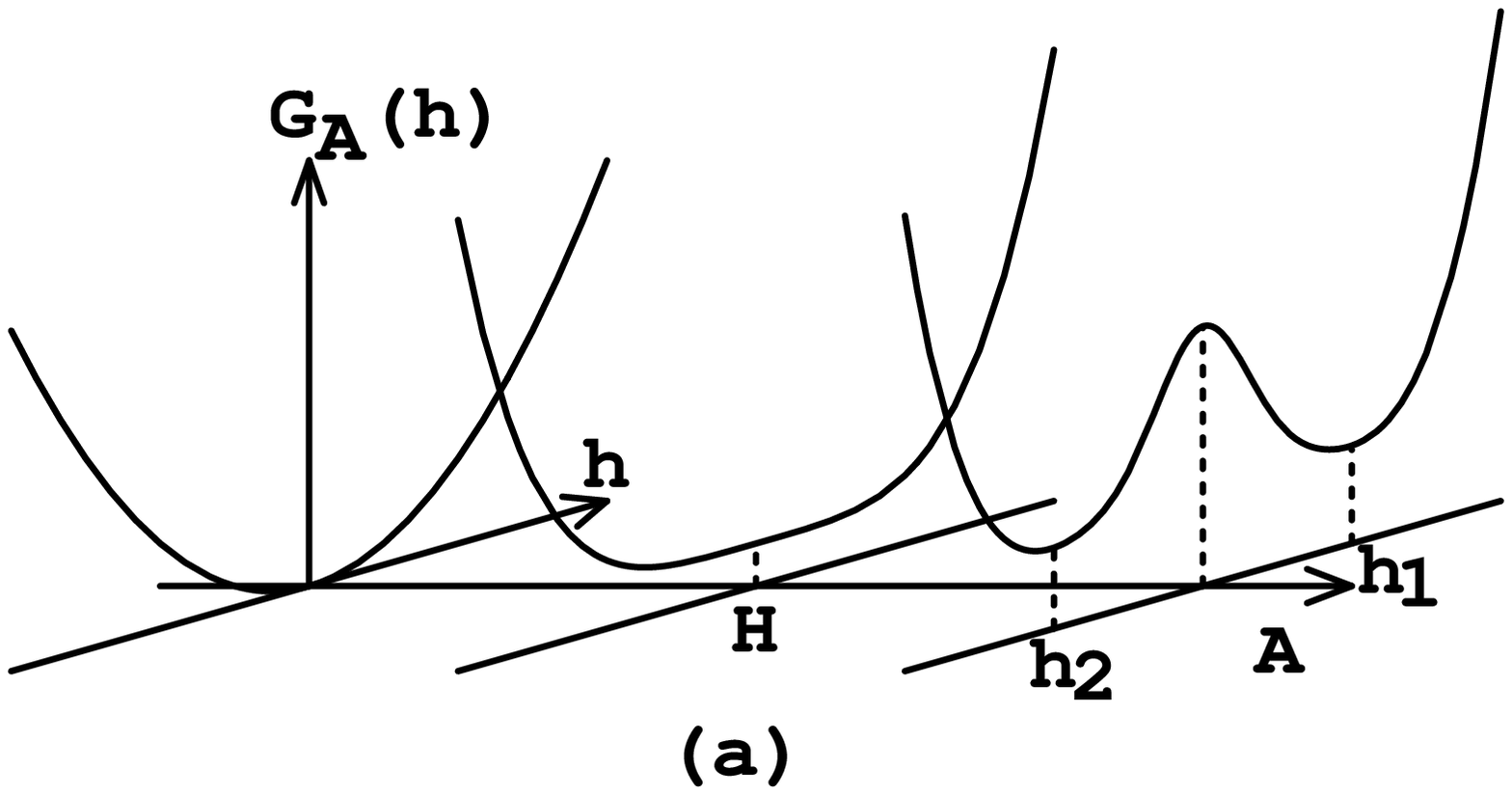}
\includegraphics{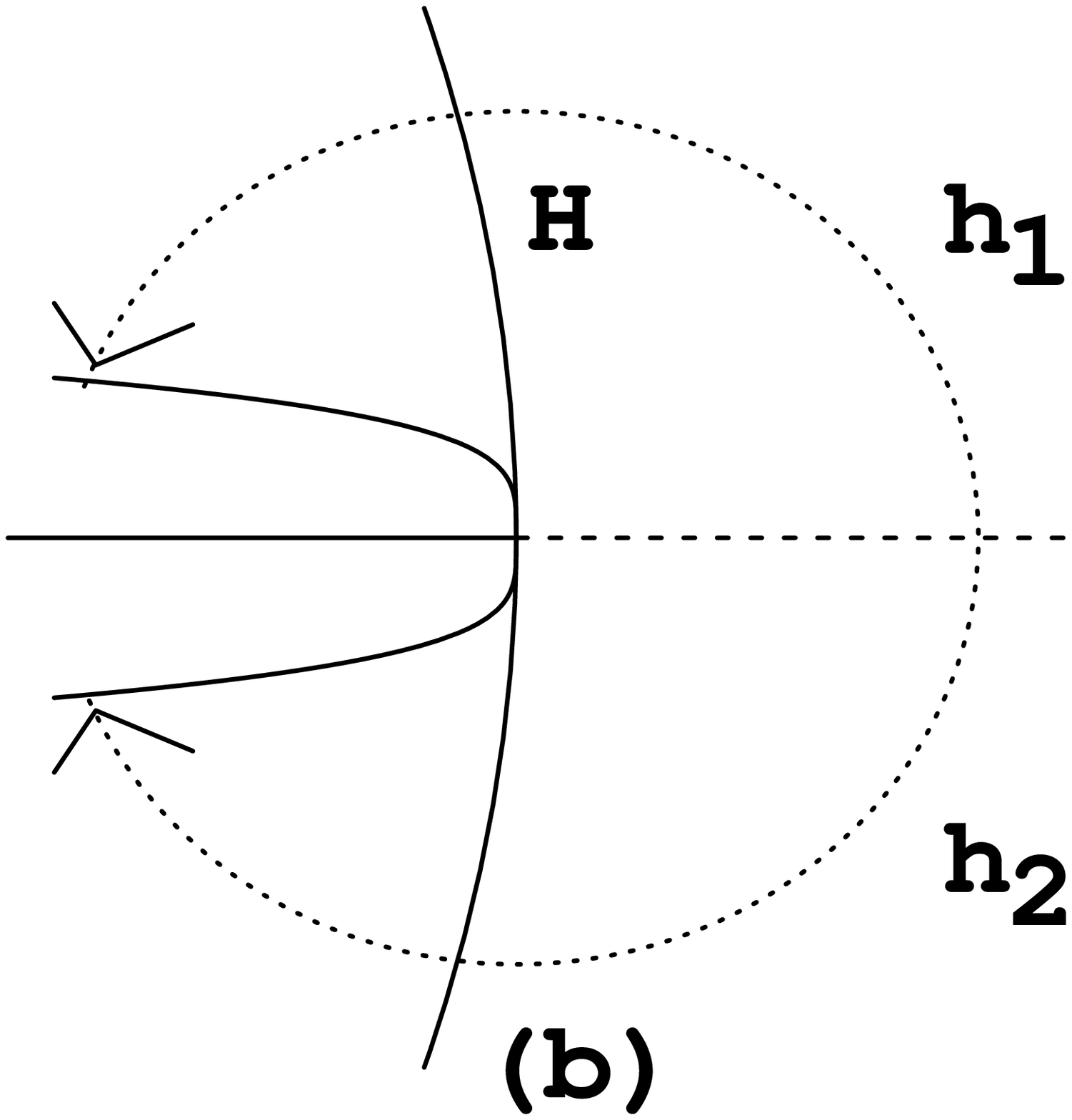}
\caption{Sketch of a singular boundary point due to a bifurcation of
the norm functional. It can be used to show that there are homotopically
trivial gauge copies inside the Gribov horizon ({\rm H}).}
\label{fig:bif}
\end{figure}

Unfortunately, one can argue that singular boundary points are to be 
expected.\cite{Vba1} Generically, at singular boundary points the norm 
functional undergoes a bifurcation moving from inside to outside the 
fundamental (and Gribov) region. The absolute minimum turns into a saddle 
point and two local minima appear, as indicated in Fig.~\ref{fig:bif}. These 
are necessarily gauge copies of each other. The gauge transformation is 
homotopically trivial as it reduces to the identity at the bifurcation point, 
evolving continuously from there on. 

Also Gribov's original arguments for the existence of gauge copies\Cite{Grib}
(showing that points just outside the horizon are gauge copies of points
just inside) can be easily understood from the perspective of bifurcations in
the norm functional. It describes the generic case where the zero-mode of
the Faddeev-Popov operator arises because of the coalescence of a local
minimum with a saddle point with only one unstable direction. At the Gribov
horizon the norm functional locally behaves in that case as $X^3$, with $X$
the relevant zero eigenfunction of the Faddeev-Popov operator. The situation
sketched in Fig.~\ref{fig:bif} corresponds to the case where the leading 
behavior is like $X^4$. 

The necessity to restrict to the fundamental domain, a subset of the transverse
gauge fields, introduces a non-local procedure in configuration space. This 
cannot be avoided since it reflects the non-trivial topology of this space. We 
stress again that its topology and geometry is scale independent. Homotopical 
non-trivial gauge transformations are in one to one correspondence with 
non-contractible loops in configuration space, which give rise to conserved 
quantum numbers. The quantum numbers are like the Bloch momenta in a periodic 
potential and label representations of the homotopy group of gauge 
transformations. On the fundamental domain the non-contractible loops arise 
through identifications of boundary points. Although slightly more hidden, 
the fundamental domain will therefore contain all the information relevant 
for the topological quantum numbers. Sufficient knowledge of the boundary 
identifications will allow for an efficient and natural projection on the 
various superselection sectors. Typically we integrate out the high-energy 
modes, being left with the low-energy modes whose dynamics is determined by 
an effective hamiltonian defined on the fundamental domain (restricted to 
these low-energy modes). In this it is assumed that the contributions of the 
high-energy modes can be dealt with perturbatively, generating the running 
coupling and the effective interactions of the low-energy modes. We will in 
detail discuss the results for finite volumes with a torus and sphere geometry. 

\section{Gauge Fields on the three-Torus}\label{sec:torus}

\begin{figure}[htb]
\vspace{5cm}
\includegraphics{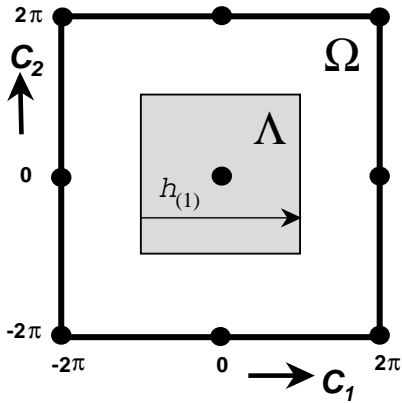}
\caption{A two dimensional slice of the vacuum valley along the $(C_1,C_2)$
plane. The fat square give the Gribov horizon, the grey square is the
fundamental domain. The dots at the Gribov horizon are Gribov copies
of the origin.}
\label{fig:torus}
\end{figure}

Probably the most simple example to illustrate the relevance of the fundamental
domain is provided by gauge fields on the torus in the abelian zero-momentum 
sector. Let us first take $G$=SU(2) and $A_j=i{C_j\over 2L}\tau_3$ ($L$ is 
the size of the torus, $\tau_j$ are the Pauli matrices). These modes are 
dynamically motivated as they form the set of gauge fields on which the 
classical potential vanishes. It is called the vacuum valley (sometimes 
also referred to as toron valley) and one can attempt to perform a 
Born-Oppenheimer-like approximation for deriving an effective hamiltonian in 
terms of these ``slow'' degrees of freedom. To find the Gribov horizon, one 
easily verifies that the part of the spectrum for $FP(A)$ that depends on 
$\vek C$, is given by $\lambda^{gh}_{\vek n}(\vek C)=2\pi\vek n\cdot(2\pi
\vek n\pm\vek C)/L^2$, with $\vek n\neq\vek 0$ an integer vector. The lowest 
eigenvalue vanishes if $C_j=\pm2\pi$. The Gribov region is therefore a cube 
with sides of length $4\pi$, centered at the origin, specified by 
$|C_j|\leq2\pi$ for all $j$, see Fig.~\ref{fig:torus}.

The gauge transformation $h_{(j)}=\exp(\pi i x_j\tau_3/L)$ maps $C_j$ to
$C_j+2\pi$, leaving the other components $C_{i\neq j}$ untouched. As $h_{(j)}$
is anti-periodic it is homotopically non-trivial (they are 't~Hooft's twisted
gauge transformations.\cite{Tho1}) We thus see explicitly that gauge copies
occur inside $\Om$, but furthermore the naive vacuum $A=0$ has (many) gauge
copies under these shifts of $2\pi$ that lie on the Gribov horizon. It can
actually be shown in the Coulomb gauge that for any three-manifold, any
Gribov copy by a homotopically non-trivial gauge transformation of $A=0$ will
have a non-trivial zero eigenvalue for the Faddeev-Popov operator.\cite{Vba1} 
Taking the symmetry under homotopically non-trivial gauge transformations 
properly into account is crucial for describing the non-perturbative dynamics 
and one sees that the singularity of the hamiltonian at Gribov copies of $A=0$,
where the wave functionals are in a sense maximal, could form a severe obstacle
in obtaining reliable results.

To find the boundary of the fundamental domain we note that the gauge copies
$\vek C=(\pi,C_2,C_3)$ and $\vek C=(-\pi,C_2,C_3)$ have equal norm. The
boundary of the fundamental domain, restricted to the vacuum valley formed by
the abelian zero-momentum gauge fields therefore occurs where $|C_j|=\pi$,
well inside the Gribov region, see Fig.~\ref{fig:torus}. The boundary 
identifications are by the homotopically non-trivial gauge transformations 
$h_{(j)}$. The fundamental domain, described by $|C_j|\leq \pi$ with all 
boundary points regular, has the topology of a torus. To be more precise, 
as the remnant of the constant gauge transformations (the Weyl group) changes 
$\vek C$ to $-\vek C$, the fundamental domain $\Lm/G$ restricted to the abelian
constant modes is the orbifold $T^3/Z_2$. As we will see, the orbifold 
singularities are associated with problems in using the harmonic approximation 
in perturbation theory for modes orthogonal to those of vacuum valley. 

Formulating the hamiltonian on $\Lm$, with the boundary identifications implied
by the gauge transformations $h_{(j)}$, avoids the singularities at the Gribov 
copies of $A=0$. ``Bloch momenta'' associated with the $2\pi$ shift, 
implemented by the non-trivial homotopy of $h_{(j)}$, label `t~Hooft's electric
flux quantum numbers\Cite{Tho1} $\Psi(C_j=-\pi)=\exp( \pi ie_j)\Psi(C_j=\pi)$. 
Note that the phase factor is not arbitrary, but $\pm 1$. This is because 
$h^2_{(j)}$ is homotopically trivial. In other words, the homotopy group of 
these anti-periodic gauge transformations is $Z_2^3$. Considering a slice of 
$\Lambda$ can obscure some of the topological features. A loop that winds 
around the slice twice is contractible in $\Lambda$ as soon as it is allowed 
to leave the slice. Indeed including the lowest modes transverse to this slice 
will make the $Z_2$ nature of the relevant homotopy group evident.\cite{Kovb} 
It should be mentioned that for the torus in the presence of fields in the 
fundamental representation (quarks), only periodic gauge transformations are 
allowed.\cite{Ferm} This will be discussed in Sec.~\ref{subsec:quarks}.

In weak coupling L\"uscher showed unambiguously that the wave functionals are 
localized around $A=0$, that they are normalizable and that the spectrum is 
discrete.\cite{Bjor,Lue1} In this limit the spectrum is insensitive to the 
boundary identifications (giving rise to a degeneracy in the topological
quantum numbers). This is manifested by a vanishing electric flux energy,
defined by the difference in energy of a state with $|\vek e|=1$ and the
vacuum state with $\vek e=\vek 0$. Although there is no classical potential
barrier to achieve this suppression, it comes about by a quantum induced
barrier, in strength down by two powers of the coupling constant. 
\beq
V(\vek C)=2L^{-1}\sum_{\vek p\in \zahlen^3}(|2\pi\vek p+\vek C|-|2\pi\vek p|)
=\frac{4}{L\pi^3}\sum_{\vek n\neq\vek 0}\frac{\sin^2(\half\vek n\cdot\vek C)}{
(\vek n\cdot\vek n)^2}.\label{eq:VSU2}
\eeq
The second identity is obtained by Poisson resummation,\cite{Lue1} and the 
periodicity reflects the gauge symmetry discussed above. This quantum induced 
barrier leads to a suppression\Cite{Vba0} with a factor $\exp(-S/g)$ instead 
of the usual factor of $\exp(-8\pi^2/g^2)$ for instantons.\cite{Bpst,Tho2} 
The associated tunneling configuration was called a {\em pinchon}.\cite{Vba0}
At stronger coupling the wave functional spreads out over the vacuum valley 
and drastically changes the spectrum.\cite{Kov0} At this point the energy 
of electric flux suddenly switches on.

Integrating out the non-zero momentum modes, for which Bloch degenerate 
perturbation theory\Cite{Bloc} provides a rigorous framework,\cite{Lue1} one 
finds an effective hamiltonian. Near $A=0$, due to the quartic nature of the 
potential energy $V(A)=-\half\int d^3x\tr(F^{\,}_{ij})$ for the zero-momentum 
modes (the derivatives vanish, such that the field strength is quadratic in 
the field), there is no separation in time scales between the abelian and 
non-abelian modes. Away from $A=0$ one could further reduce the dynamics to 
one along the vacuum valley, but near the origin the adiabatic approximation 
breaks down. This gives rise to a conic singularity, $V(\vek C)=2|\vek C|/L+
\Order{C^2}$, due to fluctuations of the non-abelian zero-momentum modes 
(contributing the $\vek p=\vek 0$ term in Eq.~\Ref{eq:VSU2}).

The effective hamiltonian is expressed in terms of the coordinates $c_j^a$,
where $i=\{1,2,3\}$ is the spatial index ($c_0^a=0$) and $a=\{1,2,3\}$ the 
SU(2)-color index. It will be convenient to introduce $f_{jk}^a\equiv
-\eps_{abd}c_j^bc_k^d$ and $r_j\equiv\sqrt{\sum_a c_j^ac_j^a}$. The latter are 
gauge-invariant ``radial" coordinates, that will play a crucial role in 
specifying the boundary conditions. The zero-momentum gauge field and field 
strength read $A_j(\vek x)=ic_j^a\frac{\tau_a}{2L}$ and $F_{jk}=if_{jk}^a
\frac{\tau_a}{2L^2}$. For dimensional reasons the effective hamiltonian is 
proportional to $1/L$. It will furthermore depend on $L$ through the 
renormalized coupling constant ($g(L)$) at the scale $\mu=1/L$. To one loop 
order (for small $L$) $g^2(L)=12\pi^2/[-11\ln(\Lambda_{MS}L)]$. One expresses 
the masses and the size of the finite volume in dimensionless quantities, like 
mass-ratios and the parameter $z=mL$. In this way, the explicit dependence of 
$g$ on $L$ is irrelevant. This is also the preferred way of comparing results 
obtained within different regularization schemes (e.g. dimensional and lattice
regularization). The effective hamiltonian is now given by
\bea
L H_{\eff}(c)&=&{-g^2\over 2(1+\alpha_1 g^2)}\sum_{i,a}{\partial^2\over 
(\partial c_i^a)^2}+{1\over 4}({1\over g^2}+\alpha_2)\sum_{ij,a}(f_{ij}^a)^2
+\gamma_1\sum_i r_i^2\nonumber\\ 
&&+\gamma_2\sum_i r_i^4+\gamma_3\sum_{i>j} r_i^2r_j^2+\gamma_4\sum_i r_i^6+
\gamma_5\sum_{i\neq j} r_i^2 r_j^4+\gamma_6\prod_i r_i^2\nonumber\\
&&+\alpha_3\sum_{ijk,a}r_i^2(f_{jk}^a)^2+\alpha_4\sum_{ij,a}r_i^2(f_{ij}^a)^2+
\alpha_5{\det}^2c\label{eq:Heff}\\
&&+\gamma_7 \sum_i r_i^8+\gamma_8\sum_{i\neq j}r_i^6 r_j^2+
\gamma_9\sum_{i>j} r_i^4 r_j^4+\gamma_{10}\sum_i r_i^2(r_1^2r_2^2r_3^2).
\nonumber
\eea

We have organized the terms according to the importance of their contributions.
The first two terms give (when ignoring $\alpha_{1,2}$) the lowest order 
effective hamiltonian,\cite{Bjor} whose energy eigenvalues are $\Order{g^{2/3}
}$, as can be seen by rescaling $c$ with $g^{2/3}$. Thus, in a perturbative 
expansion $c=\Order{g^{2/3}}$. L\"uscher's computation\Cite{Lue1} to $\Order{
g^{8/3}}$, determined by $\alpha_1,\alpha_2,\gamma_1,\gamma_2$ and $\gamma_3$, 
was extended to $\Order{g^4}$, to ensure that the terms parametrized by 
$\gamma_i$ included the vacuum-valley effective potential (i.e. the part that 
does not vanish on the set of abelian configurations) to sufficient numerical 
accuracy.\cite{Kovb} The order $r^8$ terms were just included to check 
stability of the results.  The vacuum-valley effective potential could 
actually be computed at two loops to all orders in the field, up to an 
irrelevant constant
\beq
V_{\rm 2-loop}(\vek C)=\frac{g^2L}{32}(\Delta V(\vek C)+\Delta V(\vek 0))^2,
\quad \Delta V(\vek C)=\sum_i\frac{\partial^2V(\vek C)}{\partial C_i\partial 
C_i},\label{eq:V2loop}
\eeq
but more cumbersome to compute were terms of the form $g^4c^2\partial^2_c$, 
also of two loop order.\cite{Vba4} Their effect on the spectrum can, however, 
be neglected and we have not listed these terms. The coefficients appearing in
$H_{\eff}$ have the following numerical values
\bea
\gamma_1=-3.0104661\cdot10^{-1}-3.0104661\cdot10^{-1}(g/2\pi)^2,&
\alpha_1=+2.1810429\cdot10^{-2},\hskip-3mm\nonumber\\
\gamma_2=-1.4488847\cdot10^{-3}-9.9096768\cdot10^{-3}(g/2\pi)^2,&
\alpha_2=+7.5714590\cdot10^{-3},\hskip-3mm\nonumber\\
\gamma_3=+1.2790086\cdot10^{-2}+3.6765224\cdot10^{-2}(g/2\pi)^2,&
\alpha_3=+1.1130266\cdot10^{-4},\hskip-3mm\nonumber\\
\gamma_4=+4.9676959\cdot10^{-5}+5.2925358\cdot10^{-5}(g/2\pi)^2,&
\alpha_4=-2.1475176\cdot10^{-4},\hskip-3mm\nonumber\\
\gamma_5=-5.5172502\cdot10^{-5}+1.8496841\cdot10^{-4}(g/2\pi)^2,&
\alpha_5=-1.2775652\cdot10^{-3},\hskip-3mm\nonumber\\
\gamma_6=-1.2423581\cdot10^{-3}-5.7110724\cdot10^{-3}(g/2\pi)^2,&\nonumber\\
\gamma_7=-9.8738947\cdot10^{-7}-5.1311245\cdot10^{-6}(g/2\pi)^2,&\nonumber\\
\gamma_8=+9.1911536\cdot10^{-6}+9.1452409\cdot10^{-5}(g/2\pi)^2,&\nonumber\\
\gamma_9=-2.7911565\cdot10^{-5}-2.5203366\cdot10^{-5}(g/2\pi)^2,&\nonumber\\
\!\gamma_{10}\!\!=+1.8208802\cdot10^{-5}+6.0939067\cdot10^{-5}(g/2\pi)^2.&
\label{eq:coefs}
\eea

A challenge was to rigorously determine the semiclassical expansion for the 
energy of electric flux due to the tunneling through a {\em quantum induced} 
potential barrier. The so-called Path Decomposition 
Expansion,\cite{AuKi} was an important tool that could be tailored to 
this situation.\cite{VbAu,Adri} One required matching the perturbative tail 
of the wave function derived from L\"uscher's effective hamiltonian to the 
semiclassical contribution in the classically forbidden region. The resulting
asymptotic expansion for the energy of one-unit ($|\vek e|=1$) of electric 
flux, $\Delta E=E_0(\vek e)-E_0(\vek 0)$ was found to be\Cite{Vbk1,AnnP} 
\beq
\Delta E=2L^{-1}\lambda B^2g^{5/3}(L)\exp\left[-S/g(L)+\veps_1T/g^{2/3}(L)
\right]\bigl\{1+f(g(L))\bigr\}.\label{eq:delE}
\eeq
Here $S=12.4637$ is the tunneling action and $T=3.9186$ is related to the 
tunneling time, and comes from the classical turning points of the quantum 
induced potential $V(\vek C)$, which also determines $\lambda=0.6997$, due to 
its transverse fluctuations along the tunneling path. It was shown\cite{Kovb} 
that $f(g)=\Order{g}$. The constant $\veps_1=4.116719735$ determines to lowest 
order the mass of the scalar glueball, $m_0=L^{-1}\veps_1 g^{2/3}(L)$, which 
was already computed by L\"uscher and M\"unster.\cite{LuMu} Finally, the 
quantity $B=0.206$ is taken from the asymptotic expansion of the lowest order 
perturbative wave function (in the direction of the vacuum valley). To compute 
$\veps_1$ and $B$ it suffices to consider $H_{\eff}$ in lowest non-trivial 
order, by putting all constants $\alpha_i$ and $\gamma_i$ equal to zero. 

Of more practical importance is to go to the domain where the typical energies 
are above those of the quantum induced potential barrier, when the wave 
functional has spread out over the full fundamental domain, see 
Fig.~\ref{fig:torus}. In this case one will become sensitive to the boundary 
identifications on the fundamental domain. The influence of the boundary 
conditions on the low-lying glueball states is felt as soon as $z=m_0L>0.9$. 
We summarize below the ingredients that enter the calculation.

The choice of boundary conditions, associated with each of the irreducible
representations of the cubic group $O(3,\zahlen)$ and the electric flux quantum
numbers,\cite{Tho1} is best described by observing that the
cubic group is the semidirect product of the group of coordinate permutations
$S_3$ and the group of coordinate reflections $Z_2^3$. We denote the parity
under the coordinate reflection $c_i^a\rightarrow -c_i^a$ ($\forall a$) by 
$p_i=\pm 1$. The electric flux quantum number for the same direction will
be denoted by $q_i=\pm 1$.  This is related to the more usual additive
(mod 2) quantum number $e_j$ by $q_j=\exp(i\pi e_j)$. Note that for SU(2)
electric flux is invariant under coordinate reflections. If not all of the
electric fluxes are identical, the cubic group is broken to $S_2\times
Z_2^3$, where $S_2(\cong Z_2)$ corresponds to interchanging the two
directions with identical electric flux (unequal to the other electric flux).
If all the electric fluxes are equal, the wave functions are irreducible
representations of the cubic group. These are the four singlets $A_{1(2)}^\pm$,
which are completely (anti-)symmetric with respect to $S_3$ and have
$p_1=p_2=p_3=\pm 1$. Then there are two doublets $E^\pm$, also with
$p_1=p_2=p_3=\pm 1$ and finally one has four triplets $T_{1(2)}^\pm$. Each 
of these triplet states can be decomposed into eigenstates of the coordinate 
reflections. Explicitly,\cite{Kovb,Voh1} for $T_{1(2)}^\pm$ we have one state 
that is (anti-)symmetric under interchanging the two- and three-directions, 
with $p_1=-p_2=-p_3=\pm 1$. The other two states are obtained through cyclic 
permutation of the coordinates. Thus, any eigenfunction of the effective 
hamiltonian with specific electric flux quantum numbers $q_i$ can be chosen 
to be an eigenstate of the parity operators $p_i$. The boundary conditions 
of these eigenfunctions $\Psi_{\vek q,\vek p}(c)$ are given by 
\bea
\Psi_{\vek q,\vek p}(c)|_{_{r_i=\pi}}=0\quad,&\quad{\rm if}\quad
p_iq_i=-1 \nonumber\\
{\partial\over\partial r_i}(r_i\Psi_{\vek q,\vek p}(c))|_{_{r_i=\pi}}=0\quad,
&\quad{\rm if}\quad p_iq_i=+1\label{eq:bc}
\eea
and one easily shows that with these boundary conditions the hamiltonian is 
hermitian with respect to the inner product $<\Psi,\Psi^\prime>=\int_{r_i\leq
\pi}d^9c\,\Psi^*(c)\Psi^\prime(c)$. In Sec.~\ref{subsec:SUN}, when discussing
the case of SU(3), we show in more detail how one arrives at these boundary 
conditions. For negative parity states ($\prod_i p_i=-1$) this description is, 
however, not accurate\Cite{Voh1} as parity restricted to the vacuum valley 
is equivalent to a Weyl reflection (a remnant of the invariance under constant 
gauge transformations). One can use as a basis for $\Psi_{\vek p,\vek q}(c)$
\beq
<\vek c|\vek l,\vek n;\vek e>=\sum_{\vek m}W(\vek l,\vek m)\prod_{i=1}^3r_i^{-1}
\chi_{n_i,l_i}^{(e_i)}(r_i)Y_{l_i,m_i}(\theta_i,\phi_i),\label{eq:base}
\eeq
with $Y_{l,m}$ spherical harmonics, and $W(\vek l,\vek m)$ the Wigner
coefficients for adding three angular momenta $\sum_i\vek L_i=\vek 0$
(ensuring gauge invariance). The angular quantum numbers $l_i$ are restricted
to be even or odd, for resp. $p_i=1$ or $p_i=-1$, and the radial wave function
$\chi_{n,l}^{(e)}(r)$ either vanishes, or has vanishing derivative at $r=\pi$
(for resp. $l+e$ even or odd, see Eq.~\Ref{eq:bc}). One computes the matrix 
elements of the effective hamiltonian (Eq.~\Ref{eq:Heff}) and solves for the 
low-lying spectrum by Rayleigh-Ritz (providing {\em also} lower bounds from 
the second moment of the hamiltonian). The first two lines in Eq.~\Ref{eq:Heff}
are sufficient to obtain the mass-ratios to an accuracy of better than 5\%. 

\begin{figure}[htb]
\vspace{4.5cm}
\includegraphics{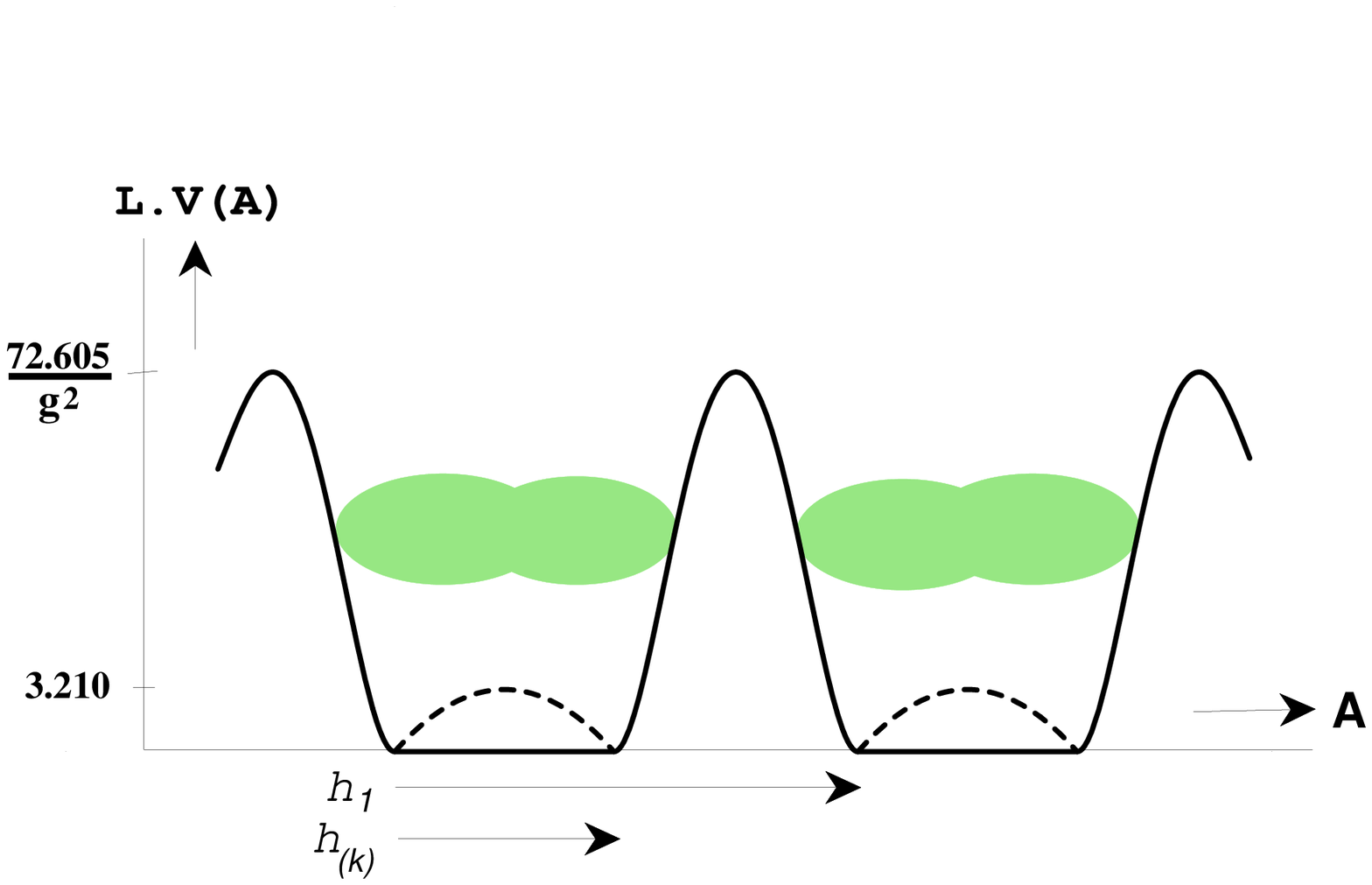}
\caption{Sketch of the potential for the torus. Shown are two vacuum valleys, 
related to each other by a gauge transformation $h_1$, with winding number 
$\nu(h_1)=1$.  The induced one-loop effective potential, of height $3.210/L$, 
has degenerate minima related to each other by the anti-periodic gauge 
transformations $h_{(k)}$. The classical barrier, separating the two valleys, 
has a height $72.605/Lg^2$.}
\label{fig:pot}
\end{figure}

At larger volumes extra degrees of freedom will behave non-perturbatively. We 
know from the existence of gauge transformations with non-trivial winding 
number 
\beq
\nu(h)={1\over 24\pi^2}\int_M\tr((h^{-1}dh)^3),\label{eq:wind}
\eeq
that there exist gauge equivalent vacuum valleys, that can be reached only 
by crossing a saddle point (called the finite volume sphaleron), see 
Fig.~\ref{fig:pot}. Typically this saddle point lies on the tunneling path 
(the instanton), with the euclidean time of the instanton solution playing the 
role of the path parameter. We expect, as will be shown for the three-sphere, 
that the boundary of the fundamental domain along this path in field space 
across the barrier occurs at the saddle point in between the two minima. The 
degrees of freedom along this tunneling path go outside of the space of 
zero-momentum gauge fields and if the energy of a state flows over the barrier,
its wave functional will no longer be exponentially suppressed below the 
barrier and will in particular be non-negligible at the boundary of the 
fundamental domain. Boundary identifications in this direction of field space 
now become dynamically important too. The relevant ``Bloch momentum'' is in 
this case the $\theta$ parameter. Wave functionals pick up a phase factor 
$e^{i\theta}$ under a gauge transformation with winding number one. 

\begin{figure}[htb]
\vspace{7.3cm}
\includegraphics{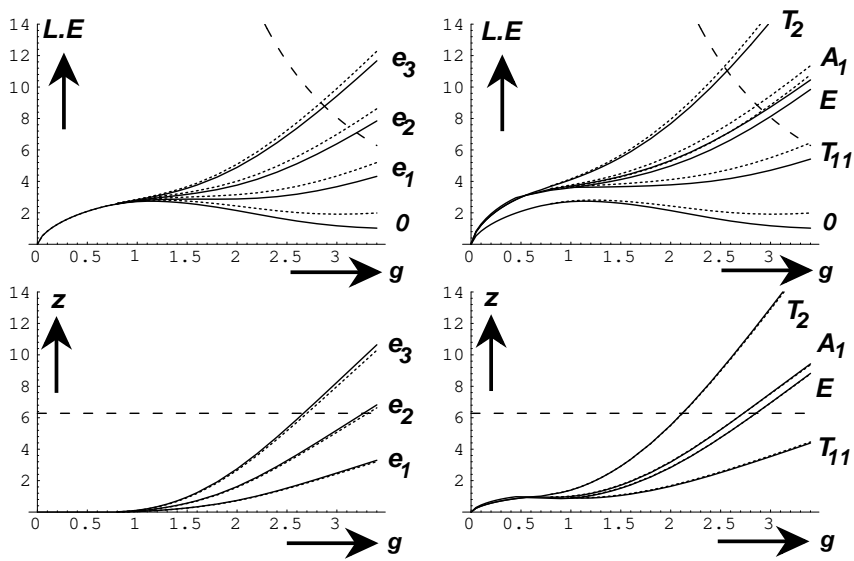}
\caption{The top figures show $L E(g)$ for the relevant (positive 
parity) representations. The dotted lines are without the two loop correction 
included. The dashed curve denotes the barrier height $L E_{sph}(g)$. 
The bottom figures show $z(g)=L m(g)=L(E(g)-E_0(g))$, and the dashed 
line is at $z=2\pi$.} \label{fig:Espec}
\end{figure}

To estimate for which volumes the extra degrees of freedom start to contribute
non-perturbatively, the minimal barrier height that separates two vacuum 
valleys was found to be $E_{sph}(g)=72.605/Lg^2$, using the lattice 
approximation and carefully taking the continuum limit.\cite{Gavb} As long as 
the states under consideration have energies below this value, the transitions 
over this barrier can be neglected (or treated semi-classically if there is no 
perturbative contribution) and the zero-momentum effective hamiltonian provides
an accurate description. One can now find for which volume the energy starts to
be of the order of this barrier height. For this purpose we have collect the 
energy levels as a function of $g$ in Fig.~\ref{fig:Espec}. In the top two 
figures we have plotted $L E(g)$ for the various representations of interest 
(0 stands for the vacuum state, which is in the $A_1^+$ representation of the 
cubic group). The dotted curves are {\em without} the two loop corrections 
included. In the same figures the dashed curve shows the height of the 
instanton barrier. In the lower two figures we show $z=L m\equiv L(E-E_0)$. 
The dashed line is at $z=2\pi$. We see that the sensitivity to the two loop 
corrections almost entirely drops out for the energy differences, at the scale 
of this figure. We now read off that instantons only become important for 
$z_0=5$ to 6, i.e. for $L$ roughly 5 to 6 times the correlation length set by 
the scalar glueball mass. Certain states may actually be less sensitive to 
instantons, in case the representation is such that the wave functional in the 
direction of the barrier is suppressed. This may for example be the case for 
the $T_2^+$ state. It should also be noted that $z=2\pi$ is associated with 
the energy scale $2\pi/L$ of the non-zero momentum field modes that have been 
integrated out. Therefore, $z$ should also not be much bigger than $2\pi$, 
although also here the particular representation may matter.

On the three-torus we have therefore achieved a self-contained picture of the 
low-lying glueball spectrum in intermediate volumes from first principles with 
{\em no free parameters}, apart from the overall scale. For very small volumes
the energy of electric flux vanish and there is an accidental rotational
symmetry, only split by the $\Order{g^{8/3}}$ terms in Eq.~\Ref{eq:Heff}. The 
$2^+$ tensor glueball splits in a nearly degenerate doublet $E^+$ and a triplet
$T_2^+$. Both are lighter than the scalar $0^+$, also denoted by $A_1^+$ as the
scalar singlet representation of the cubic group. Going to larger volumes, 
$L>0.1$ fermi, the energy of electric flux per unit length, which in an 
infinite volume would be the string tension $K$, is surprisingly constant in 
intermediate volumes, whereas the splitting of the tensor states becomes quite 
large. In intermediate volumes the doublet $E^+$ and triplet $T_2^+$ have 
respectively masses of roughly 0.9 and 1.7 times the scalar mass. This doublet 
$E^+$, {\em lighter} than the scalar, was first observed in the lattice studies 
of Berg and Billoire,\cite{BeBi} and caused some stir at the time. 

The lattice data for the triplet\Cite{Mitt} were obtained only after 
our continuum results first appeared\Cite{Kov0,Kovb} and did not confirm the 
predictions for this state. This was resolved by Vohwinkel\Cite{Voh1} by 
observing that the state we had initially identified as $T_2^+$ (taking 
$p_1=p_2=p_3=1$ instead of $p_1=-p_2=-p_3=1$) actually carried two units of 
electric flux (cmp. Eq.~\Ref{eq:bc}), making it even more of a surprise that 
it was found to be even lighter than the doublet $E^+$ in intermediate volumes.
In the infinite volume limit it is pushed out of the low-lying spectrum. 
Around the same time this state (named $T_{11}$) was as such measured on the 
lattice,\cite{KrM2} confirming the proper interpretation of these states.

Electric flux energies (for the trivial representation) are labelled by $e=e_1$ 
for $\vek e=(1,0,0)$, $e_2$ for $(1,1,0)$ and $e_3$ for $(1,1,1)$. For $e_n$ 
we speak of $n$ units of electric flux, sometimes called ``torelon" energies. 
As was argued by 't~Hooft,\cite{Tho1} if a confining string would have formed,
$E_e=K L$ (or $z_e=L E_e=K L^2$), it would be energetically favorable to run 
along the direction of $\vek e$, giving $R_n\equiv E(e_n)/E(e_1)=\sqrt{n}$, 
instead of splitting in separate strings, each winding in the direction for 
which $e_j=1$, which would give $R_n=n$. In intermediate volumes it is the 
latter behavior that we found,\cite{Kov0} confirmed by Monte Carlo 
results.\cite{Berg} One has to go to quite large volumes to start to observe 
the expected $\sqrt{n}$ behavior.\cite{Tore} The same is found for SU(3), in 
a study where a different lattice Monte Carlo method was used to get the 
electric flux energies.\cite{Ha2N}

\begin{figure}[htb]
\vspace{5cm}
\includegraphics{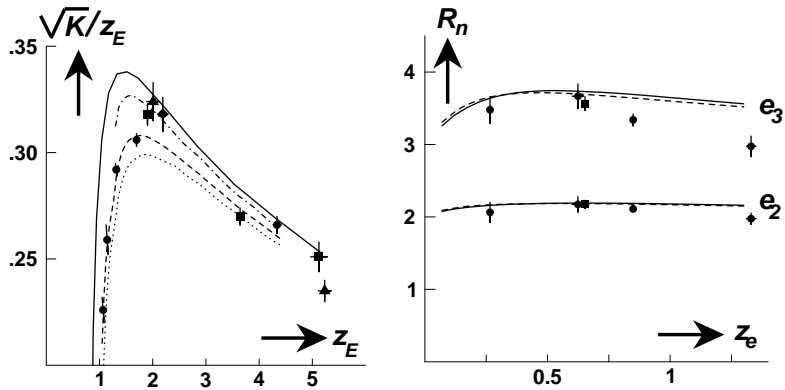}
\caption{Left: The ratio $\protect\sqrt{K(L)}/m_E$ as a function of $z_E=m_E(L)
L$. Full (continuum), dashed-dotted ($6^3$ lattice) and dashed ($4^3$ lattice) 
curves give the hamiltonian results. Rayleigh-Ritz diagonalization errors 
are smaller than the thickness of the lines. The sensitivity to the two loop 
correction can be read off from the dotted curve ($4^3$ lattice), where 
this correction was {\bf not} included. Note the blown-up scale. Right: 
electric flux ratios $R_n$ as a function of $z_e=E_e(L)L$. Monte Carlo 
data\protect\Cite{Berg,BeBR} obtained with the Wilson action on a 
$N^3\times N_t$ lattice ($N_t>6N$), for $N=4$ (dots), 6 (squares), 
8 (triangles) and 10 (diamond).}\label{fig:ratiosBB}
\end{figure}

The lattice Monte Carlo data of Berg and Billoire\Cite{Berg,BeBR} are compared
with the hamiltonian results in Fig.~\ref{fig:ratiosBB}. On the left we show
the ratio $\sqrt{K(L)}/z_E$ as a function of $z_E$. Their methods had 
considerable difficulty in dealing with the scalar glueball state, since it
has the same quantum numbers as the vacuum. To push the results to low values
of $z_E$, they used lattices $N^3\times N_t$ with $N_t=256$ for $N=4$. Please 
note that $z=m_EL$ around 0.95 is nearly a constant (even not single valued) 
function of $g(L)$, see Fig.~\ref{fig:Espec}. This over-emphasizes the steep 
behavior where the wave function starts to spread over the whole vacuum valley,
leading to non-zero energies of electric flux. We took the Monte Carlo data
from Table~V of their paper,\cite{BeBR} but removed those entries with
$N_t/N\leq6$. On the right in Fig.~\ref{fig:ratiosBB} we compare with the 
lattice Monte Carlo data of Berg\Cite{Berg} for the electric flux ratios, but
used where available, the more accurate results\Cite{BeBR} for $z_e=LE_e$. 
The data at $\beta=4/g_0^2=2.7$ from both papers\Cite{Berg,BeBR} were left 
out because they were not consistent with each other. 

\begin{figure}[htb]
\vspace{7.6cm}
\includegraphics{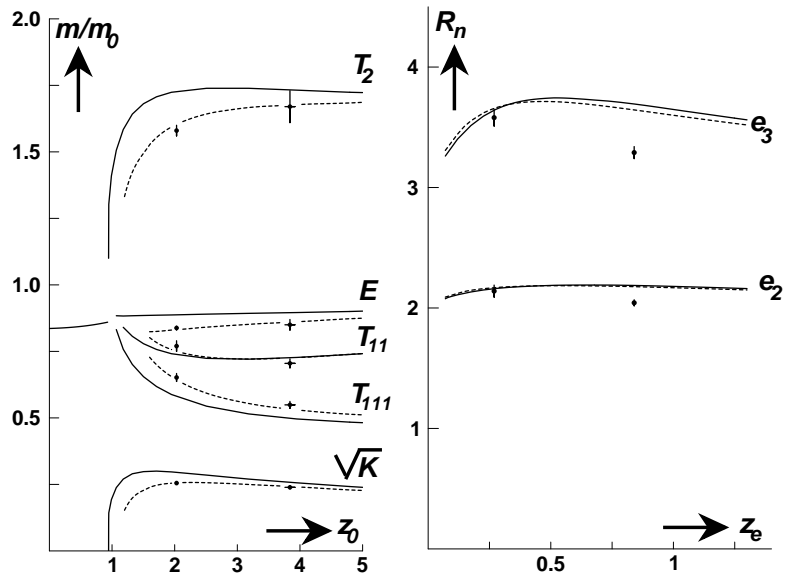}
\caption{Left: mass ratios $m/m_0$ as a function of  $z=m_0(L)L$, with $m_0$
the scalar glueball mass. Full (continuum) and dashed ($4^3$ lattice) curves 
give the hamiltonian results. Shown are string tension $\protect\sqrt{K(L)}$, 
$E^+$ and $T_2^+$ tensor masses and the exotic states $T_{11}$ with two, 
and $T_{111}=T_2(111)$ with three units of electric flux. Right: electric flux 
ratios $R_n$ as a function of $z_e$. Monte Carlo data\protect\Cite{Mich} 
obtained with Wilson action on $4^3\times N_t$ lattice ($N_t=32,\beta=2.4$
and $N_t=99,\beta=3.0$).}\label{fig:ratiosMI}
\end{figure}

Considerable progress was achieved by the so-called variational method, which
allowed much more accurate results, not only for the scalar glueball, but also
for arbitrary other representations.\cite{Mitt} In Fig.~\ref{fig:ratiosMI} we 
present a comparison with the Monte Carlo results of Michael,\cite{Mich} 
obtained for a lattice of spatial size $4^3$, confirmed in a study of improved 
lattice actions.\cite{Impr} The hamiltonian results below $z=0.95$ are due to 
L\"uscher and M\"unster,\cite{LuMu} which is where the spectrum is insensitive
to {\em any} identifications at the boundary of $\Lm$. The electric flux energy
ratios $R_2$ and $R_3$ are shown on the right, cmp. Fig.~\ref{fig:ratiosBB}. 

\begin{figure}[htb]
\vspace{3.9cm}
\includegraphics{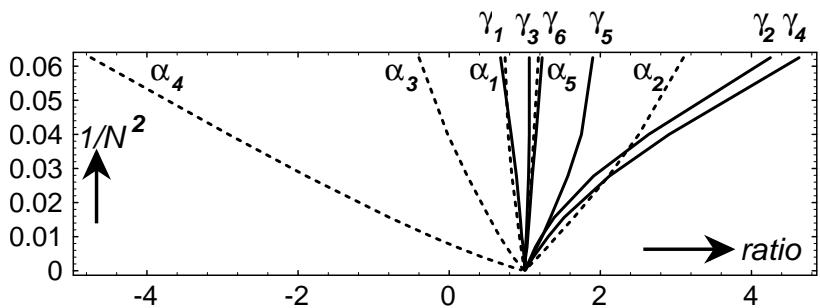}
\caption{Flow of the lattice hamiltonian coefficients to their continuum 
($N\rightarrow\infty$) values. Shown are $\alpha_i(N)/\alpha_i$ (dashed 
lines) and $\gamma_i(N)/\gamma_i$ (full lines) as a function of the square 
of the lattice spacing $(a/L)^2=1/N^2$.}\label{fig:flow}
\end{figure}

We note that the solid curves that represent the continuum results, which 
were reproduced by Berg and Vohwinkel,\cite{BeVR,Voh4} deviate significantly 
from the lattice data. Initially the lattice data were not accurate enough to 
show this deviation. Even though one should not expect a $4^3$ lattice to be 
an accurate approximation for the continuum, it was cause for some doubt that 
the approximations made in the continuum studies were not under control as 
well.\cite{BeBR} To settle this issue we redid the complete derivation of the 
effective hamiltonian starting from the lattice theory, {\em without taking the
continuum limit}.\cite{Latt,Vba4} The hamiltonian is basically of the same form 
as in Eq.~\Ref{eq:Heff}, except that the coefficients in Eq.~\Ref{eq:coefs} 
depend on the lattice spacing (some extra corrections appear, e.g. to correct 
for the discrete time evolution). One can follow the renormalization group 
flow of the hamiltonian to its continuum fixed point in this formalism in all 
detail, see Fig.~\ref{fig:flow}. Using the same analysis as in the continuum 
leads for a finite lattice to the dashed curves in 
Figs.~\ref{fig:ratiosBB},\ref{fig:ratiosMI}. The lattice data now agree 
perfectly, up to a volume of about 0.75 fermi, the regime for which we have 
shown that the effective hamiltonian in the zero-momentum modes should provide 
a good approximation. The deviations for $R_2$ and in particular for $R_3$
are in accordance with the fact that these are of relatively high energy,
and therefore expected to be sensitive to other non-perturbative effects
that invalidate integrating out all the non-zero momentum modes, see
Fig.~\ref{fig:Espec}.

\subsection{General Gauge Group}\label{subsec:SUN}

The question of extending the previous results to SU(3) is a natural one in 
the light of QCD. The perturbative expansion,\cite{Lue1,WeZi} vacuum-valley 
effective potential,\cite{Lue1,AnnP} and the semi-classical evaluation of the 
energy of electric flux\Cite{SU3E} (due to the tunneling through the quantum 
induced vacuum-valley effective potential) are more or less straightforward. 
The non-trivial problem of formulating the appropriate boundary conditions on 
the boundary of the fundamental domain for SU(3) was solved by Vohwinkel and 
qualitative agreement with the lattice data was found.\cite{Voh2,Voh4} With the
results for SU(3) in hand generalization to SU($N$) was achieved.\cite{Vba3} 
The formalism sketched in Sec.~\ref{sec:copies} was not yet developed in those 
early days. What is now called the fundamental domain was then called the unit 
cell. Although the argumentation was more cumbersome, the underlying principles
were the same. 

A simple approximation for L\"uscher's effective hamiltonian in terms of the 
zero-momentum gauge fields $A_j(\vek x)=ic_j^a T_a/L$ ($T_a$ the hermitian 
generators of SU($N$) and $f_{ij}\equiv L^2F_{ij}(\vek x)$ the dimensionless
field strength) is given by 
\beq
H_{\eff}(c)={-g^2\over 2L(1+\alpha_1 g^2)}\sum_{i,a}{\partial^2\over \partial{
c_i^a}^2}-{1\over 2L}({1\over g^2}+\alpha_2)\sum_{i,j}\tr(f_{ij}^{\,2})+V_1(c),
\label{eq:HSUN}
\eeq
with
\beq
V_1(c)={2\over L}\sum_{\vek p\neq\vek 0}\left(\sqrt{\Tr_{\ad}(2\pi\vek p+
\vek c^a\ad\,T_a)^2}-\sqrt{\Tr_{\ad}(2\pi\vek p)^2}\right),\label{eq:VSUN}
\eeq
where $g$ is the renormalized coupling constant at the scale $\mu=1/L$. In a 
perturbative expansion,\cite{Lue1,WeZi} this gives the correct result up to 
$\Order{g^{8/3}}$. Along the vacuum valley, parametrized by $A_j(\vek x)=
i\sum_{b=1}^{N-1}C_j^b T_b/L$, where the first $N-1$ generators are assumed 
to commute (forming a basis of the Cartan subalgebra $H_G$), the effective 
potential $V_1(c)$ is exact to one loop order. A more explicit result can
be found in Sec.~\ref{subsec:quarks}, Eq.~\Ref{eq:Vquark}. As for SU(2), one 
does not include the $\vek p=\vek 0$ term, since only the non-zero momentum 
modes are to be integrated out. 

The effective potential $V_1(c)$ only depends on the Casimir invariants
\beq
r_i^2=2\tr[(c_i^aT_a)^2]\,,\quad s_i=4\tr[(c_i^aT_a)^3]\,,\quad\cdots,
\label{eq:Cas}
\eeq
where the sum over {\em all} color indices is implicit. There are as many 
independent Casimir invariants as the rank of the gauge group. For SU(2) only 
$r_i$ will be non-trivial. These coordinates are uniquely related to those 
obtained by restricting the zero-momentum gauge field to be abelian. For SU(2),
$c_i^aT_a=C_i\tau_3/2$ yields $r_i^2=C_i^2$, whereas for SU(3), in terms of 
the Gell-Mann matrices, $c_i^aT_a=(C^1_i\lambda_8+C^2_i\lambda_3)/2$ gives 
$s_i=C^1_i\sqrt{3}((C^2_i)^2-(C^1_i)^2/3)$ and $r_i^2=(C^1_i)^2+(C^2_i)^2$.
In this manner the effective potential on the set of abelian zero-momentum 
modes can indeed be minimally extended in a unique way to all constant gauge 
fields. 

By adding the zero-momentum ($\vek p=\vek 0$) contribution to the one-loop 
effective potential,
\beq
V_{\eff}(c)=V_1(c)+2L^{-1}\sqrt{\Tr_{\ad}[(\vek c^aT_a)^2]}\,,\label{eq:VeffN}
\eeq
restricted to the vacuum valley gives the appropriate effective potential
when integrating out all the field modes orthogonal to the vacuum valley. Its 
symmetries are a consequence of gauge invariance, which can be divided in two 
classes. The constant gauge transformations, that leave $H_G$ invariant. This 
is represented by the Weyl group $\Ss{W}$ acting on $\vek C^b$ (for SU(2) 
$\vek C\rightarrow -\vek C$). The other class of gauge transformations that 
leaves the set $\vek C^b$ invariant are of the form $h_{\vek\Theta}(\vek x)=
\exp(2\pi i\vek x\cdot\vek\Theta/L)$, where $\Theta_i\in H_G$ such that $h$ 
does not affect the periodic boundary conditions on the gauge fields. These 
lead to shift symmetries on $\vek C^b$. The associated lattice of $\Theta_j$ 
corresponds to the dual root lattice $\tilde\Lambda_r$, as follows from the 
condition that $\exp(2\pi i\Theta_j)=1$. The vacuum valley, i.e. the moduli 
space of flat connections ($F_{ij}=0$), corresponds to the orbifold $[H_G/2\pi
\tilde\Lambda_r]^3/\Ss{W}$. Recently it has become clear that for orthogonal 
and exceptional Lie-groups there are other, disconnected, components in the 
moduli space of flat connections, which we will briefly discuss in 
Sec.~\ref{subsec:susy}.

A crucial role is played by the twisted gauge transformations, only periodic 
up to an element of the center $Z_N$ of the gauge group.\cite{Tho1} These can 
be realized with $h_{\vek\Theta}$, but now $\Theta_j$ belongs to the dual 
weight lattice $\tilde\Lambda_w$, which follows from the requirement that 
$\exp(2\pi i\Theta_j)$ is an element of the center $Z_G$ of the gauge group. 
Indeed, $\tilde\Lambda_w/\tilde\Lambda_r\cong Z_G$. These twisted gauge 
transformations generate a shift symmetry associated with $\tilde\Lambda_w$.
Their homotopy type is specified by $\vek\Theta\in(\tilde\Lambda_w/\tilde
\Lambda_r)^3\cong Z_G^3$. A particular representative for SU($N$) is given by 
$h_{\vek k}(\vek x)=\exp(2\pi i \vek k\cdot\vek x\Theta_0/L)$, where 
$\Theta_0\in H_G$ is a generator for the center, $\exp(2\pi i\Theta_0)=
\exp(2\pi i/N)$. For SU(2) we can choose $\Theta_0= \tau_3/2$ and for SU(3) 
there are two independent choices (that span the dual weight lattice $\tilde
\Lambda_w$) $\Theta_0=\lambda_8/\sqrt{3}$ and $\Theta_0'=\half(\lambda_3-
\lambda_8/\sqrt{3})$. The non-trivial homotopy is labelled by $\vek k\in Z_N^3$
and is thus in general non-trivially represented on the wave functionals. Any 
gauge transformation can be decomposed\Cite{CMP0} in $h=h_\vek k h_1^\nu(h) 
h_0$, where $h_1$ is a particular strictly periodic gauge transformation with 
unit winding number, $\nu(h_1)=1$. What is left is a homotopically trivial 
gauge function $h_0$, under which the wave functional is invariant. 
We therefore have
\beq
\Psi([h]A)=\Psi([h_{\vek k}h_1^\nu]A)=\exp(2\pi i\vek e\cdot\vek k/N)
\exp(i\theta\nu)\Psi(A),\label{eq:bloch}
\eeq
where $\theta$ is the usual vacuum parameter associated with instantons and
$\vek e$ (defined modulo $N$) is the gauge invariant definition of electric 
flux.\cite{Tho1} As for SU(2) the electric flux quantum number can be 
implemented within the zero-momentum effective hamiltonian by imposing
suitable boundary conditions on the boundary of the fundamental domain.
For this we study the action of $h_{\vek k}$ on $\vek C^b$, corresponding 
to a shift. For SU(2) $\vek C\rightarrow\vek C+2\pi\vek k$ and for SU(3) 
$(\vek C^1,\vek C^2)\rightarrow(\vek C^1+2\pi(2\vek n-\vek l)/\sqrt{3},
\vek C^2+2\pi\vek l)$, with $\vek k=\vek n+\vek l$ (mod 3) specifying the 
homotopy type of the gauge transformation.

\begin{figure}[htb]
\vspace{6.3cm}
\includegraphics{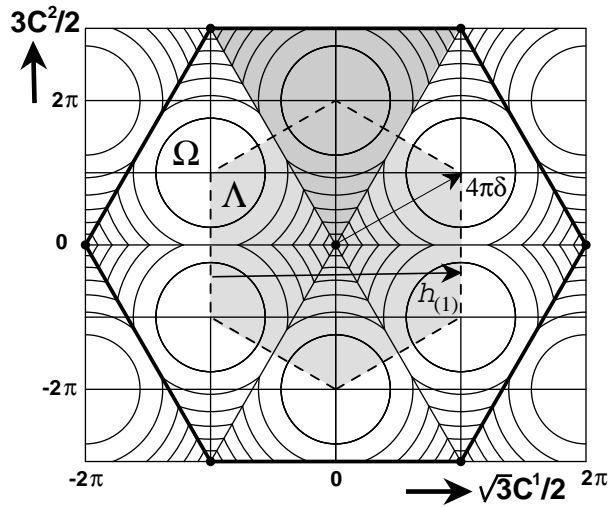}
\caption{Two dimensional cross-section of $V_{\eff}$, for $C_1^bT_b=
C_1^1\lambda_8/2+C_1^2\lambda_3/2$ and $C_{2,3}^b=0$. The circles indicate the 
equipotential lines (in increments of $V_{\eff}(4\pi\delta)/7$). The light
shaded area represents the fundamental region, with the dashed hexagon as its 
boundary, the fat hexagon indicates the Gribov horizon, with gauge copies of 
$A=0$ indicated by the dots. The darker shaded area is the so-called 
fundamental Weyl chamber.}\label{fig:webb}
\end{figure}

Fig.~\ref{fig:webb} gives for SU(3) the equipotential lines of the effective 
potential $V_{\eff}$, in a plane specified by putting $C_2^b=C_3^b=0$, the 
fundamental domain $\Lambda$ (bounded by the dashed hexagon) and the Gribov 
region $\Omega$ (bounded by the fat hexagon). The triangular lattice structure 
is due to the invariance under twisted gauge transformations and reflects the 
dual weight lattice of SU(3). Indeed, the gauge transformations $\exp(2\pi 
i\Theta_0 x_j/L)$ and $\exp(2\pi i\Theta_0' x_j/L)$ all map points on the 
boundary of the fundamental domain to the same boundary. They also map 
$\vek C^b=0$ to the Gribov horizon. The Weyl transformations are generated 
by the reflections in the three principle axes of the dual weight lattice. 

The near spherical behavior of the equipotential lines within each triangle 
is due to a remarkably accurate approximation (cmp. Fig.~\ref{fig:Vquark}) that 
holds for any SU($N$) when restricting $C_i^b$ to one dimension, say 
$C_1^b=C^b$ (normalizing as usual $\tr(T_aT_b)=\half\delta_{ab}$) 
\beq
V_{\eff}(C)\approx N\left((4\pi\delta)^2-(C-4\pi\delta)^2\right)/2\pi L,
\label{eq:Vappr}
\eeq
for $C$ in the Weyl chamber restricted to the fundamental domain, centered 
around $4\pi\delta$, where $\delta$ is the highest weight. For SU(2) $\delta=
1/4$ and for SU(3) $\delta=(\delta^1,\delta^2)=(\sqrt{3},1)/6$. 
It is also instructive from Fig.~\ref{fig:webb} to identify the orbifold 
singularities, along the axes of the dual weight lattice (edges of the Weyl 
chambers), which are fixed under the Weyl reflections. This is where the 
$U(1)^2$ symmetry in a generic point of the vacuum valley is restored to SU(2), 
whereas at the corners (all gauge copies of $A=0$) the full SU(3) symmetry is 
restored. At these locations also some of the directions in which the classical 
potential is quadratic turn quartic. It is this that gives rise to the conic 
singularities in $V_{\eff}$. The effective potential $V_1$ in Eq.~\Ref{eq:VSUN},
when restricted to the fundamental domain, is free from any of these 
singularities, since they are caused by fluctuations in the non-abelian 
constant modes, which are not integrated out in Eq.~\Ref{eq:VSUN}. The 
conic singularities of $V_{\eff}$ in $\Lambda$ are therefore all described 
by $2L^{-1}\sqrt{\Tr_{\ad}[(\vek C^bT_b)^2]}=2NL^{-1}|\delta_b\vek C^b|$, 
see Eq.~\Ref{eq:VeffN}.

To implement the symmetries on the wave functional we perform a change of 
coordinates $c_i^a\rightarrow(C^b_i,\Omega_i)$, where $\Omega_i$ stands for the
collection of SU($N$)-angular coordinates parametrizing SU($N$)/U$(1)^{N-1}$ 
and $(C^a_i)$ are restricted to a fundamental Weyl chamber, for SU(3) the 
triangular shaded region in Fig.~\ref{fig:webb}, where the relation between 
$(C^1,C^2)$ and $(r,s)$ is one to one. For SU(2) these new coordinates are the 
spherical coordinates $(r_i,\Omega_i)$, with $\Omega_i=(\theta_i,\phi_i)$ 
coordinates on $S^2=SU(2)/U(1)$. The square root of the Jacobian of this 
transformation, is given by $J=\prod_iJ_i$, with $J_i=C_i$ for SU(2) and 
$J_i=C^2_i((C^2_i)^2-3(C_i^1)^2)$ for SU(3). The wave functions can be 
decomposed as
\beq
\Psi(c)=\prod_i J_i^{-1}(C_i^b)\chi_i^\phd(C_i^b)
\Ss{Y}_i(\Omega_i).\label{eq:decom}
\eeq
The ``radial'' part $\chi$ will be antisymmetric with respect to Weyl
reflections, so as to cancel the zero's of the Jacobian. For SU(2) the angular 
wave functions $\Ss{Y}(\Omega)$ are nothing but the spherical harmonics, see 
Eq.~\Ref{eq:base}, whereas in general they are irreducible representations of 
the gauge group. Helpful in making suitable choices for $\chi$ is that the 
vacuum valley kinetic term after the rescaling with the square root $J$ of 
the Jacobian becomes $-\half g^2\sum_{i,a}(\partial/\partial C_i^a)^2$, 
compatible with the canonical flat metric on the orbifold. Furthermore, as 
is familiar from the radial reduction in three dimensions, this rescaling does 
not create an additional potential term, since $J$ is in all cases harmonic,
$\sum_{i,a}\partial^2 J/(\partial C_i^a)^2=0$.

We note that suitable combinations of a shift and Weyl reflection leave 
invariant the lines that constitute the polygon at the boundary of the 
fundamental domain, see Figs.~\ref{fig:torus} and \ref{fig:webb}. This implies 
that alternatively the properties of $\chi$ can be described in terms of 
boundary conditions at the boundary of the fundamental domain. E.g., for SU(2) 
this leads to two possible choices of boundary conditions at $r_j=\pi$, for 
each $j$, see Eq.~\Ref{eq:bc}. One can use $\chi(C)=\sin(nC/2)$, although 
spherical Bessel functions provide a more efficient choice, keeping the 
hamiltonian more sparse.\cite{Kovb} Carefully working out the consequences 
of the symmetries one can show\Cite{Voh2,Vba3} that a complete basis for 
$\chi$ in the case of SU(3) is given by
\bea
\chi(C^1,C^2)&=&\sin\left(mC^2/2\right)\exp\left(inC^1\sqrt{3}/2\right)
\label{eq:wave}\\&+&\sin\left(m(\sqrt{3}C^1-C^2)/4\right)\exp\left(in\sqrt{3}
(\sqrt{3}C^1+C^2)/2\right)\nonumber\\&-&\sin\left(m(\sqrt{3}C^1+C^2)/4\right)
\exp\left(in\sqrt{3}(\sqrt{3}C^1-C^2)/2\right),\nonumber
\eea
for each of the three coordinate directions. The quantum numbers $n$ and $m$ 
will be restricted by the electric flux and irreducible representations of 
the Weyl and cubic groups, which is the part that requires most of the care. 
Finally if one restricts $\prod_i\Ss{Y}_i$ to transform as a singlet under 
SU($N$) one obtains a complete and gauge invariant basis for the effective 
hamiltonian, that through the boundary conditions carries the information 
of electric flux. Needless to say that for SU(3) the computation of all the 
relevant matrix elements\Cite{Voh2} is rather more cumbersome than for SU(2). 
However, once the matrix of the hamiltonian for a particular basis is computed 
one performs a simple Rayleigh-Ritz analysis to determine the spectrum. The 
region where the wave functional spreads out over the whole vacuum valley, with 
the energy of electric flux no longer suppressed by the quantum induced 
barrier of $V_{\eff}$, is well described by this Rayleigh-Ritz analysis, based
on Eq.~\Ref{eq:wave}. Of course, also for SU(3) the results are valid as long 
as the classical barrier is sufficiently high as compared to the energy-levels 
studied, cmp. Fig.~\ref{fig:pot}. Like for SU(2) the approximations break down 
for $L$ large than 5 to 6 times the correlation length of the scalar glueball. 
That beyond this volume, $L\sim 0.75$ fermi, instanton effects set in can be 
seen from the rather sudden onset of the topological susceptibility.\cite{Hoek}

\subsection{Including Massless Quarks}\label{subsec:quarks}

Quarks fields are not invariant under the center of the gauge groups. This 
means that on the space of zero-momentum abelian gauge fields, which still 
form the classical vacuum valley, the twisted gauge transformations no 
longer represent a symmetry. This complicates matters since without the
equivalence under twisted gauge transformations the fundamental domain 
extends to the Gribov horizon. Some of these complications may be avoided 
by introducing on the original fundamental domain additional quark fields, 
obtained by applying a twisted gauge transformation. The boundary 
identification on the boundary of the fundamental domain are include
an operation that permutes these fermion field components. This has not 
been worked out so far.

Nevertheless, interesting statements can be made about the vacuum structure 
in small enough volumes, for which the wave functional is sufficiently localized
around the vacuum configuration. One simply adds in one loop order the quantum 
effects of the quark field fluctuations. The resulting effective potential will
no longer respect the center symmetry, but it still properly reflects 
invariance under constant and periodic gauge transformations. The quark fields 
can satisfy either periodic or anti-periodic spatial boundary conditions. 
Actually, for SU(2) (with $-1$ a non-trivial element of $Z_2$) these are 
equivalent by a twisted gauge transformation with homotopy type $\vek k=
(1,1,1)$. Under this gauge transformation the gauge field is shifted and shows 
it is not a priori clear that $A=0$ will represent the proper classical vacuum 
to expand around. As we will show, it will be the correct one {\em only} with 
anti-periodic boundary conditions for the quark fields, both for SU(2) and 
SU(3). In that case, due to the anti-periodicity, there will be no zero-energy
modes for the quark fields, and chiral symmetry is {\em unbroken} in the finite 
volume. For SU(3) no gauge equivalence of periodic to anti-periodic boundary 
conditions holds, and the vacuum structure with periodic quark fields 
actually leads to spontaneous breaking of some discrete symmetries. Yet,
no zero-energy quark modes appear, and chiral symmetry remains unbroken.
It also means that in a small volume, with quark momenta of the order $\pi/L$ 
and glueball masses of order $g^{2/3}(L)/L$, that glueballs cannot decay
in mesons. The quark degrees of freedom can be integrated out.

\begin{figure}[htb]
\vspace{2.7cm}
\includegraphics{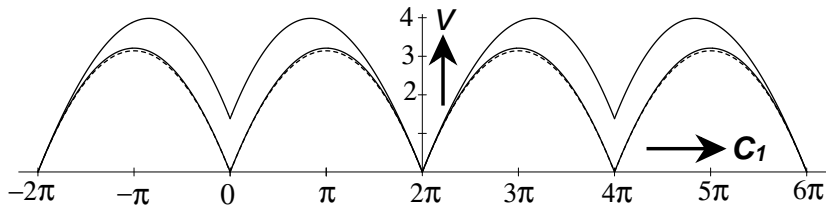}
\caption{Vacuum-valley effective potential at $C_2=C_3=2\pi$ for SU(2) and 
two flavors of periodic massless quarks (Eq.~\protect\Ref{eq:Vquark} with 
$\vek k=\vek 0$), normalized to 0 at its minimum, as well as the effective 
potential for zero flavors (Eq.~\protect\Ref{eq:VSU2}). The dashed curve 
represents Eq.~\protect\Ref{eq:Vappr}.}\label{fig:Vquark}
\end{figure}

To be more specific let us first generalize the computation of the 
vacuum-valley effective potential to include the quark field fluctuations. 
The most efficient way to represent the result is to introduce the weight 
vectors $\mu^{(i)}$, determined by the eigenvalues of the abelian generators, 
\beq
T_b=\diag(\mu^{(1)}_b,\mu^{(2)}_b,\cdots,\mu^{(N)}_b),\quad T_b\in H_G.
\label{eq:weights}
\eeq
For SU(2) one finds $\mu^{(1)}=\half(1,-1)$, whereas for SU(3) $\mu^{(1)}=
(1,1,-2)/\sqrt{12}$ and $\mu^{(2)}=\half(1,-1,0)$ (the conventions used in this
paper are $T_1=\half\lambda_8$ and $T_2=\half\lambda_3$). With $n_f$ flavors of
massless quark fields we find\Cite{Ferm} (see Fig.~\ref{fig:Vquark})
\beq
V_{\eff}^{\vek k}(\vek C^b)=\sum_{i>j}V(\vek C^b[\mu^{(i)}_b-\mu^{(j)}_b])
-n_f\sum_{i}V(\vek C^b\mu^{(i)}_b+\pi\vek k),\label{eq:Vquark}
\eeq
with $\vek k=\vek 0$ or $\vek k=(1,1,1)$, for resp. periodic or anti-periodic 
boundary conditions on the quark fields. The function $V(\vek C)$ is the SU(2)
one-loop effective potential for $n_f=0$, Eq. \Ref{eq:VSU2}. The correct
quantum vacuum is to be found at the minimum of this effective potential. 

Observe that the gauge symmetry should not be spontaneously broken, which 
implies that the Polyakov loop observables
\beq
P_j(\vek x)\equiv\frac{1}{N}\tr\left(P\exp\,(\int_0^Lds\,A_j(\vek x+s
\vek e^{(j)})\right),\label{eq:Pol}
\eeq
should be a constant center element at the vacuum configurations, or 
\beq
P_j=\frac{1}{N}\tr\left(\exp(iC_j^bT_b)\right)=\frac{1}{N}\sum_n
\exp(i\mu^{(n)}_bC^b_j)=\exp(2\pi i l_j/N).\label{eq:PVac}
\eeq
This implies that $\mu^{(n)}_b\vek C^b=2\pi\vek l/N$ (mod $2\pi$), independent 
of $n$, and gives $V_{\eff}^{\vek k}=-n_fNV(2\pi\vek l/N+\pi\vek k)$. In the 
case of anti-periodic boundary conditions, $\vek k=(1,1,1)$, this is minimal 
only when $\vek l=\vek 0$ (mod $2\pi$). This means the quantum vacuum in this 
case is the naive one, $A=0$ ($P_j=1$). In the case of periodic boundary 
conditions, $\vek k=\vek 0$, the above candidate vacua have $\vek l\neq\vek 0$,
that is $P_j$ correspond to non-trivial center elements. Both for SU(2) and 
SU(3) this means $l_j=\pm1$. For SU(2) the vacuum with $P_j=-1$ is unique, as 
also follows from the gauge equivalence argument given above. The only 
difference is that one now needs to expand around $\vek C=2\pi(1,1,1)$. For 
SU(3), however, there are now 8 possible choices $P_j=\exp(\pm 2\pi i/3)$, 
related by the three coordinate reflections. As this is a symmetry of the full 
hamiltonian, each is indeed equivalent. But it does mean that in a small volume
parity (P) and charge conjugation (C) are spontaneously broken, although CP is 
still a good symmetry.\cite{Ferm} A consequence of the spontaneous breaking of 
parity is that the mass gap, the lowest excitation above the vacuum, is 
exponentially small in a small volume. All these intricate effects would make 
this an ideal testing ground for dynamical fermion algorithms in lattice gauge 
theory. 

The minima of the effective potential are obtained from $A=0$ by the twisted 
gauge transformation $h_{\vek l}(\vek x)$. As there are no zero-energy quark 
field modes, also the effective hamiltonian can be expressed as in the bosonic 
case in terms of the zero-momentum gauge fields, after taking this shift into 
account. In the case of SU(3) with periodic boundary conditions, because of 
the spontaneous break down of parity and charge conjugation invariance, extra 
terms appear in L\"uscher's effective hamiltonian. To the order in which this 
hamiltonian was worked out the new interaction $\tr(\{A_i,A_j\}A_k)$ appears.
In addition couplings that were equal because they were related by the discrete
symmetries, now spontaneously broken, will split.\cite{Ferm} All these 
corrections are linear in the number ($n_f$) of quark flavors. The flavor 
dependence of the effective hamiltonian for the case of anti-periodic boundary 
is simpler, since no new couplings appear. For SU(2) one simply replaces 
$\alpha_i$ and $\gamma_i$ in Eq.~\Ref{eq:Heff} by $\alpha_i-2n_f\alpha_i'$ and 
$\gamma_i-2n_f\gamma_i'$, with 
\bea
\gamma_1'=-2.1272012\cdot10^{-2},&&\alpha_1'=+3.098211\cdot10^{-5},\nonumber\\
\gamma_2'=+4.2255250\cdot10^{-4},&&\alpha_2'=+1.7211922\cdot10^{-3},\nonumber\\
\gamma_3'=-7.3994300\cdot10^{-4},&&\alpha_3'=+3.0178786\cdot10^{-5},\nonumber\\
\gamma_4'=-2.8659656\cdot10^{-6},&&\alpha_4'=+3.2156523\cdot10^{-5},\nonumber\\
\gamma_5'=+1.1578663\cdot10^{-5},&&\alpha_5'=-3.2271736\cdot10^{-5},\nonumber\\
\gamma_6'=-7.9447492\cdot10^{-5}.&&\label{eq:coefq}
\eea
Independently Kripfganz and Michael calculated for SU($N$) to $\Order{g^{8/3}}$ 
the change in the coefficients of L\"uscher's effective hamiltonian, due to 
quarks with anti-periodic boundary conditions only.\cite{KrM1} They confirmed
the values in Eq.~\Ref{eq:coefq} and also introduced for SU(2) a lagrangian 
formulation of the effective hamiltonian in terms of compact group variables, 
that incorporates in a simple way the proper boundary conditions in field 
space.\cite{KrM2} After full equivalence was established,\cite{Fer2,VbKr} the 
Monte Carlo analysis of this effective lagrangian model continued to suffer 
from a technical difficulty in efficiently implementing the kinetic 
term,\cite{Mich} only fully understood by Vohwinkel a number of years 
later.\cite{Voh3} This has hampered using the lagrangian formulation as a 
reliable alternative\Cite{Tied} for the hamiltonian Rayleigh-Ritz analysis.

There is another choice of boundary conditions that strongly reduces the center
symmetry. This is called C periodic boundary conditions, where the field is 
periodic up to a charge conjugation. The boundary conditions can be used 
to avoid the system to be charge neutral, as is the case for periodic 
boundary conditions, both for magnetic and electric charge. This was the 
original motivation to introduce these boundary conditions for the abelian 
theory,\cite{PoWi} which were subsequently studied for non-abelian gauge 
theories.\cite{KrW0} For SU(2), which is pseudo real, one retrieves the 
periodic case, but for SU(3) the center symmetry is completely broken and
a number of the features we saw when including quark fields, appear here as 
well. The vacuum valley for SU(3) is reduced from six to three dimensional
in terms of the real gauge field $\vek A=\half i\vek C\lambda_2/L$, 
with\Cite{KrW1}
\beq
V_{\eff}(\vek C)=V(\half\vek C)+V(\vek C-\pi\vek l)+V(\half\vek C-\pi\vek l),
\label{eq:VCper}
\eeq
where $V(\vek C)$ is the SU(2) effective potential, see Eq.~\Ref{eq:VSU2},
and $\vek l=(1,1,1)$. The minimum of this effective potential occurs at the
four points $\vek C=(\pi,\pi,\pi)$, $(-\pi,-\pi,\pi)$, $(\pi,-\pi,-\pi)$, and 
$(-\pi,\pi,-\pi)$, which correspond to orbifold singularities with quartic 
modes (associated with the three real generators $i\lambda_A/2$, for $A=2$, 5 
and 7, forming an SU(2) subalgebra). The effective hamiltonian is again of 
the L\"uscher type, at $\Order{g^{2/3}}$ identical to it, at higher order 
additional couplings appear because of the spontaneous breaking of parity,
i.e. the cubic group is broken down to the permutation group. No attempts
have been made to study the fundamental domain for this theory, and we
may expect similar difficulties as in the presence of quark fields.

\subsection{The Renormalized Coupling}\label{subsec:running}

We have assumed there is a renormalized coupling in terms of which perturbation
theory in the field modes that are integrated out is well-behaved. By 
expressing quantities in dimensionless combinations, lattice Monte Carlo 
results and continuum (or lattice) hamiltonian results can be compared without 
being sensitive to any problems in expressing the renormalized coupling in 
terms of the bare coupling. Determining the renormalized strong coupling 
constant non-perturbatively in a reliable way is, however, an important 
problem. The integration constant of the beta-function, the so-called Lambda 
parameter, ideally should be fixed in terms of the infrared quantities of the 
theory, like the mass gap and string tension, or other observables in the 
low-lying spectrum of the theory. The running of the coupling allows one to 
compute unambiguously the strong coupling constant at, say the Z-boson mass. 
The most accurate such method is based on a finite volume study, proposed by 
L\"uscher,\cite{LuRC} long before it was feasible to be implemented.\cite{LuWW}
It makes use of a discrete version of the beta-function, the so-called step 
scaling function. The scale at which the renormalized coupling is defined is 
fixed by the volume. The volume is subsequently changed by an integer factor 
(usually, but not necessarily) of 2. So instead of an infinitesimal scale 
transformation it considers a finite one. The change in the coupling can of 
course be obtained by intergrating the beta-function, but this function is not 
available non-perturbatively. Instead, one picks a suitable definition of the 
renormalized coupling constant at the scale $L$, set by a given physical 
volume, than doubles $L$ and calculates the value of the coupling at this new 
scale $2L$. This can all be performed on the lattice (hence the integer scaling 
factor) using Monte Carlo calculations, at each step carefully extrapolating 
to the continuum limit. Also (euclidean) time is taken finite, $T=L$. Small 
volumes go together with high temperatures in such geometries. The naive 
strategy of taking $L$ large, and defining an observable set by a variable 
scale {\em within the same} finite volume, fails because the lattice spacing 
gives a limit to the shortest distance one can probe in a given volume, due to 
computer limitations.

Many definitions of the renormalized coupling could in principle be used, but 
technical requirements have led to a particular one that is related to the 
effective action in the background of a constant chromo-electric field, based 
on the so-called Schr\"odinger functional\Cite{LNWW} (SF). In the spatial 
directions the boundary conditions are periodic, but in the (finite) time 
direction one prescribes an initial and a final configuration of gauge 
fields taking values in the vacuum valley.\cite{LNWW,LSWW} 
\bea
{\rm SU(2)}:&&C_j(t=0)=2\pi\delta+2\eta,\quad C_j(t=T)=2\pi-C_j(t=0),\nonumber\\
{\rm SU(3)}:&&(C^1_j,C^2_j)(t=0)=2\pi\delta+(-\sqrt{3}\eta,0),\\
    &&(C^1_j,C^2_j)(t=T)=(4\pi/\sqrt{3},0)-(C^1_j,C^2_j)(t=0),\quad\forall j,
\nonumber
\eea
with $\delta$ the highest weight, defined below Eq.~\Ref{eq:Vappr}.
The classical equations of motion lead to a linear interpolation in time, 
giving rise to a constant chromo-electric background. The euclidean quantum 
effective action, $\Gamma(\eta)$, describes the reaction of the system to 
this background. The renormalized coupling is defined by $g^2(L)=\frac{d
\Gamma(\eta)}{d\eta}/\frac{d\Gamma_0(\eta)}{d\eta}$ evaluated at $\eta=0$, 
where $\Gamma_0(\eta)$ is the bare effective action. This background has 
been chosen to stay well away from the orbifold singularities along the vacuum 
valley for the entire classical path that interpolates between the initial and 
final configuration, to simplify the perturbative analysis (in part for 
estimating the lattice artifacts). This particular choice of coupling fits 
{\em extremely well} to the perturbative running of the coupling constant up 
to the largest volume probed,\cite{LSWW} with $L$ up to $\sim$0.35 fermi.
At large volumes the non-perturbative running is bound to deviate.

An alternative definition for the non-perturbatively defined running coupling 
for SU(2) has been based on ratios of the expectation values of suitable 
Polyakov loop operators, using twisted boundary conditions,\cite{Tho1} the 
so-called twisted Polyakov (TP) scheme.\cite{Petr} We will see next that 
twisted boundary conditions remove the zero-momentum modes, making perturbation
theory well behaved. Without these twisted boundary conditions, computing 
expectation values in the four dimensional euclidean finite volume is difficult
to control.\cite{CGJK} The TP coupling also agrees well with perturbation 
theory and after matching of the scales, with the SF coupling.\cite{Comp} Only 
the largest volume result ($L=0.28$ fermi) probed by the TP coupling lies 
slightly, but significantly, below the perturbative result. The near 
perturbative behavior seems to support the fact that non-zero momentum modes 
do behave perturbatively in intermediate volumes.

\subsection{Twisted Boundary Conditions}\label{subsec:twist}

With twisted boundary conditions, in the absence of zero-momentum modes, the 
classical vacuum at $A=0$ is isolated and the small volume behavior for the 
glueball masses is described by a perturbative series in $g^2(L)$, as opposed 
to $g^{2/3}(L)$ in absence of twist. The volume dependence will be quite 
different, this is in particular true for electric flux energies.\cite{HvBZ} 
Nevertheless, in large volumes the results should not depend on boundary 
conditions. Therefore, comparing the different boundary conditions gives 
valuable information about the transition to large volumes. In the hamiltonian 
formulation the twisted boundary conditions are most easily implemented in a 
gauge where the so-called twist matrices $\Omega_j\in$SU($N$) are constant 
\beq
\vek A(\vek x+L\vek e^{(j)})=\Omega_j^\dagger\vek A(\vek x)\Omega_j.
\label{eq:twb}
\eeq
They satisfy 't~Hooft's consistency condition,\cite{Tho1} which also gives 
the relation to the magnetic flux $\vek m\in\zahlen^3_N$,
\beq
\Omega_k^\dagger\Omega_\ell^\dagger\Omega_k^\phd\Omega_\ell^\phd=
\exp(2\pi i\eps_{k\ell j}m_j/N).\label{eq:mtwist}
\eeq
These generate a so-called Heisenberg group (the group commutator is central,
i.e. commutes with all group elements). The finite group theory allows one to 
construct in an elegant way the most general set $\Omega_k\in$SU($N$) for any 
given $\vek m$, and its generalizations to higher dimensions.\cite{Vbvg,LePo}

It seems that twisted boundary conditions spontaneously break the gauge 
invariance, due to the explicit choice of $\Omega_k$. However, this is 
of course similar to the case of periodic boundary conditions, which also
represents a gauge choice in formulating the boundary conditions. Once 
one has specified the gauge choice for the boundary conditions, Gauss's 
law tells us that local gauge transformations have to satisfy
\beq
h(\vek x+L\vek n)=\prod_j\Ad^{n_j}(\Omega_j)h(\vek x),\quad\Ad(\Omega)(h)
\equiv\Omega^\dagger h\Omega,\label{eq:mgt}
\eeq
for $\vek n\in\zahlen^3$, indicating the shifts over multiple periods. Note 
that in the adjoint representation the twist matrices commute. Thus, in a 
sense $L$ is (typically) a $1/N$ period (intimately related to the notion 
of color momentum, underlying the principle of the Twisted-Eguchi-Kawai 
one-point lattice model.\cite{GO83}) However, for finite size effects in 
large volumes, where the degrees of freedom that propagate ``around'' the
boundary are colorless, this has no consequence (see Sec.~\ref{subsec:stable}). 

\begin{figure}[htb]
\vspace{2.6cm}
\includegraphics{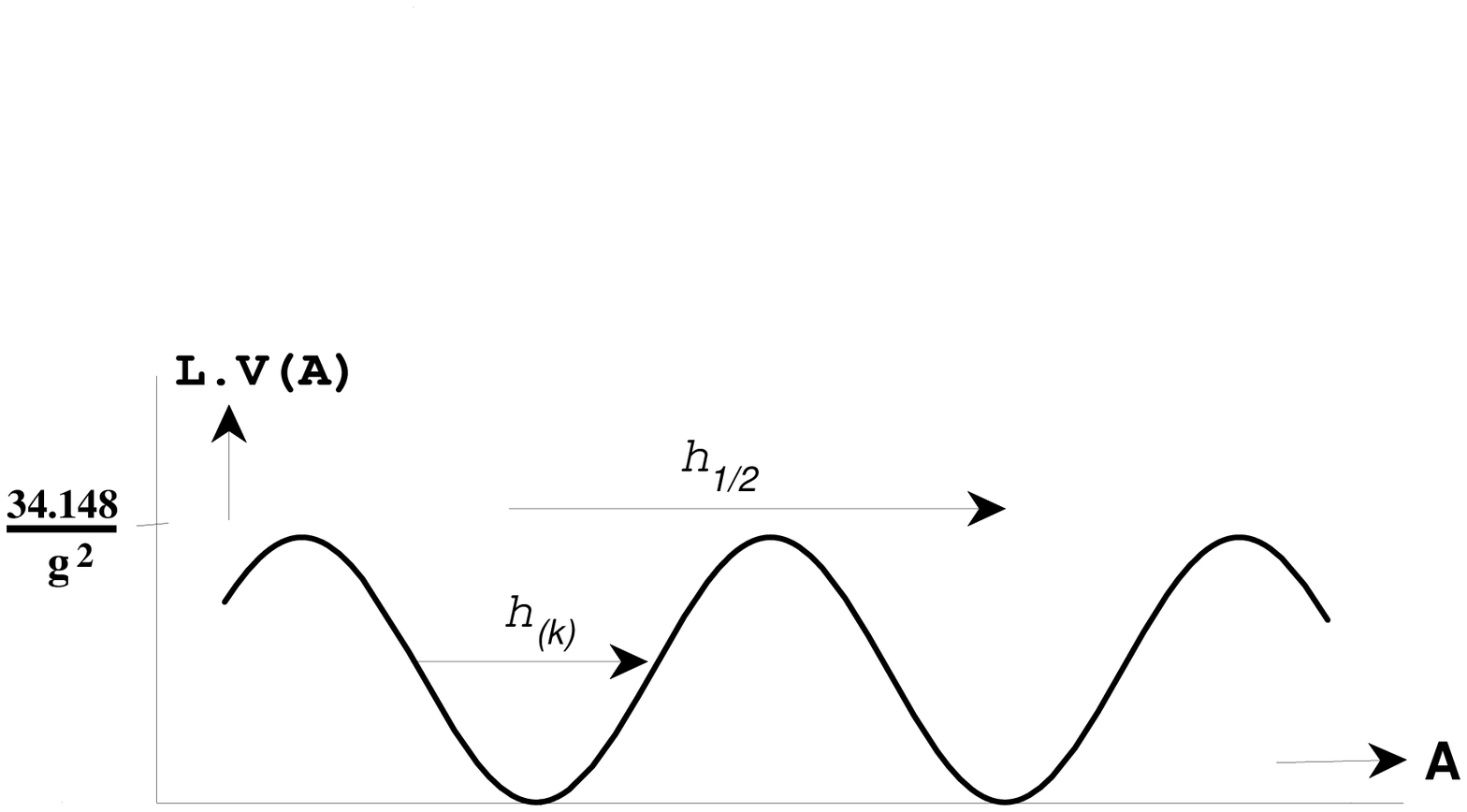}
\caption{Sketch of the SU(2) potential energy with twisted boundary conditions.
Shown are two isolated vacua, separated by a barrier with height $34.148/Lg^2$.
These vacua are related by twisted gauge transformations $h_{1/2}=h_{\vek p}$ 
with $\vek p\cdot\vek m=1$ (mod~2), having winding number $\nu(h_{1/2})=\half$.
Gauge transformations $h_{(k)}=h_{\vek k}$ with $\vek k\cdot\vek m=0$ (mod~2), 
leave the vacuum invariant.}
\label{fig:ptw}
\end{figure}

The candidate classical vacuum configuration satisfying the twisted boundary 
conditions is $A=0$, of zero classical energy despite the presence of magnetic 
flux.\cite{Ambj,GJKA} As we argued in Sec.~\ref{subsec:quarks}, the Polyakov 
loop (before taking the trace), evaluated at the classical vacuum configuration
should be invariant under gauge transformations. In the case of periodic 
boundary conditions, this implies it should be in the center of the gauge 
group, uniquely specified by $P_j\in Z_N$. To address the same question with 
twisted boundary conditions, the proper definition of the Polyakov loop 
has to be used\Cite{VB84}
\beq
P_j(\vek x)\equiv\frac{1}{N}\tr\left[\left(P\exp\,(\int_0^Lds\,A_j(\vek x+s
\vek e^{(j)})\right)\Omega_j^\dagger\right],\label{eq:tPol}
\eeq
(path ordering from left to right)
which indeed is invariant under the gauge transformations, Eq.~\Ref{eq:mgt}. 
For $A=0$ this is even so before taking the trace, such that gauge invariance 
is indeed not spontaneously broken. Thus, $P_j(A=0)=\tr(\Omega_j^\dagger)/N$ 
and using Eq.~\Ref{eq:mtwist} one finds that $P_j=0$, whenever $m_j\neq0$ 
(mod $N$). This is closely related to invariance under constant gauge 
transformations that are compatible with the allowed twisted gauge 
transformations,
\beq
h_{\vek k}(\vek x+L\vek n)=\exp(2\pi i\vek k\cdot\vek n/N)
\prod_j\Ad^{n_j}(\Omega_j)h_{\vek k}(\vek x).\label{eq:tgt}
\eeq
As for the periodic case, $\vek k\in Z_N^3$ specifies the homotopy type of the 
gauge transformation. These gauge transformations multiply $P_j$ with
$\exp(2\pi ik_j/N)$. We note that $A=0$ is left unaffected by constant gauge 
transformations, $h_{\vek k}(\vek x)\equiv\Omega_0(\vek k)$, which from 
Eq.~\Ref{eq:tgt} have to satisfy
\beq
\Omega_0^\dagger(\vek k)\Omega_\ell^\dagger\Omega_0^\phd(\vek k)
\Omega_\ell^\phd=\exp(2\pi ik_\ell/N),\label{eq:etwist}
\eeq
extending Eq.~\Ref{eq:mtwist} to four dimensions. This equation is solved for 
example by $\Omega_j$, with $k_\ell=m_i\eps_{ij\ell}$. In general solutions 
exist if and only if\Cite{CMP0,Vbvg} $\vek k\cdot\vek m=0$ (mod $N$). When 
$\vek k\cdot\vek m\neq 0$ (mod $N$), the gauge transformation $h_{\vek k}
(\vek x)$ does {\em not} leave $A=0$ invariant, but maps to a vacuum state 
with fractional Chern-Simons number, see Eq.~\Ref{eq:chsim} (equal to 
$\nu(h_{\vek k})$ as defined in Eq.~\Ref{eq:wind}). It is separated from $A=0$ 
by a classical potential barrier related to the instanton with {\em fractional}
topological charge for twisted gauge fields on the torus,\cite{Tho3,CMP0} to be
discussed in more detail in Sec.~\ref{sec:inst}. In Fig.~\ref{fig:ptw} we 
illustrate these features. Since electric flux quantum numbers are associated
with the representations of the $Z_N^3$ homotopy, this means\Cite{HvBZ} that 
some of the electric flux states will have energies that do not vanish 
in perturbation theory, whereas the electric flux energies associated 
with tunneling through the classical barrier will be suppressed by 
$\exp(-8\pi^2/Ng^2)$. Both differ from the behavior we observed when 
$\vek m=\vek 0$. For example, for SU(2) and $\vek m=(0,0,1)$, with 
$\Omega_1=i\tau_1$, $\Omega_2=i\tau_2$ and $\Omega_3=1$, one finds 
$\Omega_0(1,0,0)=\pm i\tau_2$, $\Omega_0(0,1,0)=\pm i\tau_1$ and 
$\Omega_0(1,1,0)=\pm i\tau_3$ as the only non-trivial constant gauge 
transformations that leave $A=0$ unchanged. Therefore, energies of electric 
flux with $e_1$ and/or $e_2$ non-trivial are perturbatively lifted, whereas 
states that differ only by the $e_3$ quantum number are degenerate.

For SU(2), the boundary conditions can be solved by the following Fourier 
expansion\Cite{HvBZ,DGKS}
\beq
\vek A^a(\vek x)= \sum_{\vek k\in\zahlen^3}\vek A^a(\vek k+\vek r^{(a)})
\exp(2\pi i(\vek k+\vek r^{(a)})\cdot \vek x/L),
\eeq
with the {\em color dependent} ``fractional'' momentum
\beq
\vek r^{(a)}=(\vek e^{(a)}\times\vek m)/2.
\eeq
Momentum conservation ensures that gauge invariance is not broken by these
color dependent momenta.
Using the Coulomb gauge $\partial_iA_i=0$, one checks the classical vacuum 
$\vek A=0$ is isolated with all fluctuations quadratic. To illustrate the 
perturbative analysis we consider the simpler case with $\vek m=(1,1,1)$, 
realized by $\Omega_k=i\tau_k$, because this choice of $\vek m$ does not 
break the invariance under the cubic group. Glueball states can thus be 
classified as for the periodic case. Generalizations to SU(3) (or arbitrary 
SU($N$) and magnetic flux) are easy to obtain. The allowed non-trivial constant 
gauge transformations are now as follows: $\Omega_0(0,1,1)=\pm i\tau_1$, 
$\Omega_0(1,0,1)=\pm i\tau_2$ and $\Omega_0(1,1,0)=\pm i\tau_3$. The 
one-particle state associated with the Fourier mode $\vek A_\pm^a(\vek p)$, 
with creation operator $b^\dagger(a,\vek p,\pm)$, has non-zero electric flux. 
They are further characterized by the momentum $2\pi\vek p/L$, energy 
$2\pi|\vek p|/L$ and the polarization $\pm$ of the gauge field, satisfying 
$\vek A^a_\pm(\vek p)\cdot\vek p=0$. To be precise, the electric flux 
vector belonging to this state is $\vek e=\vek e^{(a)}$. This motivates 
interpreting\Cite{DGKS} $2\pi\vek r^{(a)}/L$ as a Poynting vector. It also 
plays an interesting role in how the wave functional behaves under translations
over $L$ in the three coordinate directions.\cite{VB84} For any 
${\vert\Psi \rangle}_{\vek m,\vek e}$ with magnetic flux $\vek m$ 
and electric flux $\vek e$
\beq
{\vert\Psi(\vek x+L\vek n)\rangle}_{\vek m,\vek e}=\exp(i\vek n\cdot\vek P L)
{\vert\Psi(\vek x)\rangle}_{\vek m,\vek e}\,,\quad\vek P=
\pi(\vek e\times\vek m)/L.
\eeq

The electric flux $\vek e=\vek e^{(a)}$ is created with the gauge invariant 
Polyakov loop operator $P_a$, see Eq.~\Ref{eq:tPol}. This contains the 
one-particle state $b^\dagger(a,\vek p,\pm)|0\rangle$, the energy of one unit 
of electric flux is therefore given by the length of the Pointing vector, the 
minimal value $2\pi|\vek p|/L$ can take,
\beq
E_e=\pi\vert\vek e^{(a)}\times\vek m\vert/L=\pi\sqrt{2}/L.
\eeq
The energy of two units of electric flux (e.g. $\vek e=(0,1,1)$) is 
perturbatively degenerate with this. At higher order one has to take into
account that the two one-gluon transverse polarizations are now no longer 
degenerate. Properly creating electric flux with the gauge invariant Polyakov 
loop operator picks out the polarization in the direction of the loop; for the 
symmetric torus along $\vek e$. This causes a perturbative splitting between 
the energy of one and two units of electric flux.\cite{DGKA} The energy of 
three units of electric flux, $\vek e=(1,1,1)$, is entirely due to instanton 
effects,\cite{HvBZ,GPGS} see Fig.~\ref{fig:ptw}. Therefore, in lowest order 
$R_2=1+\Order{g^2}$ and $R_3=0+\Order{\exp(-4\pi^2/g^2)}$, quite distinct from 
what one finds with periodic boundary conditions in small and intermediate 
volumes.

To find the mass gap in the zero electric flux sector, one needs two-particle 
states, built from states with opposite (which for SU(2) is equivalent with 
identical) electric flux. They are of the form
\beq
b^\dagger(a,\vek p,\alpha)\,b^\dagger(a,\vek q,\beta)\vert 0\rangle,
\eeq
with $\alpha,\beta=\pm$ the polarizations of the one-particle states. 
These states have total momentum $\vek Q$ and energy $E$ satisfying
\beq
\vek Q=2\pi(\vek p+\vek q)/L\;,\quad E=2\pi(\vert\vek p\vert+
\vert\vek q\vert)/L.
\eeq
The minimal zero-momentum state gives the mass gap, in lowest order 
\beq
m_{gap}=2\pi\vert\vek e^{(a)}\times\vek m\vert/L=2\pi\sqrt{2}/L,
\eeq
which is twice the length of the Poynting vector. Counting the number of
ways one can form these two-particle colorless zero-momentum states from the 
one-particle states, one finds a 24 fold degeneracy. This degeneracy will be 
lifted in one loop order, arranged in irreducible representations $r$ of the 
cubic group,\cite{DGKS} for which lattice discretization effects were reported 
as well,\cite{DGKA} but no details have been published. In the continuum, the 
$\Order{g^2}$ mass and energy shifts are parametrized by the constants 
$\gamma_r$,
\beq
z_r=m_r L=2\sqrt{2}\pi+{{g^2(L)\gamma_r}\over{16\pi^2}},\quad
z_{e_n}=E_{e_n} L=\sqrt{2}\pi+{{g^2(L)\gamma_{e_n}}\over{16\pi^2}}~(n\neq3),
\eeq
which are listed in Table~\ref{tab:irrepsI}. 

\begin{table}[t]
\begin{center}
\caption{One loop coefficients\protect\Cite{DGKS,DGKA} for SU(2) with 
$\vek m=(1,1,1)$.\label{tab:irrepsI}}
\vspace{0.3cm}
\begin{tabular}{|c|c|c|c|}
\hline
irrep $r$&$\gamma_r$&irrep $r$&$\gamma_r$\\
\hline
$A_1^+$  &-92.08& $T_1^+$  & 14.74\\
$A_1^-$  &-91.93& $A_1^+$  & 25.56\\
$T_2^+$  &-41.26& $E_{}^-$ & 36.07\\
$E_{}^+$ &-22.90& $E_{}^+$ & 36.38\\
$T_2^+$  & -7.39&          &      \\
$T_2^-$  & -6.59& $e$      & -5.43\\
$T_2^+$  &  7.53& $e_2$    & -1.16${\vphantom{|}}_{\vphantom{|}}$\\
\hline
\end{tabular}\end{center}\end{table}

In comparison to the case of periodic boundary conditions we note that the 
$E^+$ tensor state is now heavier than the scalar $A_1^+$, but with the 
$T_2^+$ in between. Also, $z_0$ is here a decreasing function of the volume, 
which has to turn around at some point, when the mass stabilizes and $z_0$ 
becomes linear in $L$. A clear finite volume artifact is also the near 
degeneracy of the oddball ($A_1^-$) with the glueball ($A_1^+$). Both of these 
features we will also find for the sphere, see Sec.~\ref{sec:sphere}. There it 
will be demonstrated that taking the non-perturbative effects of instantons 
into account will lead to an appreciable splitting between the oddball and 
glueball.  Also here, like for the case of periodic boundary conditions, we can
estimate at which volume instantons become important, by equating the energy of
the scalar glueball state with the height of the barrier between two vacua, 
set by the sphaleron energy,\cite{Gavb} $34.148/Lg^2(L)$. Again one finds the 
critical value of $L$ to be of the order of 6 times the scalar glueball 
correlation length. First lattice Monte Carlo results with twisted boundary 
conditions were obtained by Stephenson and Teper.\cite{StTe,Step} They find
in very small volumes\Cite{Step} ($\beta=4/g_0^2=4.7$, on a $4^3\times 96$ 
lattice) that the $A_1^\pm$, $E^\pm$, $T_2^\pm$ and $T_1^+$ glueball masses 
indeed all become degenerate and equal to $2E_e$, with $R_2=1$ and $R_3=0$. 
The $A_2^\pm$ and $T_1^-$ states are appropriately heavier. Because the shifts 
in the masses are so small, a detailed comparison with the predictions for the 
$\Order{g^2}$ shifts is inconclusive. At larger volumes, $z_0>6$, the Monte 
Carlo results show that the differences between twisted and periodic boundary 
conditions disappear.\cite{StTe,Step} 

\begin{figure}[htb]
\vspace{6.7cm}
\includegraphics{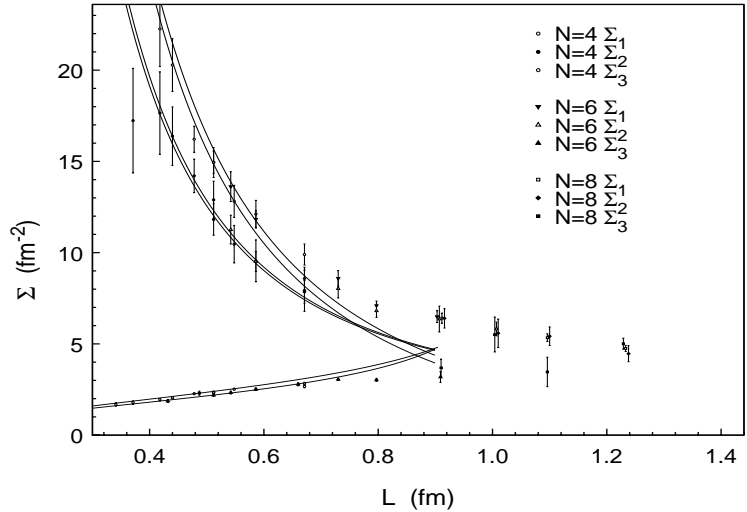}
\caption{Numerical data\protect\Cite{GAMa} for the SU(2) electric flux 
energies, $E_{e_n}=L\Sigma_n\protect\sqrt{n}$, with twisted boundary 
conditions, $\vek m=(1,1,1)$, on lattices of size $N^3\times N_t$, with 
$N=4,6,8$ and resp. $N_t=128,128,64$. The scale is set by $L=400 N\exp(
-3.38\beta)\,{\rm fm}$ and the curves show the perturbative predictions 
combined with a fit to the expected tunneling contributions.}\label{fig:twdata}
\end{figure}

Also an extensive study was made of the electric flux energies.\cite{RTNc,GAMa}
Here lattices of size $8^3\times 64$ and $N^3\times 128$ with $N=4$, 6 were
used, with $\beta=4/g_0^2$ ranging from 2.25 to 2.6, reaching volumes between 
0.3 and 1.3 fermi. In Fig.~\ref{fig:twdata} $\Sigma_n=E_{e_n}/(L\sqrt{n})$ is 
plotted as a function of $L$. This definition is such that if, as predicted by 
't~Hooft,\cite{Tho1} $R_n\rightarrow\sqrt{n}$ in the infinite volume limit, 
$\Sigma_n$ gives the infinite volume string tension $K$ independent of $n$. 
The curves test the small volume expansion. They contain the perturbative 
contribution, corrected for the lattice artifacts\Cite{DGKA} (the data is 
not accurate enough to test the $\Order{g_0^2}$ corrections), together with
const.$\times g_0^{-4}\exp(-S/g_0^2)$, the expected shift due to tunneling 
(cmp. Fig.~\ref{fig:ptw} and Sec.~\ref{subsec:EMdual}) mediated by the charge 
$\half$ instanton with classical action $S$ ($4\pi^2$ in the continuum, see 
Sec.~\ref{sec:inst}). The constant is fitted separately for $N=4$ and 6 
($N=8$ fits not shown). One finds fair agreement with the predicted behavior, 
with confirmation that the transition to the large volume starts around 0.75 
fermi.

\subsection{Supersymmetry and the Witten Index}\label{subsec:susy}

The case of supersymmetric Yang-Mills theories in a finite volume has been 
considered in the context of the Witten index\Cite{WiIn} in some detail. The 
torus geometry is crucial to preserve the gauge invariance. The Witten index 
involves counting the number of quantum states (fermionic states with a 
negative sign). At non-zero energy, because of the supersymmetry, this number 
cancels between the bosonic and fermionic states, such that the counting can 
be reduced to the vacuum sector. Here the number can be non-zero, because the 
supersymmetry generator can annihilate vacuum states. A zero Witten index is a 
sign of spontaneous breakdown of supersymmetry, where the vacuum energy is 
non-zero, explaining the physical significance of this index. A zero vacuum 
energy is a direct consequence of unbroken supersymmetry, where the hamiltonian
is given by $H=\half\{Q,Q^\dagger\}$ with $Q$, $Q^\dagger$ the supersymmetry 
generators, that annihilate the vacuum state. 

In perturbation theory bosonic loops are typically cancelled by fermionic 
loops, e.g. in the vacuum energy. Applied to the problem of non-abelian gauge 
theories in a finite volume this implies that the vacuum-valley effective 
potential vanishes. The cancellation is caused by the contribution from the 
gluino fluctuations, which are the superpartners of the gluons. They are Weyl 
fermions in the adjoint representation of the gauge group, denoted by 
$\lambda_\alpha^a$, with $\alpha$ a two-component spinor index. This means that
the wave function is no longer localized to $A=0$ or any of the other orbifold 
singularities of the vacuum valley (the moduli space of flat connections). 
It has been shown in the context of the supermembrane, when taking the 
supersymmetric Yang-Mills hamiltonian restricted to the zero-momentum modes 
only, that the spectrum is continuous, down to zero-energy. One can construct 
trial wave functions with a support arbitrarily far from $A=0$ that nevertheless
have finite energy.\cite{LNdW} The compactness of the vacuum valley is crucial 
to obtain a discrete spectrum.

The counting of quantum vacuum states was based on the assumption that for all
gauge groups the moduli space of flat connections is given in terms of the 
Cartan subalgebra, as we discussed for SU($N$). The gluonic part of the 
groundstate wave function $|0\rangle$ is assumed to be constant over the 
vacuum valley. In the reduction to the vacuum valley there are $r$ gluinos 
(associated with the generators in the Cartan subalgebra), each with two 
helicities. They are constant and carry no energy, which is the source of 
the vacuum degeneracy.  These gluinos have to be combined in Weyl invariant 
combinations, respecting Fermi-Dirac statistics. There are $r$ independent 
invariants, made from
\beq
U=\delta_{ab}\eps^{\alpha\beta}\lambda_\alpha^a\lambda_\beta^b
\eeq
and its powers. So one has $U^n|0\rangle$, $n=0,1,\cdots,r$, as the $r+1$ 
bosonic vacuum states, with no invariant fermionic vacuum states. Thus, one 
finds an index equal to the rank of gauge group plus one, $r+1$.

Because of possible problems with the adiabatic approximation near the orbifold
singularities, as we encountered in the previous sections, Witten considered 
the alternative of twisted boundary conditions.\cite{WiIn} For SU($N$) the same 
result, $r+1=N$, for the index follows. Other groups in general do not admit 
the type of twisted boundary conditions that completely remove the continuous 
vacuum degeneracy with its orbifold singularities. So it was natural to attempt
to find the exact zero-energy ground state for the supersymmetric 
generalization of L\"uscher's zero-momentum effective hamiltonian,\cite{Itoy} 
or even ignoring the time dependence, studying the path integral in the ultra 
local limit.\cite{Smil} None of these studies took the compactness of the 
vacuum valley into account and therefore fail to address the proper 
situation.\cite{LNdW} 

The problem of the adiabatic approximation remained an urgent one because of a 
discrepancy between the finite volume calculation of the Witten index, and the 
one based on the infinite volume determination of the gluino condensate through
instanton contributions.\cite{NSVZ,ShVa,AmVe} One relies on the fact that an 
index can not change under smooth perturbations, like increasing the volume. 
Also the infinite volume calculation has its problems. It uses the 
semiclassical approximation for a strongly interacting theory. This could,
however, be circumvented by first adding matter fields to introduce an 
external mass scale to control the instanton calculation and then rely on 
the index being constant under a smooth deformation (through holomorphy), that 
decouples the extra matter sector.\cite{ShVa} This resulted nevertheless in a 
discrepancy, $\sqrt{5/4}$ for SU(2), between the so-called strong and weak 
coupling calculations. Both calculations rely on the cluster decomposition 
property, since the instantons have more than two gluino zero modes, which 
seems to make the condensate $\langle\lambda\lambda\rangle$ vanish. The 
instanton calculation instead computes the appropriate power of the gluino 
condensate, $\langle(\lambda\lambda)^h\rangle$, that saturates the $2h$ gluino 
zero modes, where $h$ is the so-called dual Coxeter number of the gauge group, 
$h=N$ for SU($N$). It is this power that gives the number of vacuum states, 
\beq
\langle\lambda\lambda\rangle\equiv e^{2\pi in/h}\left(|\langle(\lambda\lambda
)^h\rangle|\right)^{1/h},\quad n=1,2,\cdots,h. 
\eeq
These arguments seem reasonable, but are not rigorous.\cite{AmVe} See for
further details a review by Shifman.\cite{Shif} Recently, 
use has been made of the constituent nature of periodic instantons 
(or calorons),\cite{KrvB,LLYi} in the context of a Kaluza-Klein reduction with 
periodic gluinos, as opposed to a high temperature reduction with anti-periodic
gluinos, which would break the supersymmetry. The constituent monopoles have 
exactly two zero-modes and saturate the condensate, $\langle\lambda\lambda
\rangle$. The strong coupling calculation now agrees with the weak coupling 
result.\cite{Khoz} The period can of course be used to control the coupling 
constant, but it is assumed the index does not change, going from a small to 
a large period. 

The mismatch in the Witten index between small and infinite volumes occurs 
for SO($N>6$) and the exceptional groups. There has, however, been a recent 
revision in counting the number of vacuum states in a finite volume. In a study
of D-brane orientifolds in string theory, Witten\Cite{WiBr} constructed for 
SO(7) an extra disconnected component on the moduli space of flat gauge 
connections, which can be embedded easily in SO($N>7$). For SO(7) and SO(8) 
this gives an isolated component of the moduli space, contributing only one 
extra vacuum state. For SO($N>8$) the extra component in the moduli space
behaves like the trivial component for SO($N$-7). Adding $r+1$ coming from the 
SO($N$) and SO($N-7$) moduli space components gives the dual Coxeter number of 
SO($N$), thereby giving the same number of vacuum states as obtained in the 
infinite volume. 

Witten's construction based on orientifolds does not work for the exceptional 
groups. This naturally led to a derivation of the extra vacuum states in a 
field theoretic context,\cite{KeRS} trivially extended to the exceptional 
group $G_2$, as a subgroup of SO(7). Three different groups have independently 
managed to solve the problem for other exceptional groups with periodic 
boundary conditions\Cite{Keur,KaSm} and for any group with twisted boundary 
conditions.\cite{BoFM} As we remarked before, twisted boundary conditions 
usually do not remove all the vacuum degeneracies, but it is important that 
the number of vacuum states is independent of the twist for all gauge groups 
that have a non-trivial center. The origin of the extra moduli space components
is actually not too hard to understand.\cite{Keur} Large gauge groups can have 
subgroups that are products of unitary groups, which each would allow for 
twisted boundary conditions. By choosing twists from all subgroups to cancel 
one obtains periodic flat connections that can not be deformed to the Cartan 
subalgebra, which supports the trivial component of flat connections. Of 
course one need not cancel these twists completely.\cite{BoFM} Needless to 
say, the group theory involved to sort out all the constraints and count the 
number of vacuum components is rather involved.

Supersymmetry does not play a role in establishing the existence of these 
extra vacuum components. Supersymmetry is, however, crucial for these extra 
components to lead to extra quantum vacua. As soon as the perturbative quantum 
fluctuations in the vacuum energy do not cancel, this will in general be 
different for different vacuum components. In a small volume, i.e. at weak
coupling, the wave functional will localize around the one with the lowest 
vacuum energy. Within such a connected component it will localize around
the minimum of the effective potential, as we discussed in the previous 
sections. It is likely, since the trivial vacuum component is the widest, 
that this is the one where the wave functional localizes. But interestingly, 
one now has potential energy barriers between vacuum components that are 
not related by a homotopically non-trivial gauge transformation (since the
different vacuum components are not isomorphic). Still these vacua can be 
characterized by fractional Chern-Simons numbers and tunneling between them 
would be described by new types of instanton solutions with fractional 
topological charge.\cite{BoFM,Keur} It considerably adds to the richness of 
non-abelian gauge theories. 

Although these new results for counting the number of vacuum states in a 
finite volume remove the urgency of addressing the problem with the adiabatic 
approximation, it does remain a sore point in the finite volume analysis, as 
also stressed recently by Witten.\cite{WiRe} One immediate problem we encounter
is that Eq.~\Ref{eq:decom} seems to imply that the vacuum-valley wave function 
has to vanish at the orbifold singularities, which seems inconsistent with it 
being constant sufficiently far from the orbifold singularities. However, one 
should take into account the behavior under Weyl reflections of the occupied 
negative energy gluino states in the Dirac sea near the orbifold singularities 
in the zero-momentum hamiltonian. Attempting to incorporate this in the 
formalism developed in the previous sections, we run into the problem that
spin $\half$ fields do not decompose in three components, which can be 
associated with each of the coordinate directions. In the bosonic sector we 
could conveniently ignore the transversality, with gauge invariance restored at 
the end by restricting to the invariant wave functions. Thereby each of the 
coordinate directions separately allowed a polar decomposition, not compatible
with the nature of the gluino fields. For SU(2) there is, however, an elegant 
polar decomposition using the spherical and gauge symmetry of the zero-momentum 
hamiltonian,\cite{Savv} which even holds for $H_{\eff}$ in Eq.~\Ref{eq:Heff} 
to $\Order{g^{4/3}}$. In the supersymmetric case one would require the wave 
function to become constant after reduction to the vacuum valley, sufficiently 
far away from $A=0$, to match to the expected behavior away from each of the 
orbifold singularities. Such a matching can in principle be controlled 
sufficiently rigorously,\cite{Adri} as was done in the semiclassical 
calculation of the energy of electric flux,\cite{Vbk1,AnnP} but here we do 
not have the benefit of a well localized zero-momentum wave function at the 
orbifold singularities,\cite{LNdW} and judging the incomplete results in the 
literature,\cite{Itoy} this seems not the way to go. It is in the light of 
the robustness of supersymmetry somewhat surprising (and frustrating) this 
problem remains so technically demanding. 

\section{Instantons and Sphalerons on the Torus}\label{sec:inst}

Instantons are associated with the tunneling through barriers that separate 
different vacuum components. We are interested in these barriers to establish 
the directions in field space in which the spreading of the wave functional 
has to be taken into account, when energies of the low-lying states no longer 
are well below the barriers. The barrier is a saddle point of the energy 
functional with one unstable mode, a so-called sphaleron.\cite{Klma} They 
exist here by virtue of the finite volume. Remarkably, on any compact four 
manifold instantons exist,\cite{Taub,DoKr} still with an arbitrary scale 
parameter, only limited by the size of the manifold. The finite volume 
sphaleron is typically associated with the instanton that achieves this 
maximal scale. 

For low topological charge, sometimes smooth instantons can {\em not} be proven 
to exist.\cite{Taub} An example of non-existence actually occurs\Cite{BrvB} on 
$T^{\,4}$. In general one takes a small localized instanton from $\real^4$ 
that is matched smoothly to the flat connection. When the moduli
space of flat connections has a continuous degeneracy, as is the case for the 
torus, there can be obstructions against this procedure. One can get 
arbitrarily close to satisfying the self-duality equations, $F_{\mu\nu}=
\pm\tilde F_{\mu\nu}\equiv\pm\half\eps_{\mu\nu\alpha\beta}F^{\alpha\beta}$, 
that saturate the bound for the {\em euclidean} action
\beq
S=-\half\int\tr(F_{\mu\nu}^{\,2})=-\quart\int\tr(F_{\mu\nu}\pm\tilde 
F_{\mu\nu})^2\pm\half\int\tr(F_{\mu\nu}\tilde F^{\mu\nu})\leq 8\pi^2|\nu|,
\label{eq:Seucl}
\eeq
with $\nu$ the topological charge. However, an exact self-dual solution is only
achieved in the limit where the scale parameter is forced to zero or the volume
is taken to infinity. For twisted boundary conditions\Cite{Tho1}, existence of
charge one instantons\Cite{BrMT} can be understood from the fact that twist 
removes the continuous degeneracy in the moduli space of flat connections.

From the hamiltonian point of view we are interested in instantons on 
$T^3\times\real$, where time is not constrained to be finite. This can be 
seen as a limit of $T^{\,4}$ where one of the periods, called $T$, is taken 
to infinity. In the $A_0=0$ gauge the field is periodic up to a gauge 
transformation $h$ in the time direction, $A(\vek x,T)=[h]A(\vek x,0)$, and
\beq
-\frac{1}{16\pi^2}\int\tr(F_{\mu\nu}\tilde F^{\mu\nu})=\int_0^TCS(A)
=CS([h]A)-CS(A)=\nu(h),\label{eq:topch}
\eeq
with $\nu(h)$ defined in Eq.~\Ref{eq:wind}. Using that $\tr(F_{\mu\nu}\tilde 
F^{\mu\nu})=\partial_\mu K_\mu(A)$, we find for the Chern-Simons functional 
\beq
CS(A)=-\frac{1}{16\pi^2}\int d^3x K_0(A)=-\frac{1}{8\pi^2}\int d^3x
\tr\left(A_i(\partial_j A_k+\twoth A_j A_k)\right)\eps_{ijk}.\label{eq:chsim}
\eeq
When $T^3$ has periodic boundary conditions there is an ambiguity {\em which} 
point in the vacuum valley to tunnel from and to. These need not be the same 
points, as is already evident with tunneling related by non-trivial twisted 
gauge transfomations (see Eq.~\Ref{eq:bloch} and \Ref{eq:tgt}), under which the 
Polyakov loop is periodic in time up to a non-trivial $Z_N$ factor. In that
case charge one instanton solutions can be shown to exist.\cite{BrMT} The 
obstruction for the periodic four-torus is reflected in the fact that the 
instanton does not want to tunnel between the {\em same} points in the vacuum 
valley. 
\begin{figure}[htb]
\vspace{10.9cm}
\includegraphics{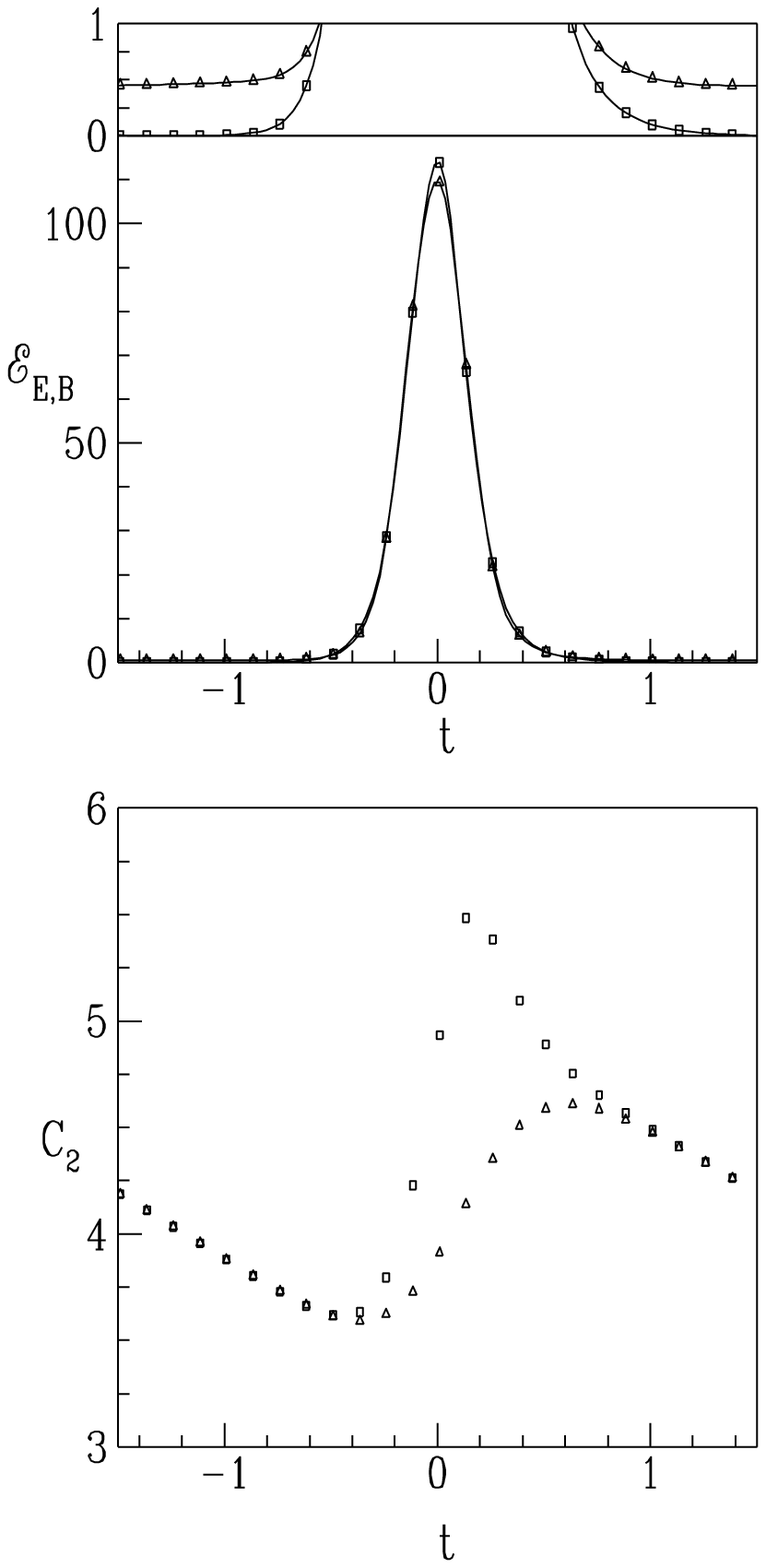}
\includegraphics{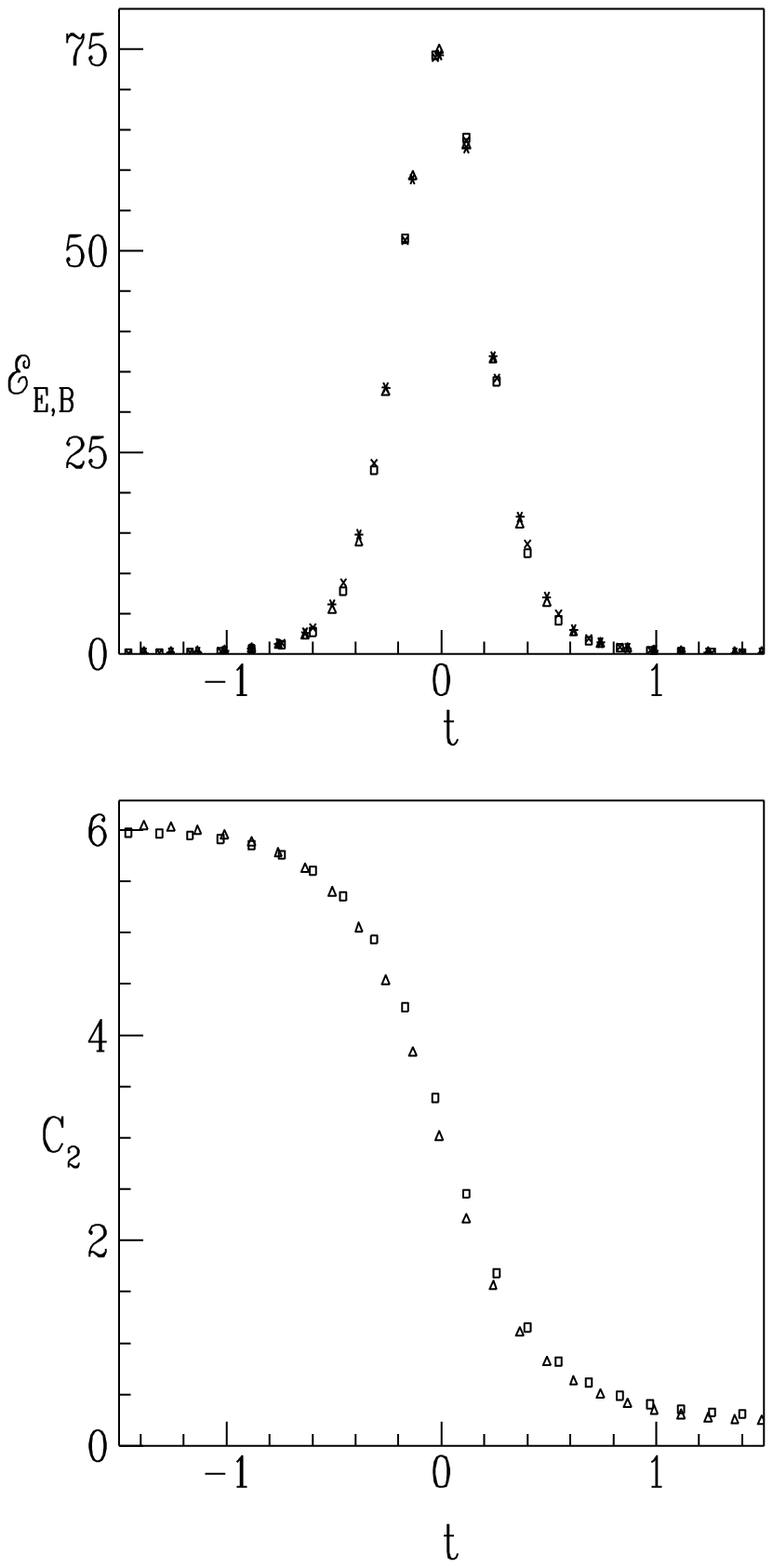}
\caption{Numerical results\protect\Cite{GGVS} obtained with cooling on $N^3
\times 3N$ lattices (scaled to $L=1$). Left is with $\vek k=\vek 0$, $N=8$ 
and right with $\vek k=(1,1,1)$, $N=7$ and 8 ($\vek k$ indicating the twist 
in time). The top figures show electric ($\Ss{E}_E(t)$ triangles) and magnetic 
($\Ss{E}_B(t)$ squares) energies ($g=1$), the inset shows the tails at an 
enlarged scale. In the lower figures we plot $C_2(t)$ through two distinct (on 
the right one) point(s).}\label{fig:num}
\end{figure}
This is illustrated in Fig.~\ref{fig:num}, obtained by a numerical procedure 
called cooling\Cite{Cool} that minimizes the lattice action in a given 
topological sector, carefully dealing with the lattice discretization 
problems.\cite{GGVS} With periodic boundary conditions the configuration first 
relaxes to a solution which is not self-dual, although it has reached the 
vacuum valleys at either end in euclidean time, judging the zero magnetic 
energy. However, it moves in accordance with the equations of motion along 
this vacuum valley with constant speed (i.e. electric field) to match the 
forced periodicity, as is also evident from the linear (space independent) 
slope for $C_2(t)$. When one would continue to lower the action, in order to 
reach a self-dual solution, the instanton is forced to shrink to zero size. 
With a twist in the time direction the configuration immediately relaxes to 
a self-dual one. 

\begin{table}[t]
\caption{The lowest eigenvalues\protect\Cite{Gavb} for the Hessian of the 
energy functional evaluated at the twisted and periodic sphaleron on a 
$N^3$ lattice for $N=4,~6$ and 8.\label{tab:hess}} 
\vspace{0.3cm}
\begin{center}
\begin{tabular}{|r|r|r||r|r|r|}
\hline
\mco{3}{|c||}{\vek m=(1,1,1)~~sphaleron${\vphantom{|}}^{\vphantom{|}}$}
&\mco{3}{|c|}{\vek m=\vek 0~~sphaleron${\vphantom{|}}^{\vphantom{|}}$}\\
\hline
 \mco{1}{|c|}{N=4}&\mco{1}{|c|}{N=6}&\mco{1}{|c||}{N=8}
&\mco{1}{|c|}{N=4}&\mco{1}{|c|}{N=6}&\mco{1}{|c|}{N=8}\\
\hline
-11.1280&-10.9003&-10.8174&-13.1960&-13.3935&-13.4432\\
  0.0074&  0.0001&  0.0000&  0.0073&  0.0006&  0.0000\\
  0.0091&  0.0001&  0.0000&  0.0124&  0.0010&  0.0000\\
  0.0101&  0.0001&  0.0000&  0.0139&  0.0011&  0.0000\\ 
 13.1667& 13.3371& 13.3934&  1.7041&  1.4151&  1.3009\\
 13.1667& 13.3371& 13.3934&  1.7070&  1.4303&  1.3012\\ 
 13.1667& 13.3372& 13.3934&  7.9677&  8.0984&  8.1260\\ 
 16.0475& 16.8663& 17.1597&  7.9682&  8.1008&  8.1261\\ 
 16.0478& 16.8663& 17.1597&  7.9684&  8.1279& 
             ${\vphantom{|}}_{\vphantom{|}}$   8.1266\\
\hline
\end{tabular}
\end{center}
\end{table}

\begin{figure}[htb]
\vspace{6.7cm}
\includegraphics{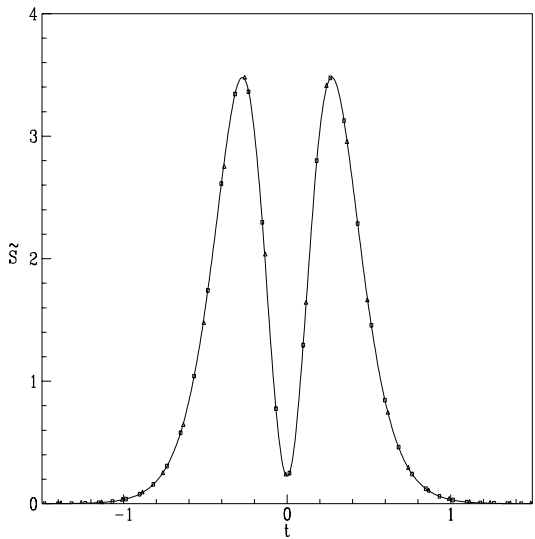}
\caption{Numerical results\protect\Cite{Gavb} for $\tilde S(t)$, the square 
of the equations of motion, along the tunneling path for the SU(2), charge 
$\half$ twisted instanton solution for $\vek m=\vek k=(1,1,1)$, evaluated 
on a $8^3\times 24$ (triangles) and a $12^3\times 36$ (squares) lattice 
(scaled to $L=1$). The dip at $t=0$ occurs at the finite volume sphaleron.}
\label{fig:eom}
\end{figure}

Shown on the right in Fig.~\ref{fig:num} is an instanton that is close to being
maximal in its size, thereby going through the lowest barrier, to maintain the 
same action. This lowest barrier is assumed to be a finite volume sphaleron. 
This was verified,\cite{Gavb} by explicitly looking for saddle-point solutions 
on $T^3$, minimizing the square of equations of motion using a similar 
numerical procedure of cooling (cmp. Fig.~\ref{fig:eom}). This saddle point is 
not invariant under translations, giving rise to zero-modes, which in addition 
to the unstable (i.e. tunneling direction) will become the degrees of freedom 
that need to be treated non-perturbatively, by imposing the proper boundary 
conditions in field space. In addition it was found that there are two 
directions {\em not} associated with zero-modes, in which the saddle point is 
nearly flat (see Table~\ref{tab:hess}). Such modes would have to be treated 
non-perturbatively as well. Without an analytic solution, it is hard to judge 
if such an analysis is feasible. This was one of the motivations to look at the
situation where the torus geometry is replaced by that of a sphere. In this 
case the finite volume sphaleron has a constant energy density and no (almost) 
flat directions, see Sec.~\ref{sec:sphere}. 

We already mentioned in the previous section that with non-zero magnetic 
flux, there exist instanton solutions with fractional topological charge. 
When first proposed by 't~Hooft,\cite{Tho3} this was met with some disbelieve. 
The mathematical classification of so-called fiber bundles, to describe 
gauge fields on compact manifolds seemed to require integer topological 
charge, specified by the second Chern number (see Eq.~\Ref{eq:chclass}). 
But due to the twist, the fiber bundle has to be considered real and the 
appropriate classification is by the first Pontryagin class. With the proper 
normalizations, this makes the would-be fractional second Chern class an 
integer first Pontryagin class.\cite{CMP0} It is the difference between 
a unitary and orthogonal fiber bundle, e.g. SU(2) versus SO(3). This 
makes also precise the intimate connection between the topological charge 
and twist, as first discovered by 't~Hooft.\cite{Tho3} As we have seen in 
Secs.~\ref{subsec:SUN} and \ref{subsec:twist} (see Eq.~\Ref{eq:tgt}, which 
also applies for $\vek m=\vek 0$, in which case $\Omega_j=1$), there are 
homotopically non-trivial gauge transformations $h_{\vek k}$, ``periodic'' up 
to an element of the center of the gauge group, that do not affect the boundary
conditions of the gauge fields. In particular $h_{\vek k}^{-1}dh^{}_{\vek k}$ 
has the right periodicity, and $\nu(h_{\vek k})$ in Eq.~\Ref{eq:wind} remains 
well-defined (in the gauge where the twist matrices $\Omega_j$ are constant 
and $A_0=0$). When $\vek m =\vek 0$ we have seen that we can always choose 
$h_{\vek k}$ abelian, and $\nu(h_{\vek k})=0$. In this case there is no 
interplay between the twist and the topological charge $\nu$, which is here 
integer. When $\vek m$ is non-trivial we stated in Sec.~\ref{subsec:twist} 
that when $\vek k\cdot\vek m=0$ (mod $N$), $h_{\vek k}$ can be chosen constant,
$h_{\vek k}=\Omega_0(\vek k)$, as can be checked for the simple case of SU(2) 
by inspection, but can be proven from first principles.\cite{Vbvg} However, 
when $\vek k\cdot\vek m\neq 0$ mod $N$ it may be that $\nu(h_{\vek k})\neq 0$. 
The additive nature of the topological invariant $\nu(h)$,
\beq
\nu(h'h)=\nu(h')+\nu(h),\label{eq:add}
\eeq
implies that $\nu(h_{\vek k}^N)=N\nu(h_{\vek k})$. This has to be an integer, 
since $h_{\vek k}^N$ is ``periodic'' and the winding number of a periodic 
function $h:T^3\rightarrow G$ is an integer. Also one can duplicate $T^3$, 
under which the topological charge is additive, as many times in the space 
directions as is necessary to get a new torus for which $\vek m=\vek 0$, 
and hence with integer topological charge. Therefore,\cite{Tho3} $\nu(h)$ is 
proportional to $N^{-1}$ and linear in $\vek k$ and $\vek m$, such that 
$\nu+\vek m\cdot\vek k/N\in\zahlen$, as was proven rigorously in the context 
of fiber bundles.\cite{CMP0} Conversely, using this result one easily sees 
that the existence of constant twist matrices in four dimensions, i.e. 
solutions to both Eq.~\Ref{eq:mtwist} and Eq.~\Ref{eq:etwist}, require 
$\vek k\cdot\vek m=0$ mod $N$. This is so because $A=0$, with $\nu=0$,
satisfies the boundary conditions. For further details on these twist-related
issues one can also consult a recent review.\cite{GoAr}

The above implies that the lowest non-trivial topological charge that can be 
reached is $1/N$. With Eq.~\Ref{eq:Seucl}, the associated euclidean action 
of this instanton solution is found to be $8\pi^2/N$, which can indeed be 
attained as one can prove existence of self-dual solutions in the appropriate
sector.\cite{Sedl} Like for topological charge one, despite much effort up 
to date no analytic solutions are known. What can be deduced from general 
principles\Cite{Schw} is that the solutions with charge $1/N$ have only 
the four translational degrees of freedom. In particular they have no free 
scale parameter, their size is fixed with respect to the torus. This fixed 
scale is a great benefit in studying these instantons on the 
lattice.\cite{BaPs,GPGS,Twis} For this fractionally charged instanton the
finite volume sphaleron was constructed as well, by finding for SU(2) with 
twist $\vek m=(1,1,1)$ on $T^3$ the appropriate saddle point. Like for 
$\vek m=\vek 0$ it is not invariant under translations, giving three flat 
directions at the saddle point, but in this case no other near zero-modes 
appear,\cite{Gavb} see Table.~\ref{tab:hess}. 

In Fig.~\ref{fig:eom} we illustrate the fact that along the tunneling path 
there is a saddle point, by plotting the square of the equations of motion, 
\beq
\tilde S(t)\equiv-\frac{1}{32}\int_{T^3}d^3x
\tr\bigl(\partial_i F_{ij}(\vek x)+[A_i(\vek x),F_{ij}(\vek x)]\bigr)^2,
\eeq
as a function of euclidean time. The fact that $\tilde S$ does not exactly 
vanish in the middle, at the top of the barrier, is due to a limited resolution
on the finite lattices employed.

\subsection{T-duality for Instantons on the Torus}\label{subsec:Nahm}

Although no explicit solutions of the basic instantons of charge $1/N$ and 
charge 1 are known, progress has been made in better understanding instantons 
on the torus. We will first discuss the case of $T^4$, before addressing 
$T^3\times \real$. For $T^4$ the only explicitly known solutions are the 
ones constructed by 't~Hooft with constant field strength.\cite{Tho4,CMP1,LePR}
These are essentially abelian and are self-dual {\em only} when the periods of 
the torus match the field strength (which is quantized due to quantization of 
flux). Recently these solutions to the self-duality equations were followed
in perturbation theory by deforming away from the special geometry,\cite{Pert} 
using as a starting point the exact fluctuation spectrum\Cite{CMP1} in the 
background of the non-selfdual constant curvature solutions.

The essential ingredient for gaining further understanding has been the 
so-called Nahm or Mukai transformation.\cite{Nahm,Muka} Nahm introduced it to 
find magnetic monopole solutions, getting inspiration from the algebraic 
Atiyah-Drinfeld-Hitchin-Manin\Cite{ADHM} (ADHM) construction of instantons in 
$\real^4$. The formalism can as a matter of fact be used to provide a {\em 
simple} proof of the ADHM construction.\cite{DoKr,CoGo} It will be convenient 
in the following to see $T^4$ as $\real^4/\Xi$, where $\Xi$ is a four 
dimensional lattice spanned by the four periods $a^{(\mu)}$, and to introduce 
the connection one-form $\omega(x)=A^\mu(x)dx_\mu$, which is invariant up to 
a gauge transformation under translation over a period. These gauge 
transformations, called cocycles, satisfy cocycle conditions\Cite{Tho1} 
(twisted boundary conditions will be discussed later), to assure the gauge 
invariant quantities are periodic, and such that one has an appropriate 
fiber bundle over the torus,\cite{CMP0}
\bea
\omega(x+a)&=&h^{-1}_a(x)(\omega(x)+d)h_a(x),\\
h_{a+b}(x)&=&h_b(x)h_a(x+b)=h_a(x)h_b(x+a),\quad a,b\in\Xi.\nonumber
\eea
It will be convenient to take U($N$) as gauge group, allowing for a U(1) 
factor. The associated vector bundle in the fundamental representation of 
U($N$) is denoted by $E$. The curvature of this bundle $E$, given by
\beq
F=d\omega+\omega\wedge\omega=\half F^{\mu\nu}dx_\mu\wedge dx_\nu,
\eeq
leads to a topological charge $\nu=\int_{T^4} (c_2(E)-\half c_1(E)\wedge 
c_1(E))$ as an integral over Chern classes (cmp. Eq.~\Ref{eq:topch} and note 
that $c_1(E)=0$ for SU($N$) because of the vanishing trace), where
\beq
c_1(E)=\tr\left(\frac{F}{2\pi i}\right),\quad
c_2(E)=\half\tr\left(\frac{F}{2\pi i}\wedge\frac{F}{2\pi i}\right).
\label{eq:chclass}
\eeq

The starting point of the Nahm transformation is that gauge fields with 
topological charge $\nu$, have $\nu$ positive chirality zero-modes for the 
massless Dirac equation,\cite{AtSi} which is most conveniently written in 
the Weyl representation. For this we introduce the unit quaternions 
$\sg_\mu$ and their conjugates $\sgbar_\mu\equiv\sg^\dagger_\mu$
\beq
\sg_\mu=(1_2,i\tau_j),\quad\sgbar_\mu=(1_2,-i\tau_j).\label{eq:quar}
\eeq
They satisfy the multiplication rules
\beq
\sg_\mu \sgbar_\nu = \eta^\al_{\mu \nu} \sg_\al, \quad
\sgbar_\mu \sg_\nu = \bar{\eta}^\al_{\mu \nu} \sg_\al,\label{eq:eta}
\eeq
where we used the 't~Hooft $\eta$ symbols,\cite{Tho2} generalized slightly to 
include $\eta^0_{\mu\nu}=\bar\eta^0_{\mu\nu}=\delta_{\mu\nu}$, also 
useful in Sec.~\ref{sec:sphere}. The Weyl-Dirac operator is given by 
\beq
D\equiv\sg_\mu D_\mu(A)=\sg_\mu(\partial_\mu+A_\mu).
\eeq
Hence, in the background of a charge $\nu$ gauge field there are $\nu$ 
independent solutions to $D\Psi=0$. For $\Psi(x)$ to be defined as a two-spinor
on the torus one requires $\Psi(x+a)=h_a(x)\Psi(x)$. Strictly speaking, the
index theorem\Cite{AtSi} only states that the number of zero-modes of positive 
chirality minus those of negative chirality (for which $D^\dagger\Psi=0$) is 
equal to the topological charge, but it will be assumed that there are no 
negative chirality zero-modes. For self-dual gauge fields we will discuss 
this assumption later. One now adds a spectral parameter $z\in\real^4$ in 
the form of a flat abelian connection\Cite{Nahm}
\beq
\omega_z=\omega+2\pi iz^\mu dx_\mu,
\eeq
(the unit generator of U(1) is implicit) which leaves the curvature unchanged, 
$F_z=F$. Hence there is a smooth family of $\nu$ normalized fermionic
zero-modes
\beq
D_z\Psi_z^{(i)}(x)=\sg_\mu(\partial_\mu+A_\mu+2\pi iz_\mu)\Psi_z^{(i)}(x)=0,
\quad\int_{T^4}d^4x~\Psi_z^{(i)}(x)^\dagger \Psi_z^{(j)}(x)=\delta_{ij}.
\eeq
From this family one constructs the connection $\hat\omega\equiv
\hat A_\mu(z)dz_\mu$ by
\beq
\hat A_\mu^{ij}(z)=\int_{T^4}d^4x~\Psi_z^{i}(x)^\dagger\frac{\partial}{
\partial z_\mu}\Psi_z^{(j)}(x).
\eeq
Using that $D_{z+\hat a}\bigl(e^{-2\pi ix\cdot\hat a}\Psi_z^{(i)}(x)\bigr)=0$ 
and the fact that $\bigl\{\Psi_{z+\hat a}^{(i)}(x)\bigr\}$ forms a complete 
orthogonal set of solutions, we find that
\beq
\Psi_{z+\hat a}^{(i)}(x)=e^{-2\pi ix\cdot\hat a}\Psi_z^{(j)}(x)
\hat h^{ji}_{\hat a}(z),
\eeq
with $\hat h_{\hat a}(z)$ a unitary $\nu\times\nu$ matrix, which defines 
the cocycle for $\hat\omega$
\beq
\hat\omega(z+\hat a)=\hat h^{-1}_{\hat a}(z)(\hat\omega(z)+\hat d)
\hat h_{\hat a}(z),\quad \hat d\equiv dz_\mu\partial_{z_\mu},
\eeq
as a U($\nu$) connection on the dual torus $\hat T^4=\real^4/\tilde\Xi$, where 
\beq
\tilde\Xi=\bigl\{\hat a\in\real^4|<\hat a,a>\in\zahlen,\ \forall a\in\Xi\bigr\}
\eeq
is the lattice dual to $\Xi$. For example, if $\Xi$ is generated by $a^{(\mu)}=
L e^{(\mu)}$, giving a hypercube with sides of length $L$, the dual lattice 
$\tilde\Xi$ is generated by $\hat a^{(\mu)}=L^{-1} e^{(\mu)}$, a hypercube with 
sides of length $1/L$. So the Nahm transformation is a T-duality.

In the absence of negative chirality zero-modes ($\coker D_z=\ker D_z^\dagger
=0$) $\hat E_z\equiv\ker D_z$ depends smoothly on $z$. It was realized by 
Braam,\cite{Lect,Sche} that one can use the Atiyah-Singer {\em family} index 
theorem\Cite{AtSi} to compute the Chern character of the bundle $\hat E$ with 
connection $\hat\omega$, as the integral over $T^4$ of the Chern character of 
$E\otimes\Ss{P}$, which is the vector bundle relevant for $\omega_z$, as a 
vector bundle over $T^4\times\hat T^4$. Here $\Ss{P}$ is the 
so-called\Cite{BrvB} Poincar\'e line bundle with the connection $2\pi i
z^\mu dx_\mu$. It was crucial this line bundle had no curvature as a bundle 
over $T^4$, but extending it to $T^4\times\hat T^4$, this is no longer the 
case and $F_{\Ss{P}}=2\pi idx_\mu\wedge dz^\mu$, with a first Chern class, 
$c_1(\Ss{P})=dx_\mu\wedge dz^\mu$ (for a line bundle all other Chern classes 
vanish). The Chern character satisfies $ch(E\otimes E')=ch(E)\wedge ch(E')$ 
and is a formal polynomial in the Chern classes,
\beq
ch(E)=c_0(E)+c_1(E)+\half c_1(E)\wedge c_1(E)-c_2(E),\quad
ch(\Ss{P})=\exp(c_1(\Ss{P})).
\eeq
Perhaps somewhat confusingly $c_0(E)=rk(E)$ is called the rank of the vector 
bundle $E$, here the dimension of the fundamental representation (so $rk(E)=N$ 
for a U($N$) fiber bundle). It now follows that 
\beq
\ch(\hat E)=\int_{T^4}\ch(E)\wedge\ch(\Ss{P}).
\eeq
The integral over $T^4$ only gives a non-zero result if the combined form is of 
degree four in $dx_\mu$, i.e. a top-form. This means that 
\begin{itemize}
\item The zero-form $c_0(E)=N$, becomes a volume form on $\hat T^4$, whose 
integral over $\hat T^4$ gives the topological charge of the dual gauge field.
\item The two-form $c_1(E)$ whose cohomology class can be specified by 
the {\em abelian} fluxes $c_1(E)=\half n^{\mu\nu}dx_\mu\wedge dx_\nu$, goes 
over in $c_1(\hat E)=\half\tilde n^{\mu\nu}dz_\mu\wedge dz_\nu$.
\item The four-form whose integral is the topological charge $\nu$ of the 
original gauge field becomes a zero-form, $c_0(\hat E)=\nu$.
\end{itemize}
Thus we see that the Nahm transformation interchanges the rank $N$ of the 
bundle with the topological charge $\nu$, and maps the first Chern class to 
its dual. In particular, an SU($N$) bundle has vanishing first Chern class and 
is thus mapped under the Nahm transformation to U($\nu$) with vanishing first 
Chern class. It is well-known that in the latter case the bundle can be gauged 
to an SU($\nu$) bundle (its U(1) part is trivial and can be gauged away).

The family index theorem only provides topological information on the dual 
gauge field. We will demonstrate the wondrous result\Cite{Nahm}, that if 
$\omega$ is a self-dual connection, than also $\hat\omega$ is self-dual.
A crucial ingredient is formed by 
\beq
D^\phd_z D^\dagger_z=-D^2_\mu(A_z)-\half\sg_{[\mu}\sgbar_{\nu]}
F^{\mu\nu},
\eeq
using $\sg_\mu\sgbar_\nu=\delta_{\mu\nu}+\sg_{[\mu}\sgbar_{\nu]}$. Since 
$\sg_{[\mu}\sgbar_{\nu]}\equiv\sg_i\eta^i_{\mu\nu}$ is an {\em anti-selfdual} 
tensor,\footnote{Note that 't~Hooft had introduced $x_4$, instead of $x_0$, as 
the imaginary time. Using his symbols\Cite{Tho2} replacing the index 4 by 0, 
flips the dualities around, since $\eps_{0123}=-\eps_{4123}$.} we see that 
$D^\phd_z D^\dagger_z=-D^2_\mu(A_z)$. We first deal as promised with the 
condition that there are no opposite chirality zero-modes. We have seen 
that if $\omega$ is self-dual, and $D_z^\dagger\Psi=0$ that 
$D_\mu^2(A_z)\Psi=D^\phd_zD_z^\dagger\Psi=0$, but this would imply that 
the connection $\omega$ has a flat factor,\Cite{DoKr} meaning that the 
bundle of rank $N$ splits in {\em the direct sum} of a bundle of rank 
$N-1$ and a flat line bundle, $E=E'\oplus L_\xi$, where $L_\xi$ has the 
connection $2\pi\xi^\mu dx_\mu$. Such a flat factor would exist for any
$z$ (simply shifting $\xi$ proportionally), and so a sufficient condition
to impose the absence of opposite chirality zero-modes is to require $\omega$
to be without flat factors (WFF, sometimes also called 1-irreducible). 

A direct corollary is now that regular SU($N$) charge one instantons cannot 
exist on $T^4$. Suppose they would exist. The Nahm transformation gives rise 
to a U(1) bundle of charge $N$, which is impossible as the first Chern 
class vanishes and U(1) bundles have always vanishing second Chern class. 
Suppose flat factors prevent us from making the argument. In that case we 
can remove the flat factor and go down in rank by one (which keeps the first 
Chern class zero). This can be repeated until we are left with an SU($N-k$)
charge one instanton without flat factors, or completely reduce to flat
factors which, however, can not support topological charge.

Assuming no flat factors for the self-dual connection $\omega$, the kernel 
of $D^\phd_z D^\dagger_z$ is trivial and the well-defined Greens function 
$G_z=(D_z^\phd D_z^\dagger)^{-1}$ commutes with the quaternions $\sg_\mu$. 
It is now straightforward to show that $\hat\omega$ is self-dual as well
\beq
\hat F^{ij}(z)=\hat d\hat\omega^{ij}+\hat\omega^{ik}\wedge\hat\omega^{kj}
=<\hat d\Psi_z^{(i)}|1-P_z|\hat d\Psi^{(j)}_z>,\label{eq:Ncurv0}
\eeq
where $P_z$ is the projection on $\ker D_z$,
\beq
P_z\equiv \sum_{k=1}^\nu|\Psi_z^{(k)}><\Psi_z^{(k)}|=1-D_z^\dagger 
G_z D_z^\phd.
\eeq
The equality to the right-hand side is trivially true for a zero-mode, whereas 
the orthogonal complement of $\ker D_z$ is described by the image of $D_z$, on 
which both sides act as the identity. Next we use $D_z\hat d\Psi_z^{(i)}=
[D_z,\hat d]\Psi_z^{(i)}=-2\pi i dz_\mu\sg^\mu\Psi_z^{(i)}$ to find
\beq 
\hat F_{\mu\nu}^{ij}(z)=8\pi^2<\Psi^{(i)}_z|\sgbar_{[\mu}\sg_{\nu]}G_z|
\Psi_z^{(j)}>.\label{eq:Ncurv2}
\eeq
Self-duality immediately follows from the fact that $\sgbar_{[\mu}\sg_{\nu]}
=\sigma_i\bar{\eta}^i_{\mu\nu}$ is self-dual. 

For $T^4$, applying the Nahm transformation the second time, it can be 
shown that $\omega$ is retrieved identically, and the explicit form of 
$\hat \Psi_x(z)$ in terms of $\Psi_z(x)$ allows one to show that metric and 
hyperK\"ahler structures of the moduli spaces are preserved under the Nahm 
transformation. In other words, if we denote by $\Ss{M}_{N,\nu}$ the moduli 
space of SU($N$) charge $\nu$ instantons, the Nahm transformation induces a 
map between moduli spaces, $\Ss{N}:\Ss{M}_{N,\nu}\rightarrow\Ss{M}_{\nu,N}$, 
which is an involution that preserves the natural metric and hyperK\"ahler 
structure of the moduli space.\cite{BrvB} The dimension of the moduli space, 
$4N\nu$, is indeed symmetric under interchanging $\nu$ and $N$.  

\begin{figure}[htb]
\vskip3.5cm
\includegraphics{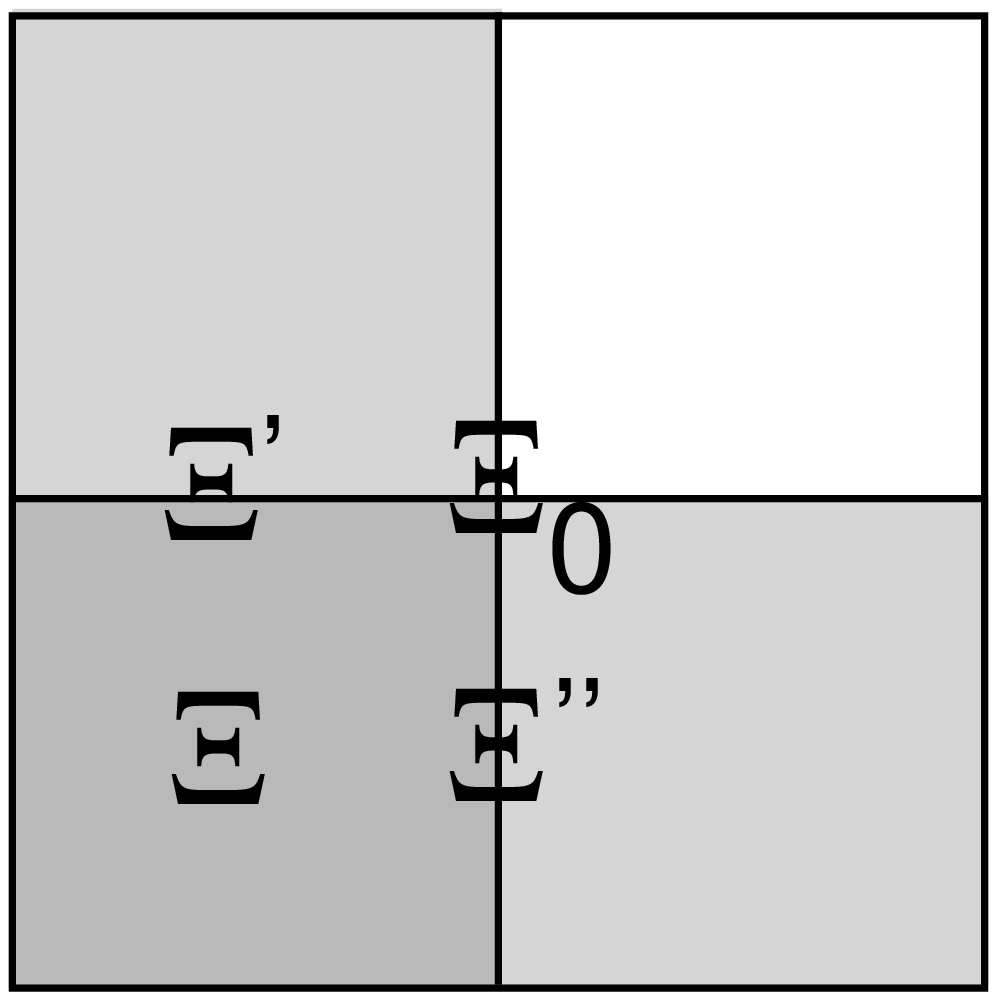}
\includegraphics{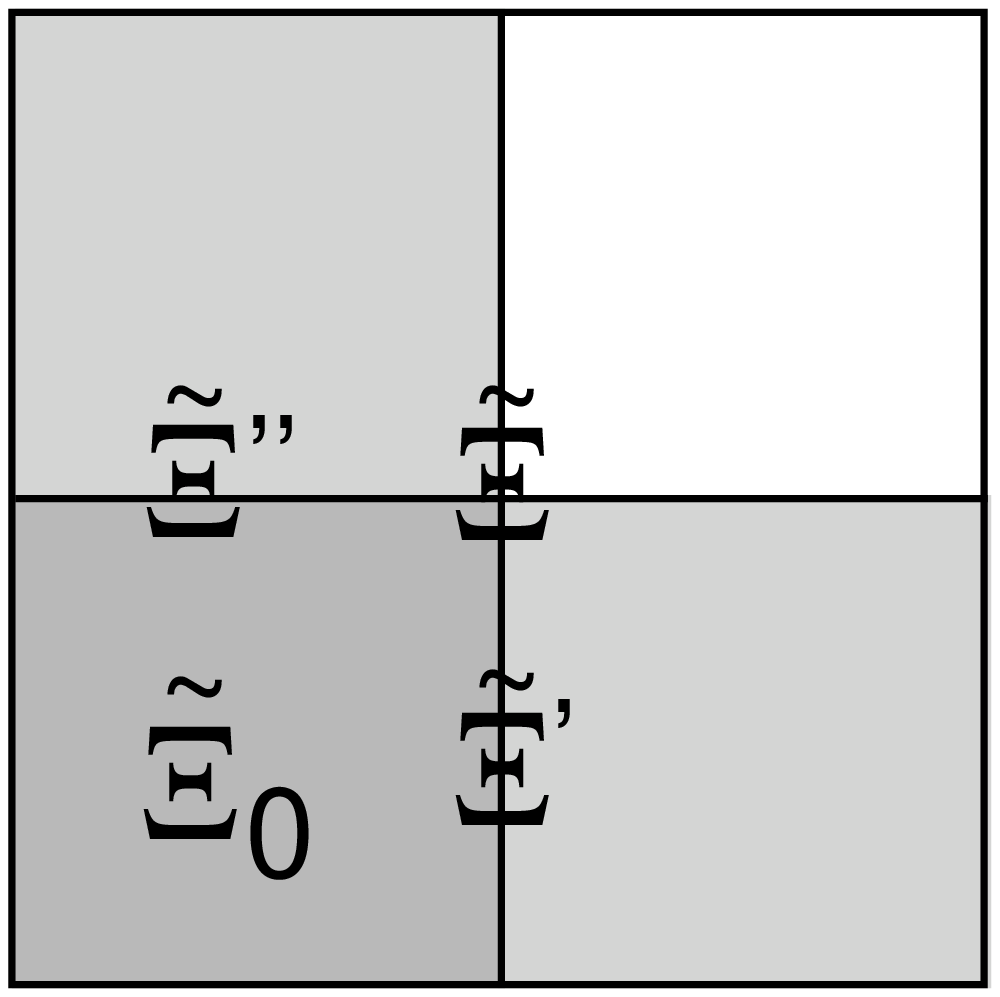}
\caption{Shown is a cross section through the 1-2 plane of the various
unit cells that appear in the Nahm transformation for SU(2) with twisted 
boundary conditions $n_{12}=-n_{21}=1$ on $R^4/\Xi$. We have chosen here
$||a^{(1)}||=||a^{(2)}||=\half\protect\sqrt{2}$. Duplicated unit cells 
without twist are $\Xi'$ and $\Xi''$, their union is denoted by $\Xi_0$. Its 
dual, $\tilde\Xi_0$, gives the torus on which the Nahm transformed gauge 
field lives, with the same gauge group and twist as for the gauge field 
we started with. The unit cell $\tilde\Xi_0$ is also obtained by 
intersecting $\tilde\Xi'$ and $\tilde\Xi''$, the duals of the duplicated 
unit cells. When a symbol overlaps with different cells it belongs to all 
of those cells as a whole.}\label{fig:Ntwist}
\end{figure}

The Nahm transformation on $T^4$ can be defined to include twisted boundary 
conditions, as Gonz\'alez-Arroyo has shown recently.\cite{Tony} The basic idea 
is simple: One duplicates the unit cell, also denoted by $\Xi$, in the various 
directions as many times as is necessary to remove the twist. Let us assume 
for SU(2) $n_{12}=-n_{21}=m_3=1$ and the rest of the twist factors trivial. We 
can duplicate $\Xi$ either in the 1- or in the 2-direction in order to remove 
the twist, giving $\Xi'$ and $\Xi''$. In both cases we can apply the Nahm 
to the enlarged unit cells $\Xi'$ or $\Xi''$, see Fig.~\ref{fig:Ntwist}.
In each case its dual $\tilde\Xi'$ or $\tilde\Xi''$ is shorter by a factor 
2 in the direction we originally duplicated. These two choices of unit cell 
intersect in a cell we call $\tilde\Xi_0$ that is twice shorter in each of 
the two directions. Due to the intersection it contains all the gauge 
invariant information of $\hat\omega'$ and $\hat\omega''$. The gauge field 
therefore has {\em half-periods} associated with twisted boundary conditions
on $\real^4/\tilde\Xi_0$. In this particular example the resulting twist 
is the same as we started with. The argument can be generalized, either along 
the above lines, or using so-called flavor multiplication,\cite{Tony,Synt}
to arbitrary SU($N$) and twist. The example shown in Fig.~\ref{fig:Ntwist} 
(with identical behavior in the 0-3 plane), maps the charge $\half$ instanton 
to the {\em same} solution. Another example of self-similarity occurs 
for SU(2) in the charge 2 sector. When applied to the special solutions of 
't~Hooft,\cite{Tho4} this can be used as an example where the Nahm 
transformation can be worked through analytically. The resulting Nahm 
transformed connection is again of the special form.\cite{Buck} 

\begin{figure}[htb]
\vspace{4.7cm}
\includegraphics{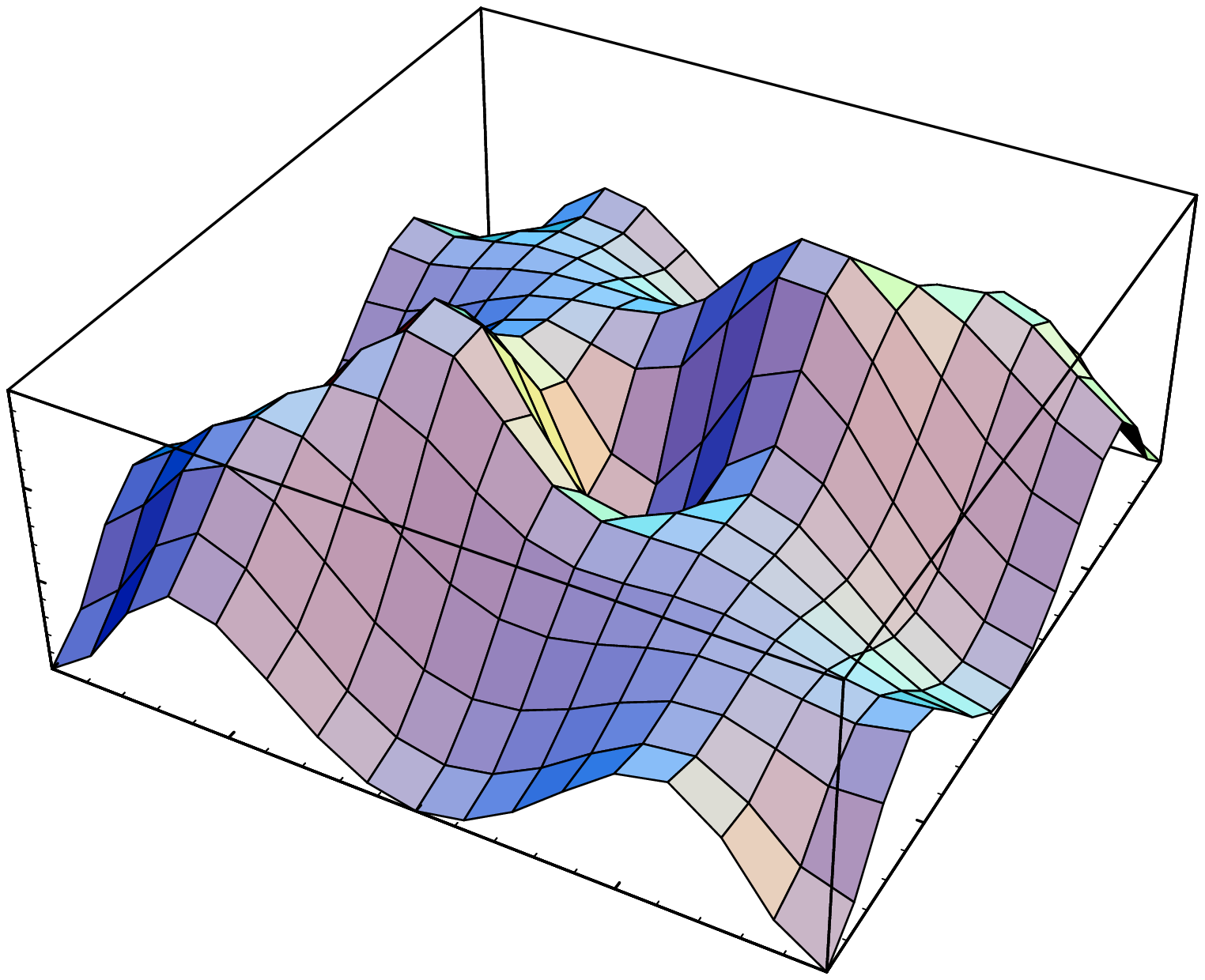}
\includegraphics{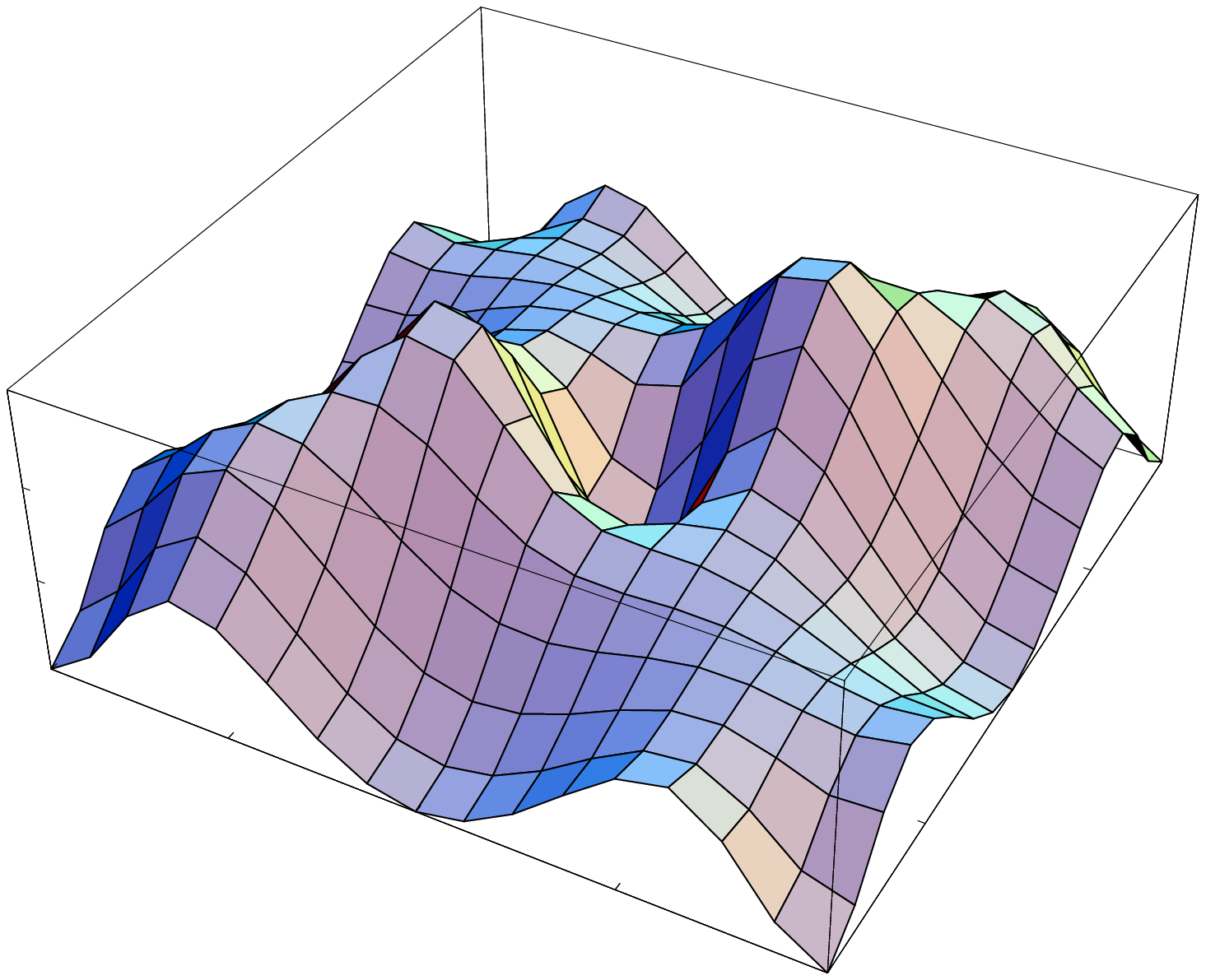}
\caption{Comparison\protect\Cite{Synt} of $\hat\vek E^2$ in the plane 
$z_3=0.5$ for the Nahm transformation of an SU(2) instanton on an $8^3
\times 40$ lattice (scaled to $L=1$) and twist $\vek k=(1,1,1)$ in the 
time direction (left), with the analytic result on $T^3\times\real$ for
$P_j^-=(0.86,0.76,0.08)=-P_j^+$ (right).}\label{fig:NT}
\end{figure}

To have the Nahm transformation help us finding explicit instanton solutions, 
simplifications have to be arranged for. These happen to be perfectly geared to 
finding charge one instanton solution on $T^3\times\real$. For charge one the 
Nahm transformed gauge field is abelian and with one period infinite, the dual 
period collapses to zero. The decompactification limit $T\rightarrow\infty$ is 
associated with dimensional reduction. This reaches a dramatic height for the 
case that all periods are send to infinity, relevant for instantons on 
$\real^4$. The dual is now a single point, explaining why the ADHM 
construction\Cite{ADHM} is in essence algebraic. However, under the 
decompactification limit, a partial integration is required in going from 
Eq.~\Ref{eq:Ncurv0} to Eq.~\Ref{eq:Ncurv2},
\bea
\hat F_{\mu\nu}^{ij}(z)&=&4\pi i\bar\eta^k_{\lambda\,[\mu}\oint 
d^3_\lambda x~\frac{\partial}{\partial z_{\nu]}}\left(\Psi^{(i)}_z
\right)^\dagger\!(x)~\sg_k\left(G_z\Psi_z^{(j)}\right)(x)\nonumber\\
&&+8\pi^2\bar\eta^k_{\mu\nu}<\Psi^{(i)}_z|\sg_k G_z|\Psi_z^{(j)}>.
\label{eq:Ncurv1}
\eea
These boundary terms destroy self-duality of $\hat F$, which can be
repaired\Cite{Nahm,CoGo} for instantons on $\real^4$ and for calorons
(periodic instantons) on $\real^3\times S_1$. Except for $\real^4$, for which 
there is no freedom, the location of the singularities are fixed by the 
asymptotically flat connections at $t\rightarrow\pm\infty$, which are required 
to occur to keep the action finite. The Weyl-Dirac hamiltonian in the 
background of a flat connection has generically a mass gap. The mass gap 
vanishes, however, in particular when the flat connection is pure gauge. The
flat connections are characterized by $\vek A_z=2\pi i~\diag\bigl(\vek w^{(1)}
+\vek z,\cdots,\vek w^{(N)}+\vek z\bigr)$, and for the mass gap to vanish it is
sufficient that $||\vek w^{(k)}+\vek z||=0$ for any of the $N$ values of $k$. 
With two asymptotically flat connections, their are $2N$ corresponding values of
$\vek z$ (which are defined modulo $\tilde\Xi$) where the mass gap vanishes. 
Only here the boundary terms in Eq.~\Ref{eq:Ncurv1} can be non-vanishing. A 
simple steepest descent analysis in their neighborhood shows that they act as 
point sources with {\em abelian} electric {\em and} magnetic charges $\pm\pi$, 
enforced by charge quantization (positive charges associated with the flat 
connection at $t=\infty$). This ensures the magnetic sources are Dirac 
monopoles with unobservable Dirac strings.\cite{PLB0} Outside of the 
singularities the {\em abelian} field is self-dual and, noting the $z_0$ 
independence, $\hat B_j(\vek z)=\hat E_j(\vek z)=-\partial\hat A_0(\vek z)/
\partial z_j$, where
\beq
\hat A_0(\vek z)=\frac{i}{2}\sum_{j=1}^N\sum_{\vek n\in\tilde\Xi}\left(
||\vek w_+^{(j)}+\vek z+\vek n||^{-1}-||\vek w_-^{(j)}+\vek z+\vek n||^{-1}
\right).
\eeq
For the usual normalization of abelian fields one multiplies this with $-i$. 
The sum over the periods $\tilde\Xi$ on $\hat T^3$, although formally 
divergent, can be resummed in a rapidly converging series and one has an exact 
result\Cite{PLB0} for $\hat A_0(\vek z)$ (see Fig.~\ref{fig:NT}). 
One may argue from this that indeed time-periodic instanton solutions 
do not exist, as for $\vek w_+^{(j)}=\vek w_-^{(j)}$ we clearly find 
$\hat A_0(\vek z)=0$, which under the Nahm transformation leads to a 
trivial connection on $T^3\times\real$. With twisted boundary conditions 
in the time direction one has $\vek w_+^{(j)}=\vek w_-^{(j)}+\vek k/N$. 
Although it is not straightforward to solve the Weyl-Dirac equation on 
$\hat T^3$ in the background of an abelian self-dual gauge field generated 
by point charges, one may hope explicit solutions of charge one instantons 
on the torus can be found this way.

The situation with the twisted instantons of fractional charge is, perhaps
somewhat paradoxically, a bit more complicated. However, much understanding
has been gained through the numerical implementation of the Nahm transformation
on a lattice.\cite{GAPe} It has been found\Cite{Synt} that, in staying away 
from the exact decompactification limit, the Dirac monopole singularities are 
resolved into the fully non-abelian constituent monopoles that appear in the 
calorons,\Cite{KrvB} with the only difference that these calorons are now 
modified by the finite volume,\cite{GGMV} for which no exact analytic solutions
are available. They nevertheless fit very well to the infinite volume 
solutions, whereas away from the non-abelian cores of the constituent 
monopoles there is a nearly perfect fit to the abelian field of the singular 
Dirac monopoles. This aspect is illustrated in Fig.~\ref{fig:NT}, comparing 
the analytic result in the decompactification limit with the numerical result 
on a lattice. For further details one should consult the literature.\cite{Synt}
It is natural to conjecture, in the light of these results that the 
fractionally charged instanton is mapped to a single point charge on a torus 
with (abelian) C periodic boundary conditions\Cite{PoWi} in the 
decompactification limit.

\section{Gauge Fields on the three-Sphere}\label{sec:sphere}

The most important reason to study gauge fields on the three-sphere is that 
conformal equivalence of $S^3\times\real$ to $\real^4$ gives a very simple and
explicit construction for the instantons.\cite{Vbda,Hoso} This allows one to 
formulate in a precise way how the $\theta$ dependence can be encoded in the 
boundary conditions on the fundamental domain.\cite{Vbda,Vbvd} Due to the 
curvature of the sphere, at large volumes the corrections to the glueball 
masses are in powers of the inverse radius, as opposed to an exponential 
approach for the torus (to be discussed in Sec.~\ref{subsec:stable}). But 
ultimately, any geometry can be scaled-up to an infinite volume, and should 
in this limit give the same results. Therefore, by comparing different 
geometries we may indirectly get some useful information. 

We embed $S^3$ in $\real^4$ by considering the unit sphere (the radius $R$ 
can be reinstated on dimensional grounds where required), parametrized by a 
unit vector $n_\mu$. Alternative formulations, useful for diagonalizing the 
Faddeev-Popov and fluctuation operators, were developed by Cutkosky.\Cite{Cutk}
We can use the 't~Hooft tensors $\eta$ and $\bar{\eta}$ to define orthonormal 
framings\Cite{Lue2} of $S^3$, which were motivated by the particularly simple 
form of the instanton vector potentials in these framings. The framing for 
$S^3$ is obtained from the framing of $\real^4$ by restricting in the following
equation the four-index $\al$ to a three-index $j$ (for $\al=0$ one obtains the
normal on $S^3$, see Eq.~\Ref{eq:eta}),
\beq
e^\al_\mu=\eta^\al_{\mu\nu}n_\nu,\quad\bar{e}^\al_\mu=\bar{\eta}^\al_{\mu\nu}
n_\nu.\label{eq:frame}
\eeq
Note that $e$ and $\bar{e}$ have opposite orientations. Each framing defines 
a differential operator with associated (mutually commuting) angular momentum 
operators $\vek{L}_1$ and $\vek{L}_2$:
\beq
\pr^i = e^i_\mu \frac{\pr}{\pr x^\mu},\quad L_2^i = \frac{i}{2}~\pr^i\quad,
\quad\bar{\pr}^i = \bar{e}^i_\mu \frac{\pr}{\pr x^\mu},\quad
L_1^i = \frac{i}{2}~\bar{\pr}^i.\label{eq:diff}
\eeq
It is easily seen that $\Lkw = \vek{L}_2^2$, which has eigenvalues
$l(l+1)$, with $l=0,\half,1,\cdots$.

By identifying the logarithm of the radius in $\real^4$ as time in the geometry
$S^3 \times \real$, the (anti-)instantons are easily obtained from 
those\Cite{Bpst} on $\real^4$
\beq
A_0=\frac{\veps^\mu\cdot\sg_\mu}{2(1+\veps^\mu n_\mu)},\quad
A_i=\frac{\eps_{ijk}\sg^j\veps^k-(v+\veps^\mu n_\mu)
\sg_i}{2(1+\veps^\mu n_\mu)}.\label{eq:vecA}
\eeq
Here $\veps^k$, $\sg_j$ and $A_j$ are defined with respect to the framing 
$\bar{e}^j_\mu$ for instantons and with respect to the framing $e^j_\mu$ for 
anti-instantons, and we introduced
\beq
u=\frac{2 s^2}{1+b^2+s^2},\quad\veps^\mu=\frac{2sb^\mu}{1+b^2+s^2},\quad
s=\lm \exp(t).\label{eq:par}
\eeq
The unit quaternions $\sg_\mu$ were given in Eq.~\Ref{eq:quar}. The instanton 
describes tunneling from $A=0$ at $t=-\infty$ 
to $A_j=-\sg_j$ at $t=\infty$, over a potential barrier at $t=0$ that is lowest
when $b\equiv 0$. This configuration corresponds to a sphaleron,\cite{Klma}
i.e. the vector potential $A_j=-\half\sg_j$ is a saddle point of the energy 
functional with one unstable mode, corresponding to the direction ($u$) of 
tunneling. At $t=\infty$, $A_j=-\sg_j$ has zero energy and is a gauge copy of 
$A_j=0$ by a gauge transformation $h=n_\mu\sg^\mu$ with winding number one.

We will be concentrating our attention to the 18 modes that are degenerate
in energy to lowest order with the modes that describe tunneling through
the sphaleron and ``anti-sphaleron". The latter is a gauge copy by a gauge
transformation $h'=n_\mu\sgbar^\mu$ with winding number $-1$ of the sphaleron.
The two dimensional space containing the tunneling paths through these
sphalerons is consequently parametrized by $u$ and $v$ through
\beq
A_\mu(u,v)=\left(-u\bar{e}^a_\mu-v e^a_\mu\right)\frac{\sg_a}{2}.
\label{eq:Auv}
\eeq
The gauge transformation $h$ with winding number one is easily seen to map 
$(u,v)=(w,0)$ into $(u,v)=(0,2-w)$. The 18 dimensional space is defined by
\beq
A_\mu(c,d)=\left(c^a_j\bar{e}^j_\mu+d^a_j  e^j_\mu\right)\frac{\sg_a}{2}=
A_j(c,d)\bar{e}_\mu^j.\label{eq:Acd}
\eeq
The $c$ and $d$ modes are mutually orthogonal and satisfy the Coulomb gauge 
condition, $\pr_j A_j(c,d)=0$. This space contains in it the $(u,v)$ plane 
through $c^a_j=-u\dl^a_j$ and $d^a_j=-v\dl^a_j$. The significance of this 18 
dimensional space is that the energy functional\Cite{Vbda}
\bea
\Ss{V}(c,d)&\equiv&-\int_{S^3}\frac{1}{2}\tr(F_{ij}^{\,2})=\Ss{V}(c)+\Ss{V}(d)+
\frac{2 \pi^2}{3}\left\{ (c^a_i)^2 (d^b_j)^2-(c^a_i d^a_j)^2\right\},
\nonumber\\ \Ss{V}(c)&=&2\pi^2\left\{2(c^a_i)^2+6\det c+\frac{1}{4}
[(c^a_i c^a_i)^2-(c^a_i c^a_j)^2]\right\},\label{eq:Spot}
\eea
is degenerate to second order in $c$ and $d$. Indeed, the quadratic
fluctuation operator \Ss{M} in the Coulomb gauge, defined by
\bea
-\int_{S^3}\frac{1}{2}\tr(F_{ij}^{\,2})&=&\int_{S^3}\tr(A_i\Ss{M}_{ij}A_j)+
\Order{A^3},\nonumber\\ \Ss{M}_{ij} &=&2\vek{L}_1^2\dl_{ij}+2\left(
\vek{L}_1+\vek{S}\right)^2_{ij},\quad S^a_{ij}=-i\eps_{aij},\label{eq:fluct}
\eea
has $A(c,d)$ as its eigenspace for the (lowest) eigenvalue $4$. These
modes are consequently the equivalent of the zero-momentum modes on the
torus, with the difference that their zero-point frequency does not vanish.

To find the fundamental region by minimizing the norm functional,
Eq.~\Ref{eq:gAnorm}, we can use $FP_f(A)$ in Eq.~\Ref{eq:FPf} as a hermitian 
operator acting on the vector space $\Ss{L}$ of functions $h$ over $S^3$ with 
values in the space of the quaternions $\quat=\{q^\mu\sg_\mu|q\in\real^4\}$. 
The gauge group $\Ss{G}$ is contained in $\Ss{L}$ by restricting to the unit 
quaternions: $\Ss{G}=\{h\in\Ss{L}|h=h^\mu\sg_\mu,h\in\real^4,h_\mu h^\mu=1\}$. 
This allows us to extract detailed information about the Gribov and fundamental
region within the 18 dimensional space $(c,d)$. When minimizing the norm 
functional over $\Ss{L}$, instead of $\Ss{G}$, one obviously should find a 
smaller space $\tilde{\Lm}\subset\Lm$. With $\Ss{L}$ a linear space, $\tilde
\Lm$ can now also be defined by the condition that $FP_f(A)$ be positive,
\beq
\tilde{\Lm}=\{A\in\Gm|\ip{h,FP_f(A)~h}\geq0,\ \forall h\in\Ss{L}\}.
\label{eq:Lmt}
\eeq
Similar to the Gribov horizon, the boundary of $\tilde\Lm$ is now determined
by the location where the lowest non-trivial eigenvalue of $FP_f(A)$ vanishes. 
For the $(c,d)$ space it can be shown\Cite{Vbvd} that the boundary $\partial
\tilde\Lm$ will touch the Gribov horizon $\partial\Om$. It establishes the 
existence of singular points on the boundary of the fundamental domain due to 
the inclusion $\tilde{\Lm}\subset\Lm\subset\Om$. (By showing that the fourth 
order term in Eq.~\Ref{eq:Xexp} is positive,\cite{Vbvd} this is seen to 
correspond to the situation as sketched in Fig.~\ref{fig:bif}.)

\begin{figure}[htb]
\vspace{6.2cm}
\includegraphics{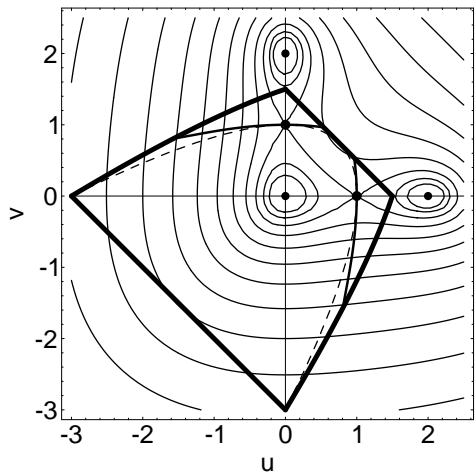}
\caption{Location of the sphalerons (large dots), classical vacua (smaller 
dots), the Gribov horizon (fat sections), the boundary of $\tilde\Lm$
(dashed parabola) and part of the boundary of the fundamental domain (full
curves, through the two sphalerons). Also indicated are the lines of equal 
potential in units of $2^n$ times the sphaleron energy.}\label{fig:bol}
\end{figure}

In order to simplify the notation, write $FP_\half\equiv FP_f$ and $FP_1\equiv 
FP$, with the indices related to the isospin. The associated generators are 
\beq
\vek T_\half=\half\vec\tau\qquad\mbox{and}\qquad\vek T_1=\half\ad\vec\tau.
\eeq
We can now make use of the SU(2)$\times$SU(2)$\times$SU(2) symmetry generated 
by $\vek L_1$, $\vek L_2$ and $\vek T_t$ to calculate explicitly the spectrum 
of $FP_t(A)$. One has
\beq
FP_t(A(c,d))=4\Lkw-\frac{2}{t}c^a_iT_t^aL_1^i-\frac{2}{t}d^a_iT_t^aL_2^i,
\eeq
which commutes with $\Lkw=\vek{L}_2^2$, but for arbitrary $(c,d)$ there are in 
general no other commuting operators (except for a charge conjugation symmetry 
when $t=\half$). Restricting to the $(u,v)$ plane one finds that
\beq
FP_t(A(u,v))=4\Lkw+\frac{2}{t}u\vek{L}_1\cdot\vek{T}_t+\frac{2}{t}v\vek{L}_2 
\cdot \vek{T}_t,\label{eq:FPuv}
\eeq
which in addition commutes with $\vek{J}_t=\vek{L}_1+\vek{L}_2+\vek{T}_t$,
the total angular momentum, and is easily diagonalized. Fig.~\ref{fig:bol} 
summarizes the results for this $(u,v)$ plane and also shows the 
equipotential lines, as well as exhibiting the multiple vacua and the 
sphalerons. As it is easily seen that the two sphalerons are gauge copies 
(by a unit winding number gauge transformation) with equal norm, they 
lie on $\partial\Lm$, which can be extended by perturbing around these 
sphalerons.\cite{Vbcu}

To obtain the result for general $(c,d)$ one can use the invariance under 
rotations generated by $\vek{L}_1$ and $\vek{L}_2$ and under constant gauge 
transformations generated by $\vek{T}_t$, to bring $c$ and $d$ to a standard 
form, or express $\det\left(FP_t(A(c,d))|_{l = \half}\right)$, which 
determines the locations of $\partial\Om$ and $\partial\tilde\Lm$, in terms 
of invariants. We define the matrices $X$ and $Y$ by $X^a_b=c^a_j c^b_j$ and
$Y^a_b=d^a_j d^b_j$, in terms of which
\bea
\det\left(FP_\half(A(c,d))|_{l=\half}\right)&=&\bigl[81-18\Tr(X+Y)+24(\det c+
\det d)\nonumber\\& &-(\Tr(X-Y))^2+2\Tr((X-Y)^2)\bigr]^2.\label{eq:fcd}
\eea
A two-fold multiplicity (the square) is due to charge conjugation symmetry. The
expression for $t=1$, that determines the location of the Gribov horizon in the
$(c,d)$ space,\cite{Vbvd} is somewhat more complicated. If we restrict to $d=0$
the result simplifies considerably. In that case one can bring $c$ to a 
diagonal form $c_i^a=x_i\delta_i^a$. Rotations and gauge transformations reduce 
to permutations of the $x_i$ and simultaneous changes of the sign of two of 
the $x_i$. One now easily finds the invariant expression ($\Tr(X)=\sum_i x_i^2$ 
and $\det c=\prod_i x_i$)
\beq
\det\left(FP_1(A(c,0))|_{l=\half}\right)=\left(2\det c-3\Tr(X)+27\right)^4.
\label{eq:det0}
\eeq

\begin{figure}[htb]
\vspace{4.6cm}
\includegraphics{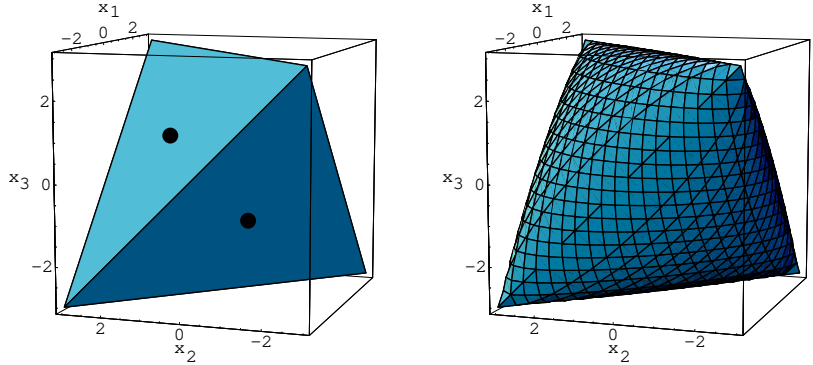}
\caption{The fundamental domain (left) for constant gauge fields on $S^3$, 
with respect to the framing $\bar{e}_\mu^j$, in the diagonal representation 
$A_j=x_j\sg_j$ (no sum over $j$). The dots on the faces indicate the 
sphalerons. On the right we show the Gribov horizon, which encloses the 
fundamental domain, coinciding with it at the singular boundary points 
along the edges of the tetrahedron.}\label{fig:tetra}
\end{figure}

In Fig.~\ref{fig:tetra} we present the results for $\Lm$ and $\Om$. In this 
particular case, where $d=0$, $\Lambda$ coincides with $\tilde\Lambda$, a 
consequence of the convexity and the fact that both the sphalerons (indicated 
by the dots) and the edges of the tetrahedron lie on $\partial\Lambda$, the 
latter also lying on $\partial\Om$. It is essential that the sphalerons do not 
lie on the Gribov horizon and that the potential energy near $\partial\Om$ is
relatively high. This is why we can take the boundary identifications near
the sphalerons into account without having to worry about singular boundary
points, as long as the energies of the low-lying states will be not much
higher than the energy of the sphaleron. It allows one to study the
glueball spectrum as a function of the CP violating angle $\theta$, but
more importantly it incorporates for $\theta=0$ the noticeable influence
of the barrier crossings, i.e. of the instantons. 

An effective hamiltonian for the $c$ and $d$ modes is derived from the one-loop
effective action.\cite{Vdhe} To lowest order it is given by 
\beq
H=-\frac{g^2(R)}{4\pi^2R}\left(\left(\frac{\partial}{\partial c_i^a}\right)^2+
\left(\frac{\partial}{\partial d_i^a}\right)^2\right)+\frac{1}{g^2(R)R}
\Ss{V}(c,d)+\frac{1}{R}\Ss{V}^{(1)}_{\rm eff}(c,d),\label{eq:Hbol}
\eeq
where $g(R)$ is the running coupling constant (related to the MS running
coupling by a finite renormalization, such that the kinetic term in 
Eq.~\Ref{eq:Hbol} has no corrections). The one loop correction to the 
effective potential is given by 
\bea
\Ss{V}^{(1)}_{\rm eff}(c,d)&=&\Ss{V}^{(1)}_{\rm eff}(c)+\Ss{V}^{(1)}_{\rm eff}
(d)+\kappa_7(c^a_i)^2 (d^b_j)^2+\kappa_8(c^a_i d^a_j)^2,\nonumber\\ 
\Ss{V}^{(1)}_{\rm eff}(c)&=&\kappa_1(c^a_i)^2+\kappa_2\det c+\kappa_3
(c^a_i c^a_i)^2+\kappa_4(c^a_i c^a_j)^2\label{eq:lpot}\\
&&+\kappa_5(c^a_i)^2\det c+\kappa_6(c^a_i c^a_i)^3.\nonumber
\eea
with the following numerical values for the coefficients
\bea
\kappa_1=-0.24534599851796,&&\kappa_5=-0.84996541224534,\nonumber\\
\kappa_2=+3.66869179814223,&&\kappa_6=-0.06550330854836,\nonumber\\
\kappa_3=+0.50070320309661,&&\kappa_7=-0.36171221599671,\nonumber\\
\kappa_4=-0.83935963341300,&&\kappa_8=-2.29535686135471.\label{eq:Scoefs}
\eea
Along the tunneling path (e.g. $c_i^a=-u\delta_i^a$ and $d=0$) the effective
potential can be calculated with simpler methods, providing an important check.

\begin{figure}[htb]
\vspace{5.0cm}
\includegraphics{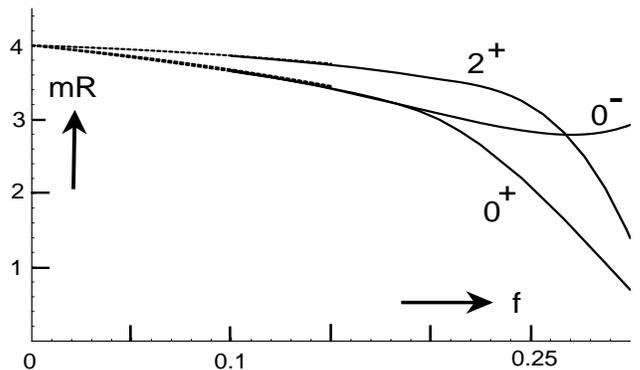}
\caption{The full one loop results for the masses of scalar, tensor and odd
glueballs on $S^3$ as a function of $f=g^2(R)/2\pi^2$ for $\theta=0$. The
dashed lines correspond to the perturbative result.}\label{fig:Sspec}
\end{figure}

Unlike for the torus, where in lowest order all excitations in the 
zero-momentum modes are degenerate and Bloch perturbation theory\Cite{Bloc}
provides a rigorous definition of the effective hamiltonian, one has to rely 
here on an adiabatic approximation not controlled by the coupling constant. 
The low lying excitations in the $c$ and $d$ modes are well below the 
excitations of the modes that were integrated out, justifying the adiabatic 
approximation.\cite{Vbda,Vdhe} It provides enough room to achieve a 
satisfactory understanding of the non-perturbative dynamics due to spreading 
of the wave functional to the boundary of the fundamental domain. The boundary 
conditions are chosen such that the gauge and (left and right) rotational 
invariances are preserved and that they coincide with the appropriate boundary 
conditions near the sphalerons. Projection on the irreducible representations 
of these symmetries is essential to reduce the size of the matrices to be 
diagonalized in a Rayleigh-Ritz analysis. All this could be implemented in a 
tractable way,\cite{Vdhe} see Fig.~\ref{fig:Sspec}. 

In perturbation theory the $0^-$ ``oddball" is slightly {\em lighter} than the 
$0^+$ glueball state. In lowest order the $0^+$, $0^-$ and $2^+$ glueball 
states are all degenerate, lifted in the next order in perturbation theory, 
by an amount independent of the one loop corrections
\bea
(m_{0^-}-m_{0^+})R&=&-g^2(R)/4\pi^2+\Order{g^4},\nonumber\\
(m_{2^+}-m_{0^+})R&=&5g^2(R)/6\pi^2+\Order{g^4},
\eea
whereas $m_0^+R=4-(1.25-\kappa_1)g^2(R)/2\pi^2+\Order{g^4}$. This perturbative 
result was also found by Diekmann.\cite{Diek} (His one loop corrections, 
however, do not survive the test of reproducing the effective potential along 
the tunneling path.) Clearly, the $0^-$ glueball being the lightest state is an
artifact of the finite volume, and it is an important test to see if 
non-perturbative effects that set in when going to larger volumes are able to 
correct for this unwanted feature. Indeed, when including the effects of the 
boundary of the fundamental domain, the $0^-/0^+$ mass ratio rapidly increases 
from slightly below 1 to above. At the same time the slow rise of the $2^+/0^+$
mass ratio becomes more rapid. This behavior is observed both with and without 
the one loop corrections included. In the latter case the non-perturbative 
effects set in at $g(R)\sim 2$. Beyond $g(R)\sim 2.8$ it can be shown that the 
wave functionals start to feel parts of the boundary of the fundamental domain 
which the present calculation is not representing properly. This value of 
$g(R)$ corresponds to a circumference of roughly 1.3 fm, when setting the scale
as for the torus, assuming the scalar glueball mass in both geometries at this 
intermediate volume to coincide. That instanton effects are largely responsible
for the spin splittings is further confirmed by studies in the instanton liquid 
model,\cite{Shur} which concludes that instantons lead to an attractive, 
neutral and repulsive force in respectively the scalar, tensor and pseudo 
scalar glueball channels.\cite{ShSh}

\section{Large Volume Results}\label{subsec:infty}

We divide $\Ss{A}$ by the set of {\em all} gauge transformations $\Ss{G}$, 
including those that are homotopically non-trivial, to get the physical 
configuration space. All the non-trivial topology is then retrieved by the 
identifications of points on the boundary of the fundamental domain. This
becomes important when the wave functional spreads out in configuration space, 
which happens at large volumes, whereas at very small volumes the wave 
functional is localized around $A=0$ and one need not worry about these 
non-perturbative effects. That these effects can be dramatic, even at 
relatively small volumes (above a tenth of a fermi across), was demonstrated 
for the case of the torus. It has historically been very important that the 
torus with periodic boundary conditions provides a relatively wide window, 
from a tenth of a fermi to three quarters of a fermi, where these 
non-perturbative effects can {\em reliably} be taken in to account, and were 
carefully verified in lattice Monte Carlo studies. It shows the intricate 
dynamics due to the intrinsic non-linearities of the theory and the fact that 
the issue of gauge copies can not be ignored and is important for the infrared 
dynamics of the theory.\cite{Grib}

The hierarchy of boundary effects is effectively described by the various saddle
points, that typically lie at the boundary of the fundamental domain. An 
important lesson can be learned from comparing different geometries, between 
which the structure of the fundamental domain deviate considerably. As we 
stressed before, the shape of the fundamental domain is independent of $L$ 
if the gauge field is expressed in units of $1/L$. It is only the strength 
of the coupling constant that controls the spreading of the wave functional. 
Suppose that the coupling constant will grow without bound. This would make 
the potential irrelevant and makes the wave functional spread out over the 
whole of field space (which could be seen as a strong coupling expansion). 
If the kinetic term would have been trivial, the wave functionals would be 
``plane waves'' on a space with complicated boundary conditions. In that 
case it seems unavoidable that the infinite volume limit would depend on 
the geometry (like $T^3$ or $S^3$) that is scaled-up to infinity. Due to 
the non-triviality of the kinetic term this conclusion cannot be readily 
made and our present understanding only allows comparison in volumes around 
one cubic fermi. However, one way to avoid this undesirable dependence on the 
geometry is that the vacuum is unstable against domain formation. As periodic 
subdivisions are space filling on a torus, this seems to be the preferred 
geometry to study domain formation. It is hard to formulate this in a precise 
way, let alone find an order parameter for this domain formation. But let us 
assume domain formation does occur.

Since the ratio of the square root of the string tension to the scalar glueball
mass shows no structure around $L=0.75$ fermi, we assume that within a domain 
both reach their large volume value. The color electric string now arises from 
the fact that flux that enters a domain has to leave it at the opposite 
side. Flux conservation with these building blocks automatically leads to 
a string picture, with a string tension as computed within a single domain and 
a transverse size of the string equal to the average size of a domain. The 
tensor glueball in an intermediate volume is heavily split between the doublet 
($E^+$) and triplet ($T^+_2$) representations of the cubic group, with resp. 
0.9 and 1.7 times the scalar glueball mass. This implies that the tensor 
glueball is at least as large as the average size of a domain. Rotational 
invariance in a domain-like vacuum comes about by averaging over all 
orientations of the domains. This is expected to lead to a mass which is the 
multiplicity weighted average of the doublet and triplet, yielding a mass of 
1.4 times the scalar glueball mass. In the four dimensional euclidean context, 
$O(4)$ invariance makes us assume that domain formation extends in all four 
directions. The deconfining temperature would again be set by the average
domain size. As is implied by averaging over orientations, domains will not 
neatly stack. There will be dislocations which most naturally are gauge 
dislocations. A point-like gauge dislocation in four dimensions is an 
instanton, lines give rise to monopoles and surfaces to vortices. In the latter
two cases most naturally of the  $Z_N$ type. We estimate the density of these 
objects to be one per average domain size. We thus predict an instanton density
of $3.2\,{\rm fm}^{-4}$, with an average size of $1/3$ fermi. For monopoles 
we predict a density of $2.4\,{\rm fm}^{-3}$. The SU(2) spectra in volumes 
around the domain size, for the sphere and the torus, are compatible with 
$m(2^+)/m(0^+)\sim 1.5$ and $m(0^-)/m(0^+)\sim 1.7$. In this picture, we would 
further predict $\sqrt{K}/m_{0^+}\sim 0.24$. All these results seem to give the
right order of magnitude, close to the values one measures in a large volume, 
\bea
&&m(2^+)/m(0^+)= 1.46\pm0.09,\quad\sqrt{K}/m_{0^+}=0.267\pm0.009,\nonumber\\
&&m(0^-)/m(0^+)=1.78\pm0.24,
\eea
taken from a recent review by Teper.\cite{Tepe} It would be useful to have
{\em high-precision} lattice Monte Carlo data around $L=1$ fermi. In this 
intermediate volume range, the nature of the states changes from being 
dominated by zero-momentum fields to genuine particle states. It is important 
to take this into account in choosing the proper variational basis for the 
operators in the analysis of the Monte Carlo data. It is not sufficiently
appreciated that it is exactly this change of the nature of the states
that can provide fundamental physical insight in the {\em formation} of the 
mass gap and the confining string. 

The studies reported so far in small and intermediate volumes were from first
principles. The volume plays the role of the control parameter to keep the 
strength of the interactions in check. The challenge is to find which are 
the effective degrees of freedom that drive the formation of the mass gap. 
The lesson we learned is that large field fluctuations are essential. They 
enter as sphaleron configurations in the hamiltonian formulation, and are thus 
associated with instantons. Nevertheless, it seems inescapable that instantons, 
monopoles and vortices {\em all} have to be part of the equation,\cite{Edin}
in one way or another.

\subsection{Volume Dependence of Stable Particle Masses}\label{subsec:stable}

Once a mass gap is being formed, this gives of course the relevant degrees of 
freedom to describe the large distance behavior. For the torus, isolating the 
volume dependence to come from the propagator winding around the boundary of 
the box, has led L\"uscher to derive the exponential approach of stable 
particle masses.\cite{Lue0} The method makes cunning use of the fact that 
the finite volume propagator is given by $\Delta_L(x)=\sum_{n\in\zahlen^4}
\Delta(x+nL)$, where $\Delta(x)$ is the propagator in the infinite volume.
With a mass gap the sum over periods will converge, since $\Delta(x)\approx 
e^{-m|x|}$. Physically $\Delta(x+nL)$ can be identified with propagation from 
$0$ to $x$, going in addition $n$ times around the box. Going around only once 
gives the leading finite volume correction of the order $e^{-ML}$. The value 
of $M$ depends on the mass gap, and which type of virtual corrections 
contribute to the self-energy of the particle under consideration. Obviously, 
without interactions their would be no volume dependence. 

Applying these ideas to QCD, one assumes that the long distance behavior is 
described by an effective theory. In the presence of light quarks this is 
of course the chiral effective lagrangian. For the calculation of the volume 
dependence of the glueball, it is assumed to be an effective scalar theory. 
Confinement implies that propagation of a quark around the boundary, unlike 
in a small volume, is not an option for large $L$ due to the confining string 
it has to stretch. Similarly, gluons are confined and thus cannot propagate 
around the volume. It is only the physical colorless states that can propagate 
around the box at large $L$. Therefore, the fractional periods that appear in 
the perturbative expansion with twisted boundary conditions, see 
Sec.~\ref{subsec:twist}, have no effect on the finite size corrections in a 
large volume. Indeed twisted boundary conditions are devised such that all 
gauge invariant quantities are strictly periodic.\cite{Tho1}

Vertex functions are not affected by the finite volume, and the expression 
of a Feynman diagram in a finite volume is exactly of the same form as in 
an infinite volume, except that integrating over the vertex positions is 
restricted to the finite volume. The exponential suppression is caused by the 
finite volume modification in the propagator, specific for periodic boundary 
conditions. For the sphere, due to the curvature of the manifold, there will be
algebraic corrections, and the volume corrections are much harder to establish.
To keep track of the volume dependence universal properties of one-particle
irreducible vertex functions are studied. A general Feynman diagram $\Ss{D}$ 
will correspond to the amplitude
\beq
\Ss{J}_L(\Ss{D})=\!\!\!\!\!\prod_{v\in\Ss{V}\backslash v_0}\prod_{\mu=0}^d
\int_0^{L_\mu}\!\!\rd x_\mu(v)\exp(i\sum_ep(e)\cdot x(e))\vv
\prod_{\ell\in\Ss{L}}\Delta_L(x(f(\ell))-x(i(\ell))),
\eeq
where $p(e)$ are external momenta. The set of vertices $v$ ($v_0$ those that 
are connected to an external line) and propagators $\ell$ are denoted by 
$\Ss{V}$ and $\Ss{L}$. The product of all vertex factors (coupling 
constants) is given by $\vv$. 

To single out those contributions that correspond to propagating $n$ times
around the box, one has to take into account the invariance
\beq
x(v)\rightarrow x(v)+m(v)L,\quad n(\ell)\rightarrow n(\ell)+m(f(\ell))-
m(i(\ell)),
\eeq
where $m_\mu(v)\in\zahlen$ and $n(\ell)$ occurs in the expansion of 
$\Delta_L(x)$. With graph theory L\"uscher\Cite{Lue0} showed that the sum over
$n(\ell)$ splits in a sum over the orbits $[n]$ and a sum over the integers 
$m(v)$. Obviously one has $\sum_m\int_0^Ldx=\int_{-\infty}^{\infty}dx$ and
only the sum over the orbits remains, 
$\Ss{J}_L(\Ss{D})=\sum_{[n]}\Ss{J}_L(\Ss{D},[n])$,
\bea
\Ss{J}_L(\Ss{D},[n])=\prod_{v\in\Ss{V}\backslash v_0}\prod_{\mu=0}^d
\int_{-\infty}^{\infty}\rd x_\mu(v)\exp(i\sum_ep(e)\cdot x(e))\vv&&\nonumber\\
\times\prod_{\ell\in\Ss{L}}\Delta(x(f(\ell))-x(i(\ell))+n(\ell)L).&&
\eea
The advantage of this decomposition is that for the orbit $[n]=[0]$, this is 
precisely the infinite volume expression, i.e. $\Ss{J}_L(\Ss{D},[0])=
\Ss{J}_\infty(\Ss{D})$. When $[n]\neq[0]$ there is net winding around
the boundary, the length of the extra winding given by $W([n])$, thereby
leading to an exponential suppression,
\beq
\Ss{J}_L(\Ss{D},[n])\approx \exp\bigl[-mL\kappa(\{p(e)\})W([n])\bigr].
\eeq
The constant $\kappa(\{p(e)\})$ measures, in units of the mass gap $m$, the 
lowest momentum flowing through the propagators, constrained by the fixed 
external momenta. For a self-energy graph $\kappa(\{p(e)\})=\half\sqrt{3}$,
with the external momenta on-shell $p(e)=(im,\vek 0)$, the only case needed 
here. Furthermore, $W([1])=1$ for a ``simple'' orbit $[1]$ (where $|n(\ell)|
=1$, for one ling and 0 for all others) and $W([n])\geq\sqrt{2}$ for all other 
cases.\cite{Lue0} 

To determine the finite volume corrections for the lightest stable one-particle
state, one takes $L_0=\infty$ and $L_j=L$ (this simply implies we always have 
$n_0=0$). The mass is measured using $\langle\phi(t)\phi(0)\langle_L=
\exp(-M(L)t)$, where $\phi(t)$ is an interpolating field for this state. This 
definition of the mass, also used in lattice calculations, coincides with the 
pole in the finite volume two-point function, $G_L^{-1}(iM(L),0)=0$. The 
two-point function $G_L(p)$ is defined by $G_L^{-1}(p)=p^2+m^2-\Sigma_L(p)$, 
where the normalization of the self-energy $\Sigma$ is such that $\Sigma_\infty
(im,\vek 0)={\partial^2\over\partial p^2}\Sigma_\infty(im,\vek 0)=0$, which 
implies that in leading order $\delta M(L)=-{1\over 2m}(\Sigma_L-\Sigma_\infty)
(im,\vek 0)=-{1\over 4m}\Gamma^{(2)}((im,\vek 0),[1])$. This can be expressed 
in terms of the full four-point Greens function by
\beq
\Gamma^{(2)}(p,[1])=\int{\rd^4q\over(2\pi)^3}\left[6e^{iq_1L}G(q)\right]
G^{(4)}(p,q,-p,-q),
\eeq
where $\left[6e^{iq_1L}G(q)\right]$ comes from summing $\sum_{[n]=[1]}
\Ss{J}_L(\Ss{D}_2,[1])$ over the Feynman diagrams $\Ss{D}_2$ that 
contribute to the full two-point function, and making use of the cubic
invariance, to equate the contributions with $n_j=\pm1$ to $n_1=1$.

The full four-point function can be related with the help of Schwinger-Dyson
equations to the one-particle irreducible three- and four-point vertex
functions. By shifts of integration contours, taking the analyticity domains 
of the vertex functions properly into account, one finds that the dominating
contribution comes from the on-shell three-point vertex functions, denoted by 
$\lambda$, a physical three particle coupling constant. With $p=(im,\vek 0)$,
and both $q$ and $p+q$ ``on-shell", $q_1=i\sqrt{q_\perp^2+3m^2/4}$ and the 
integral over $q$, assuming the three point coupling does not vanish, can be 
shown to give\Cite{Lue0}
\beq
\delta M(L)=-{3\lambda^2\over 16\pi m^2L}\exp(-\half\sqrt{3}mL)+
\Order{e^{-mL}}.
\eeq
The subleading term, of order $e^{-mL}$, is due to the contribution coming
from the four-point function $\Gamma^{(4)}$. 

The derivation has been given for the volume dependence of the mass gap, e.g. 
for the scalar glueball mass in pure gauge theory. The results also apply for 
the cases of a stable boundstate particle, with mass $m_B=2m-E_B$, and for 
a nucleon that couples to a pion. One only needs to adjust the ``on-shell" 
condition, $q_1=i\sqrt{q_\perp^2+\mu_B^2}$, for the different kinematical 
situation. For the bound state one finds $\mu_B^2\equiv m^2-m_B^2/4$, which 
leads to $\delta M_B(L)=-3\lambda^2\exp(-\mu_B^\phd L)/(16\pi m_B^2
L)$, whereas for the nucleon $\mu_N^2\equiv m_\pi^2-m_\pi^4/4m_N^2$, giving 
$\delta M_N(L)=-9m_\pi^2g_{N\pi}^2\exp(-\mu_N^\phd L)/(16\pi m_N^2
L)$, where $\lambda$ has been expressed in the appropriate dimensionless 
pion-nucleon coupling constant $g_{N\pi}^\phd$. For the finite volume
correction to the pion mass in QCD, the pion three-point vertex vanishes and 
one gets a slightly more complicated results of order $\exp(-m_\pi L)$, 
involving an integral over the forward scattering amplitude.\cite{Lue0} 

\subsection{Volume Dependence from Scattering Phase Shifts}\label{subsec:scatt}

Also it was shown by L\"uscher how to extract scattering phase shifts from 
volume dependence.\cite{Lue4,Lue5} A less rigorous method, based on the notion 
of pseudo potentials, allows for a transparent way of understanding this all 
order result.\cite{La90} Consider the reduced hamiltonian $H=-m^{-1}
\partial_{\vek x}^2+V(\vek x)$ for two interacting particles in the center of 
mass in ordinary quantum mechanics with a (Bose) symmetric reduced wave 
functions, $\psi(\vek x)=\psi(-\vek x)$. When the potential has a finite 
range $\lambda$ ($V(\vek x)=0$ for $|\vek x|>\lambda$), the wave function is 
a plane-wave outside of the interaction region. Scattering theory tells us 
there is a unique relation between the incoming wave $e^{-ip|x|}$ and the 
outgoing wave $e^{ik|x|}$ in terms of the scattering phase shift $\delta(k)$
\beq
\psi(x)=e^{-ik|x|}+e^{2i\delta(k)}e^{ik|x|},\quad |x|>\lambda.\label{eq:2dwave}
\eeq
These states form a complete basis of scattering states. In a finite volume 
the periodic boundary condition, $\psi(x+L/2)=\psi(x-L/2)$, implies the 
following implicit equation for the momenta
\beq
e^{2i\delta(k)}e^{ikL}=1,\label{eq:2deq}
\eeq
which holds as long as $L>2\lambda$, showing how the volume dependence of 
two-particle states are related to the phase shift. For a vanishing potential 
the phase shift vanishes, $\delta(k)=0$, and one recovers the standard 
discretization of the momenta, $k=2\pi n/L$.

In more than one dimension a complete set of scattering wave functions is 
given by
\beq
\psi^{(\vek k)}_{\ell m}(\vek x)=[j_\ell(|\vek k|r)-\tan(\delta_\ell(\vek k))
n_\ell(|\vek k|r)]Y_{\ell m}(\hat\vek k), \label{eq:4dwave}
\eeq
where $\hat\vek k=\vek k/|\vek k|$, $r=|\vek x|$ and $j_\ell$, $n_\ell$ are the
spherical Bessel-functions. The energy is given by $E=\vek k^2/m$. It appears 
that the spherical nature of the scattered waves makes it impossible to impose 
periodic boundary conditions. One can, however, introduce the notion of a 
pseudo potential as was introduced for the hard-sphere bose gas by Huang and 
Yang.\cite{HuYa} The essential idea is equally simple as powerful. Replace 
$V(\vek x)$ by the simple pseudo potential $V_\delta(\vek x)$ such that for 
$|\vek x|>\lambda$, eq.~(\Ref{eq:4dwave}) is still an exact solution of the 
relevant Schr\"odinger equation. Subsequently one solves the Schr\"odinger 
equation with $V_\delta$ as its (energy dependent) potential, using periodic 
boundary conditions. We illustrate this by making the simplified assumption 
that all phase shifts vanish, except for $\delta_0$. In that case one easily 
verifies that the pseudopotential is given by
\beq
V_\delta(\vek x)=-{4\pi\over mk}\tan(\delta_0(\vek k))
\hat\delta_3(\vek x). \label{eq:pseudopot}
\eeq
Since we have to allow for wave functions that are singular at the origin, 
we can extend our class of functions to those that, when averaged over the 
angles, have a Laurent expansion $\sum_{n=-N}^\infty c_n r^n$. We define 
$\hat\delta_3(\vek x)\psi(\vek x)=c_0\delta_3(\vek x)$. Alternatively, 
$\hat\delta_3(\vek x)\psi(\vek x)={\partial\over\partial r}(r\psi(\vek x))
\delta_3(\vek x)$. We expand the wave function in plane waves, suitable for 
implementing periodic boundary conditions, $\psi(\vek x)=L^{-3}\sum b_{\vek n}
\exp(2\pi i\vek x\cdot\vek n/L)$. Substituting this in the relevant 
Schr\"odinger equation
\beq
(-{1\over m}{\partial^2\over\partial\vek x^2}+V_\delta(\vek x))\psi(
\vek x)={\vek k^2\over m}\psi(\vek x),\label{eq:helmholtz}
\eeq
is easily seen to give 
\beq
b_{\vek n}=-{4\pi\tan(\delta_0(\vek k))\over|\vek k|(\vek k^2-(2\pi
\vek n/L)^2)}c_0.\label{eq:fourier}
\eeq
With $c_0=\sum b_{\vek n}$ we thus find a relation between the momentum in 
the center of mass and the scattering phase shift {\em at this} momentum
\beq
{\tan(\delta_0(\vek k))\Ss{Z}_{00}(1;\vek q)\over 2\pi^2|\vek 
q|}=1,\quad\vek q={\vek k L\over 2\pi}.\label{eq:4deq}
\eeq
The zeta-function $\Ss{Z}_{00}(s;\vek q)\equiv\sum_{\vek n\in\zahlen^3}
(\vek n^2-\vek q^2)^{-s}$ is defined through analytic continuation in $s$.

To obtain this result $c_0$ should be non-vanishing. There can be ``singular'' 
solutions\Cite{Lue5} for which $c_0=0$ at momenta $\vek k=2\pi\vek n/L$, if 
there exists $\vek n^\prime\in\zahlen^3$, such that $|\vek n|=|\vek n^\prime|$.
In that case $\psi(\vek x)=\exp(2\pi i\vek n\cdot\vek x/L)-\exp(2\pi i\vek 
n^\prime\cdot \vek x/L)$ is regular and vanishes in the origin. If $\psi$ is in
the scalar representation ($A_1$) of the cubic group, this singular behavior 
only occurs when $\vek n^\prime$ is not related to $\vek n$ by a cubic 
transformation, which will make the momentum where this occurs quite large. 
Furthermore, restricting to the $A_1$ sector, eq.~(\Ref{eq:4deq}) will also be 
valid if the phase shifts only vanish for angular momenta $\ell\geq 3$. This is
because a spin 2 wave function decomposes in the $E$ and $T_2$ representations 
of the cubic group and hence does not couple to the scalar sector (note that 
due to the Bose symmetry all odd angular momentum phase shifts will vanish).

In field theory it can be shown that the reduced hamiltonian is replaced by an 
effective Schr\"odinger equation, that can be derived from the Bethe-Salpeter
equation for the four-point function.\cite{Lue4} Its energy-dependent 
``potential'' is proportional to the so-called Bethe-Salpeter kernel, with the 
range $\lambda$ determined by the polarization cloud (which is why we restrict 
the analysis to field theories with a mass gap). The fully relativistic 
two-particle energy $W$ is given by $W=2\sqrt{\vek k^2+m^2}=2\sqrt{m(m+E)}$, 
where $\vek k$ is the center of mass momentum of the scattering pair.
L\"uscher's\Cite{Lue5} all order analysis is based on studying the solutions 
of the Helmholtz equation $(\partial^2/\partial\vek x^2+\vek k^2)\psi(\vek x)
=0$ in a finite volume, allowing for power-like singularities at the origin. 
He demonstrates that truncating to a finite number of phase shifts 
$\delta_\ell$, ($\ell<\ell_{max}$) in general converges rapidly with 
$\ell_{max}$. For the simplified case that only $\delta_0$ is non-vanishing, 
the result is as in Eq.~\Ref{eq:4deq}.

One of the applications is computing the energy of the two-particle states
as a function of $L$. Using
\bea
\Ss{Z}_{00}(1;\vek q)&=&-{1\over\vek q^2}+Z_{00}(1;\vek 0)+
\vek q^2Z_{00}(2;\vek 0)+\Order{q^4},\nonumber\\
\tan(\delta_0(\vek k))&=&a_0|\vek k|+\Order{k^3},\label{eq:expansionZ}
\eea
introducing $Z_{00}(s;\vek q)=\Ss{Z}_{00}(s;\vek q)-(-\vek q^2)^{-s}$
and $a_0$, the so-called the scattering length. One can now straightforwardly 
iterate eq.~(\Ref{eq:4deq}) and find
\bea 
\vek k^2&=&-{4\pi a_0\over L^3}(1+{c_1 a_0\over L}+{c_2 a_0^2\over L^2})
+\Order{L^{-6}},\label{eq:expansion4deq}\\
c_1&=&{Z_{00}(1;\vek 0)\over \pi},\quad c_2={Z_{00}^2(1;\vek 0)-
Z_{00}(2;\vek 0)\over\pi^2}.\nonumber
\end{eqnarray}
The numerical values\Cite{Lue4,HuYa} are $c_1=-2.837297$ and $c_2=6.375183$. 
When the two-particle energies come close to the mass of a resonance, like the 
$\rho$ or $K$ meson coupling to two-particle pion states, the phase shifts get 
large and it becomes imperative to use the all order results.\cite{Lue5}

Many of the aspects of the large volume expansions for the one-particle and 
two-particle masses have been verified successfully for the two dimensional
non-linear sigma\Cite{LuWo} and Ising model\Cite{GaLa}, and for the four 
dimensional Ising\Cite{MoWe} and O(4) $\phi^4$ model in the symmetric 
phase.\cite{Fric} Recently the method has been revisited by Lellouch and 
L\"uscher\Cite{LeLu} for studying on the lattice non-perturbatively the 
decay $K\rightarrow\pi\pi$, important for understanding direct CP violation. 
By studying the decay in a finite volume, the transition amplitude of the kaon
to the discrete set of finite volume two-pion final states can be related to 
the infinite volume decay rate. With the advance in computer power this 
calculation will be feasible in the future.

\subsection{Goldstone Modes and Chiral Perturbation Theory}\label{subsec:chiral}

The finite size corrections to the string tension $K(L)$ also rely on an 
effective description, namely that of the bosonic relativistic string. The 
finite size behavior is described in terms of the force by
\beq
F(L)=\frac{dE(L)}{dL}=\frac{dLK(L)}{dL}=K+\frac{\pi}{3}L^{-2}+\Order{L^{-3}},
\eeq
with $K$ the infinite volume string tension. The so-called L\"uscher 
term\Cite{Lute} with the $1/L^2$ correction is universal. A simple way to 
understand this correction is in terms of the Casimir energy in one space 
dimension (the string) on a periodic interval of length $L$ (the length
of the string) for two massless scalar fields (the two independent transverse 
fluctuations of the string), $2\cdot\half\sum_k2\pi|k|/L=4\pi\zeta(-1)/L$ (in 
dimensional regularization, with $\zeta$ the Riemann zeta function). 
L\"uscher's original derivation was for a string with fixed end-points, for 
which the Casimir energy is a quarter of the one with periodic boundary 
conditions. The transverse fluctuations are examples of Goldstone bosons, that 
here arise due to the spontaneous breakdown of the transverse translational 
invariance in the background of a string.\cite{Lute} 

Goldstone bosons do of course also play a very important role in the low-energy
description of QCD in the form of pions. They occur with massless quarks due to
the spontaneous breakdown of chiral symmetry. With an explicit mass term for 
the quarks, the pions acquire a mass and the results of 
Secs.~\ref{subsec:stable} and \ref{subsec:scatt} apply as long as $m_\pi L\gg 
1$. For massless pions this can of course not be realized, and instead chiral 
perturbation theory becomes an effective tool. It also applies for $m_\pi\neq 
0$ provided $L$ is in the so-called mesoscopic range, which is $m_\pi^{-1}\gg 
L\gg \Lambda_{QCD}^{-1}$. The principle behind chiral perturbation theory is 
that the symmetries of the low-energy theory strongly constrain the effective 
lagrangian. The resulting sigma model has its symmetry determined by the number
of light quark flavors and the gauge group. For SU(3) with two light flavors, 
the up and the down quark, this is the O(4) sigma model spontaneously broken 
down to O(3) by the chiral condensate. Finite size effects occur in powers of 
$1/L$, but are determined by infinite volume quantities, computed essentially 
through replacing the continuous by discrete momenta.\cite{HaLe} 

From the practical point of view this gives access to these infinite volume 
quantities through finite size effects. An important example is the chiral 
condensate, which can be related through the Leutwyler-Smilga sum 
rules\Cite{LeSm} to eigenvalue distributions of the Dirac operator (averaging 
over the gauge fields). It is the average spacing of these eigenvalues near 
zero that through the Banks-Casher formula\Cite{BaCa} gives access to the 
chiral condensate $\langle\Delta\lambda\rangle=\pi(L^3T|\langle\overbar\psi
\psi\rangle|)^{-1}$. The roughness of the gauge field causes $\langle\Delta
\lambda\rangle$ to be inversely proportional to the volume ($L^3T$) of 
space-time. At weak coupling, like in a small volume, the typical eigenvalue 
spacing would be $1/L$, but in addition we saw in Sec.~\ref{subsec:quarks} 
there are no near zero eigenvalues and chiral symmetry is manifest. 

Remarkably, in the mesoscopic domain the low-lying eigenvalue distribution 
$\langle\rho(\lambda)\rangle$ is related to a universal function called the 
microscopic spectral density,\cite{ShVe} $\rho_S(u)=\lim_{V\rightarrow\infty}
(V\Sigma)^{-1}\langle\rho\left((V\Sigma)^{-1}u\right)\rangle$. It can be 
calculated in Random Matrix Theory, leaving the chiral condensate $\Sigma=
|\langle\overbar\psi\psi\rangle|$ as a single free parameter. We refer to 
a review by Verbaarschot\Cite{Verb} for details. 

\subsection{Electric-Magnetic Duality}\label{subsec:EMdual}

Much of the finite volume work in gauge theories has started with 't~Hooft's
paper on the torus with twisted boundary conditions.\cite{Tho1} We come back 
to it one more time to show how it leads to an electric-magnetic duality, 
which can be used to derive finite size results for the energy of magnetic 
flux in terms of the string tension, assuming one has a confining string.

Consider the gauge group SU($N$) and fix the magnetic flux $\vek m$ by twisted 
boundary conditions. In the $A_0=0$ gauge this can be specified as in 
Eq.~\Ref{eq:twb}. Given these boundary conditions, the remaining gauge freedom 
is specified by the twisted gauge transformations $h_\vek k$, defined in 
Eq.~\Ref{eq:tgt}, where $\vek k$ classifies the element of the homotopy group
$H_2(T^3,Z_N)\sim Z_N^3$. Two twisted gauge transformations with the same value
of $\vek k\in Z_N^3$ differ by a gauge transformation with integer winding 
number. A particular representative of $h_{\vek k}$ can be chosen such 
that\Cite{CMP0} $\nu(h_{\vek k})=-\vek k\cdot\vek m/N$, as we argued below 
Eq.~\Ref{eq:add}. The generalization of Eq.~\Ref{eq:bloch} to 
$\vek m\neq\vek 0$ can now be written as
\beq
\Psi([h_{\vek k}h_1^\nu]\,A)=e^{i\theta\nu}\exp\left(\frac{2\pi i 
\vek e^\phd_\theta\cdot\vek k}{N}\right)\Psi(A),\quad\vek e^\phd_\theta
\equiv\vek e-\frac{\theta}{2\pi}\vek m.\label{eq:mbloch}
\eeq
The non-trivial $\theta$ dependence implies that magnetic flux will carry in 
general a fractional electric flux. This result is intimately related to the 
observation of Witten\Cite{WiCh} that the $\theta$ dependence can be 
implemented at the hamiltonian level, by replacing the canonical momentum 
belonging to the gauge field, i.e. the electric field $E_i(\vek x)$, by 
$E_i(\vek x)-\frac{\theta}{4\pi^2}B_i$, with $B_i(\vek x)\equiv\half\eps_{ijk}
F_{jk}(\vek x)$ the magnetic field. This behavior, of magnetic charge obtaining
a $\theta$ dependent electric charge, also plays a role in 't~Hooft's abelian 
projection and oblique confinement.\cite{Tho7} 

\begin{figure}[htb]
\vspace{5.5cm}
\includegraphics{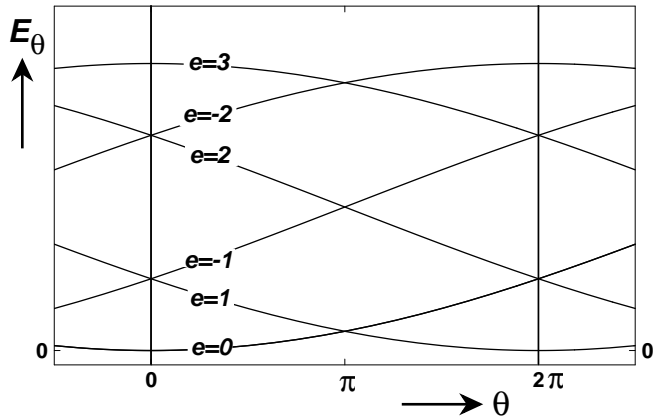}
\caption{The SU(6) spectral flow\protect\Cite{VB84} of $E_\theta^\protect\phd$ 
in a small volume, Eq.~\protect\Ref{eq:specflow}, representing the energy of 
electric flux for $\vek m=(0,0,1)$ and $e_1=e_2=0$. The spectrum is periodic 
with a period $2\pi$, the flow only with a period of $12\pi$.}
\label{fig:specflow}
\end{figure}

It seems that the $2\pi$ periodicity in $\theta$ needs to be replaced by $2\pi 
N$. Nevertheless, the spectrum itself is periodic with period $2\pi$. There is,
however, a non-trivial spectral flow, which in the infinite volume may be 
non-analytic and give rise to oblique confinement.\cite{Tho7} The spectral 
flow can neatly be illustrated in a finite volume, where energies are required 
to be analytic in $\theta$. For this we consider $\vek m=(0,0,1)$ and recall 
from Sec.~\ref{subsec:twist} that the energy splittings for $\vek e=(0,0,e_3)$ 
are due to tunneling with fractionally charged instantons. One easily computes 
the resulting energies through standard instanton calculations,\cite{Cole,Raja}
using the fact that the twisted instanton has only four zero-modes due to 
translations (cmp. Sec.~\ref{sec:inst}),
\beq
E^\phd_\theta(\vek e=e\vek m=e\vek e^{(3)})=\Ss{C}\sin^2(\pi|\vek e^\phd_\theta
|/N)\exp\bigl[-8\pi^2/Ng^2(L)\bigl]/Lg^4(L).\label{eq:specflow}
\eeq
The behavior of $E^\phd_\theta$ is illustrated, for $N=6$, in 
Fig.~\ref{fig:specflow} from which it is immediately clear that the spectrum 
is periodic with a period of $2\pi$, but that following the energy levels 
adiabatically, one only returns to the same situation after $N=6$ periods.

We now consider the euclidean partition function at finite temperature, 
specified by $1/\beta$, where $\beta$ is the period in the euclidean time 
direction. For simplicity we will also assume the period lattice $\Xi$ is 
orthogonal, $a^{(\mu)}=L_\mu e^{(\mu)}$ with $L_0\equiv\beta$. The free 
energy $F(\vek e,\vek m,L_\mu)$ is defined for the different superselection 
sectors as 
\beq
\exp\left[-\beta F_\theta(\vek e,\vek m,L_\mu)\right]=
\Tr^\phd_{\theta,\vek e}\left[\exp(-\beta H^\phd_\vek m)\right],
\eeq
where the trace is here over the Hilbert space with definite $\theta$, electric
and magnetic quantum numbers. To relate the finite temperature partition 
function to the euclidean path integral, $\theta$ and the electric flux have 
to be implemented as a sum over the homotopy classes ($\nu\in\zahlen$ and 
$\vek k\in Z_N^3$) of the gauge transformations. In the $A_0=0$ gauge, the
gauge field at $t=0$ is related to that at $t=\beta$ by $A(\vek x,\beta)=
[h_{\vek k}h_1^\nu]\,A(\vek x,0)$. For a given choice of $\nu$ and $\vek k$ 
we denote the euclidean path integral by $W_\nu(\vek k,\vek m,L_\mu)=
W_\nu(n,\Xi)$, where $n$ denotes the antisymmetric integer twist tensor, 
with $n_{ij}=\eps_{ijl}m_l$ and $n_{0i}=k_i$ specifying the twisted boundary 
conditions, which can now be formulated in any gauge. The relation between 
the free energy and the euclidean path integral reads
\beq
\exp[-\beta F_\theta(\vek e,\vek m,L_\mu)]=N^{-3}\sum_{\vek k\in Z_N^3}
\sum_{ \nu\in\zahlen}e^{i\nu\theta}\zN^{-\vek k\cdot\vek e_\theta}
W_\nu(\vek k,\vek m,L_\mu),\label{eq:ekfour}
\eeq
where we introduced the $N$-th root of unity, $\zN\equiv e^{2\pi i/N}$. 

The alternate notation, $W_\nu(n_{\mu\nu},\Xi)$, makes manifest that from the 
point of view of the euclidean path integral there is no essential distinction 
between the twisted boundary conditions in the space and the time directions. 
As observed by 't~Hooft, this path integral is invariant under a joint SO(4) 
rotation of the period lattice $\Xi$ and the twist tensor $n$. This nearly 
trivial fact, however, results in a duality between the electric and magnetic 
sectors.\Cite{Tho1} A simultaneous rotation in the 1-2 and the 3-0 planes over 
90 degrees, interchanges $(k_1,k_2)$ with $(m_1,m_2)$, leaves $k_3$ and $m_3$ 
unchanged and maps $L_\mu$ to $\tilde L_\mu=(L_3,L_2,L_1,\beta)$. Substituting 
this in Eq.~\Ref{eq:ekfour} leads to the {\em exact} duality\Cite{Tho1}
\beq
\exp[-\beta F_\theta(\vek e,\vek m,L_\mu)]=N^{-2}\!\!\!\!\!\!\sum_{\tilde\vek m,
\tilde\vek e\in Z_N^3}\!\!\!\!\!\delta_{\tilde m_3m_3}\delta_{\tilde e_3e_3}\,
\zN^{\tilde\vek e\cdot\vek m-\tilde\vek m\cdot\vek e}\exp[-L_3 
F_\theta(\tilde\vek e,\tilde\vek m,\tilde L_\mu)].\label{eq:dual}
\eeq
It is because $\tilde\vek e_\theta\cdot\vek m-\tilde\vek m\cdot\vek e_\theta$,
is independent of $\theta$, that $\theta$ does not explicitly appear in this
equation.\cite{Tho3} Hence, we will in the following ignore the $\theta$ label.

Let us normalize $F(\vek 0,\vek 0,L_\mu)=0$, and call those flux states for
which $L_\mu F(\vek e,\vek m,L_\mu)$ tends to zero {\em light fluxes}. That
they cannot all be light is easily seen by summing Eq.~\Ref{eq:dual} over 
$m_1$, $m_2$, $e_1$ and $e_2$,
\beq
N^{-2}\sum_{e_1,e_2,m_1,m_2}\exp[-\beta F(\vek e,\vek m,L_\mu)]=
\exp[-L_3 F(e_3\vek e^{(3)},m_3\vek e^{(3)},\tilde L_\mu)].\label{eq:sum}
\eeq
If $(e_3\vek e^{(3)},m_3\vek e^{(3)})$ produces a light flux state, there must 
be $N^2-1$ {\em additional} light flux states (with the same $e_3$ and $m_3$). 
On the other hand, since the euclidean path integral is positive, irrespective
the twist, once the flux $(\vek e,\vek m)$ is light, so must be the flux 
$(\vek 0,\vek m)$ (see Eq.~\Ref{eq:ekfour}). For SU(2) and SU(3) it can thus 
be concluded\Cite{Tho1} that, in the confined phase, the magnetic fluxes have 
a vanishing free energy in the infinite volume limit. 

It will be assumed that for $\beta\gg L_i$ the free energy factorizes in a 
magnetic and electric component, $F(\vek e,\vek m,L_\mu)=F_m(\vek m,L_\mu)+
F_e(\vek e,L_\mu)$. This can be justified by the fact that the electric flux 
is squeezed into a string with a finite width (to give rise to a linear 
potential), which at high $\beta$ takes negligible space, whereas magnetic flux
spreads out over the whole volume (for the free energy to vanish). We now show 
how to compute the volume dependence of the magnetic energy, $E_m(\vek m,L_i)
\equiv\lim_{\beta\rightarrow\infty}(F_m(\vek m,L_\mu)-F_m(\vek 0,L_\mu))$, for 
the case where $m_3=0$. Let us choose $e_3=0$ and (for simplicity) $L_1=L_2=L$,
with $\beta,L_3\gg L$. From Eq.~\Ref{eq:dual} it follows that
\beq
\Ss{C}\exp[-\beta F_m(\vek m_\perp,L_\mu)]=N^{-2}\sum_{\tilde\vek e_\perp
\in Z_N^2}\zN^{\tilde\vek e_\perp\cdot\vek m_\perp}\exp[-L_3 F_e(\tilde
\vek e_\perp,\tilde L_\mu)].\label{eq:magn}
\eeq
The normalization constant $\Ss{C}$ (obtained by putting $\vek m=\vek 0$) can 
be absorbed in $F_m$. One can compute the electric free energies from the 
statistical distribution of flux strings of energy $K L$, running in either
the 1- or 2-direction, to build up the fluxes $e_i$ from $n_i+e_i$ strings 
running in the $i$-direction and $n_i$ in the opposite direction. For SU(2), 
fluxes running in opposite directions are equivalent, and one instead sums over
$n_i$ even for $e_i=0$ and $n_i$ odd for $e_i=1$. Entropy allows us to ignore 
higher units of electric flux (including the ones running diagonally). Strings 
running in the {\em long} 3-directions can be neglected as well. The Boltzmann 
weight of one string equals $L L_3\exp(-\beta K L)/\Ss{A}$, where $\Ss{A}$ is a
constant related to the ``proper area" of the string. It is now straightforward
to compute the electric free energy, substitute the result in Eq.~\Ref{eq:magn}
and obtain\Cite{Tho1} the magnetic energy by extracting the limiting behavior 
in $\beta\rightarrow\infty$,
\beq
E_m=2(2-\delta_{N,2})\bigl[\sin^2(\pi m_1/N)+\sin^2(\pi m_2/N)\bigr]
\Ss{A}^{-1}L\exp(-LL_3K).
\eeq
Thus the magnetic energy falls off in leading order with the same area law 
as the Wilson loop expectation value.

\section{Conclusions}

I have tried to convince the reader of the usefulness of a finite volume as 
a control in studying four dimensional non-abelian gauge theories, with for 
the torus the practical benefit of providing a guide for lattice Monte Carlo 
results. There are many issues I have not addressed here. Needless to say 
the choices have been determined by the things I have worked on in the past, 
or that were close to me. I hope to have at least succeeded in developing 
this theme here in a logical and pedagogical fashion. It is my belief that 
probing the non-perturbative dynamics through spectral properties is the only 
reliable way to gain insight. This is not to say it is only through finite 
volume studies we may make progress. There is of course the large $N$ 
expansion,\cite{Tho6,Make} and the hope for a string representation of 
QCD.\cite{Poly} It should also be said we have followed the standard paradigm, 
that confinement is first to be understood in the pure gauge sector, instead 
of taking the attitude that the light quarks provided by nature is what mostly 
matters.\cite{Volo} 

On the basis of this review we list a number of open problems that are
worthwhile addressing.
\begin{enumerate}
\item Solve the problem with the adiabatic approximation in the calculation
      of the Witten index in small volume supersymmetric gauge theories.
\item Find analytic solutions for the basic instantons on the torus, with
      and without twist. 
\item Isolate the associated sphaleron configurations and non-perturbatively 
      important degrees of freedom.
\item Map out with {\em high-precision} the low-lying spectrum around 1 fermi.
\item Establish in a {\em reliable} way the physical nature of the finite 
      volume cross-over near 1 fermi, and find the intermediate-distance 
      effective degrees of freedom, that are responsible for the mass gap 
      and string formation.
\end{enumerate}
The first two items are interesting technical problems in their own, which one 
may hope are close to being solved. They have a wider range of applications 
than in the context of QCD alone. The fourth item does not have to wait for 
analytic predictions, and can be addressed by Monte Carlo methods with present 
day resources. The last item is of course what it is all about. It is unlikely 
this can be solved by a single mathematical equation, or with mathematical 
rigor.\cite{Clay} Fortunately, we do have experiment to guide us here. Despite 
the many phenomenological models, victory can not be claimed yet. The scenario 
of Mandelstam and 't~Hooft, for describing QCD in terms of a dual 
superconductor,\cite{Mand,Tho5} remains an appealing one and gets some support 
from supersymmetric gauge theories exhibiting Seiberg-Witten 
duality,\cite{SeWi} even though their matter content is uncomfortably far from 
that of QCD.

The challenge of {\em really} understanding QCD non-perturbatively remains wide
open, even though many in the particle physics community escape to other 
dimensions. Some of them hope that from this higher vantage point we can look 
down upon this problem that was left over from the twentieth century, and solve
it with twenty-first century techniques. Anything is allowed, as long as we 
come up with a practical and reliable solution.

\section*{Acknowledgements}

I thank Misha Shifman for inviting me to contribute to this volume. I somewhat
hesitatingly accepted, but while working on it started to enjoy bringing these 
results together in one place. I hope he is satisfied with the final result. Of
course I also hope Boris Ioffe, whom I only met briefly on one of my visits to 
ITEP, is happy with this chapter in his honor. I have covered an aspect of QCD 
that has not touched on many of his deep contributions to the field. Let me 
just say that I have always felt a special tie to my Russian friends, for whom 
physics is a way of life, something Ioffe was very much a part of. May I 
express the hope that in a changing world, this way of life does not completely
disappear. Finally, I am grateful to the many people I worked with and who have
all contributed in an essential way to developing the material presented.


\section*{References}
\begin{thebibliography}{132}
\bibitem{YaMi}C.N. Yang and R.L. Mills, \Journal{\PR}{96}{191}{1954}.
\bibitem{Chle}N.M. Christ and T.D. Lee, \Journal{\PRD}{22}{939}{1980}.
\bibitem{Bavi}O. Babelon  and C. Viallet, \Journal{\CMP}{81}{515}{1981}.
\bibitem{Grib}V.N. Gribov, \Journal{\NPB}{139}{1}{1978}.
\bibitem{Sefr}M.A. Semenov-Tyan-Shanskii and V.A. Franke, {\em Zapiski 
              Nauchnykh Sem. Leningradskogo Otdeleniya Mat. Inst. im. V.A. 
              Steklov AN SSSR} {\bf 120}, 159 (1982), Translation (Plenum,
              New York, 1986) p~999.
\bibitem{Bjor}J.D. Bjorken, ``Elements of Quantum Chromodynamics", in 
              {\em Slac Summer Institute on Particle Physics}, ed. A. 
              Mosher (SLAC, Stanford, 1980). 
\bibitem{Lue1}M. L\"uscher, \Journal{\NPB}{219}{233}{1983}. 
\bibitem{LSWW}M. L\"uscher, R. Sommer, U. Wolff and P. Weisz,\\
              \Journal{\NPB}{389}{247}{1993} (hep-lat/9207010);\\
              \Journal{\NPB}{413}{481}{1994} (hep-lat/9309005).
\bibitem{Tho1}G. 't~Hooft, \Journal{\NPB}{153}{141}{1979}.
\bibitem{WiIn}E. Witten, \Journal{\NPB}{202}{253}{1982}.
\bibitem{Nahm}W. Nahm, \Journal{\PLB}{90}{413}{1980}; in {\em Monopoles in 
              Quantum Field Theory}, eds. N. Craigie {\em et al} (World 
              Scientific, Singapore, 1982); in {\em Lect. Notes in Phys.} vol 
              {\bf 201}, eds. G. Denardo {\em et al} (Springer, Berlin, 1985).
\bibitem{Lue0}M. L\"uscher in {\em Progress in Gauge Field Theory}, eds.
              G. 't~Hooft {\em et al} (Plenum, New York, 1984) p~451;
              \Journal{\CMP}{104}{177}{1986}.
\bibitem{Leut}H. Leutwyler, {\em Chiral Dynamics}, this volume (hep-ph/0008124).
\bibitem{Del1}G. Dell'Antonio and D. Zwanziger, \Journal{\CMP}{138}{291}{1991};
              in {\em Probabilistic Methods in Quantum Field Theory and Quantum
              Gravity}, eds. P.H. Damgaard {\em et al} (Plenum, New York, 
              1990) p~107. 
\bibitem{Dezw}G. Dell'Antonio and D. Zwanziger, \Journal{\NPB}{326}{333}{1989}.
\bibitem{Zwan}D. Zwanziger, \Journal{\NPB}{378}{525}{1992}.
\bibitem{Vba1}P. van Baal, \Journal{\NPB}{369}{259}{1992}.
\bibitem{Kovb}J. Koller and P. van Baal, \Journal{\NPB}{302}{1}{1988}.
\bibitem{Ferm}P. van Baal, \Journal{\NPB}{307}{274}{1988} 
              [erratum B{\bf 312}, 752 (1989)].
\bibitem{Vba0}P. van Baal, \Journal{\NPB}{264}{548}{1986}.
\bibitem{Bpst}A. Belavin, A. Polyakov, A. Schwarz and Y. Tyupkin, 
              \Journal{\PLB}{59}{85}{1975}.
\bibitem{Tho2}G. 't~Hooft, \Journal{\PRD}{14}{3432}{1976} [erratum D{\bf 18},
              2199 (1978)].
\bibitem{Kov0}J. Koller and P. van Baal, \Journal{\PRL}{58}{2511}{1987}. 
\bibitem{Bloc}C. Bloch, \Journal{\NP}{6}{329}{1958}.
\bibitem{Vba4}P. van Baal, \Journal{\NPB}{351}{183}{1991}.
\bibitem{AuKi}A. Auerbach, S. Kivelson and D. Nicole,
              \Journal{\PRL}{53}{411}{1984}; A. Auerbach and S. Kivelson, 
              \Journal{\NPB}{257}{799}{1985}.
\bibitem{VbAu}P. van Baal and A. Auerbach, \Journal{\NPB}{275}{93}{1986}.
\bibitem{Adri}P. van Baal in {\em Lectures on Path Integration}, eds. H.A. 
              Cerdeira {\em et al} (World Scientific, Singapore, 1993) p~54. 
\bibitem{Vbk1}J. Koller and P. van Baal, \Journal{\NPB}{273}{387}{1986}. 
\bibitem{AnnP}P. van Baal and J. Koller, \Journal{\AP}{174}{299}{1987}.
\bibitem{LuMu}M. L\"uscher and G. M\"unster, \Journal{\NPB}{232}{445}{1984}.
\bibitem{Voh1}C. Vohwinkel, \Journal{\PLB}{213}{54}{1988};\\
              \Journal{\NPBPS}{9}{242}{1989}.
\bibitem{Gavb}M. Garc\'{\i}a P\'erez  and P. van Baal,
              \Journal{\NPB}{429}{451}{1994}\\ (hep-lat/9403026).
\bibitem{BeBi}B.A. Berg and A.H. Billoire, \Journal{\PLB}{166}{203}{1986}
              [erratum B{\bf 185}, 446 (1987)]; B.A. Berg, A.H. Billoire and 
              C. Vohwinkel, \Journal{\PRL}{57}{400}{1986}.
\bibitem{Mitt}C. Michael, G.A. Tickle and M.J. Teper, 
              \Journal{\PLB}{207}{313}{1988}.
\bibitem{KrM2}J. Kripfganz and C. Michael, \Journal{\NPB}{314}{25}{1989}.
\bibitem{Berg}B.A. Berg, \Journal{\PLB}{206}{97}{1988}.
\bibitem{Tore}C. Michael, \Journal{\PLB}{232}{247}{1989}.
\bibitem{Ha2N}P. Hasenfratz, A. Hasenfratz and F. Niedermayer,\\
              \Journal{\NPB}{329}{739}{1990}.
\bibitem{BeBR}B.A. Berg and A.H. Billoire, \Journal{\PRD}{40}{550}{1989}.
\bibitem{Mich}C. Michael, \Journal{\NPB}{329}{225}{1990}.
\bibitem{Impr}M. Garc\'{\i}a P\'erez, J. Snippe and P. van Baal,\\
              \Journal{\PLB}{389}{112}{1996} (hep-lat/9608036).
\bibitem{BeVR}B.A. Berg and C. Vohwinkel, \Journal{\AP}{204}{351}{1990}.
\bibitem{Voh4}C. Vohwinkel, {\em Calculation of the Mass Spectrum and 
              Deconfining Temperature in Non-Abelian Gauge Theory}, PhD 
              thesis (Tallahassee, September 1989).
\bibitem{Latt}P. van Baal, \Journal{\PLB}{224}{397}{1989};\\
              \Journal{\NPBPS}{17}{581}{1990}.
\bibitem{WeZi}P. Weisz and V. Ziemann, \Journal{\NPB}{284}{157}{1987}.
\bibitem{SU3E}J. Koller and P. van Baal, \Journal{\PRL}{57}{2783}{1986}.
\bibitem{Voh2}C. Vohwinkel, \Journal{\PRL}{63}{2544}{1989}.
\bibitem{Vba3}P. van Baal in {\em Probabilistic Methods in Quantum Field Theory
              and Quantum Gravity},  eds. P.H. Damgaard {\em et al} (Plenum,
              New York, 1990) p~31.
\bibitem{CMP0}P. van Baal, \Journal{\CMP}{85}{529}{1982}.
\bibitem{Hoek}J. Hoek, \Journal{\NPB}{332}{530}{1990}.
\bibitem{KrM1}J. Kripfganz and C. Michael, \Journal{\PLB}{209}{77}{1988}.
\bibitem{Fer2}P. van Baal in {\em Frontiers in Nonperturbative Field Theory}, 
              eds. Z. Horvath {\em et al}, (World Scientific, Singapore, 
              1989), p~204.
\bibitem{VbKr}P. van Baal and A.S. Kronfeld,\\ \Journal{\NPBPS}{9}{227}{1989}.
\bibitem{Voh3}C. Vohwinkel, \Journal{\NPB}{443}{417}{1995} (hep-lat/9410010).
\bibitem{Tied}H. Tiedemann, \Journal{\PRD}{44}{1280}{1991}.
\bibitem{PoWi}L. Polley and U. Wiese, \Journal{\NPB}{356}{629}{1991}.
\bibitem{KrW0}A. Kronfeld and U. Wiese, \Journal{\NPB}{357}{521}{1991}.
\bibitem{KrW1}A. Kronfeld and U. Wiese, \Journal{\NPB}{401}{190}{1993}\\
              (hep-lat/9210008).
\bibitem{LuRC}M. L\"uscher, {\em Project Proposal for the EMC$^2$
              Collaboration}, unpublished notes, December 1983.
\bibitem{LuWW}M. L\"uscher, P. Weisz and U. Wolff, 
              \Journal{\NPB}{359}{221}{1991}.
\bibitem{LNWW}M. L\"uscher, R. Narayanan, P. Weisz and U. Wolff,\\
              \Journal{\NPB}{384}{168}{1992} (hep-lat/9207009).
\bibitem{Petr}G.M. de Divitiis, R. Frezzotti, M. Guagnelli and R. Petronzio,\\
              \Journal{\NPB}{422}{382}{1994} (hep-lat/9312085);\\
              \Journal{\NPB}{433}{390}{1995} (hep-lat/9407028).
\bibitem{CGJK}A. Gonz\'alez-Arroyo, J. Jurkiewicz and C.P. Korthals Altes, 
              in {\em Proc. 11th NATO Summer Institute}, eds. J. Honerkamp 
              {\em et al} (Plenum, New York, 1982); A. Coste, 
              A. Gonz\'alez-Arroyo, C.P. Korthals Altes, B. S\"oderberg and 
              A. Tarancon, \Journal{\NPB}{287}{569}{1987}.
\bibitem{Comp}G.M. de Divitiis, R. Frezzotti, M. Guagnelli, M. L\"uscher,
              R. Petronzio, R. Sommer, P. Weisz and U. Wolff,
              \Journal{\NPB}{437}{447}{1995}\\ (hep-lat/9411017).
\bibitem{HvBZ}T.H. Hansson, P. van Baal and I. Zahed, 
              \Journal{\NPB}{289}{628}{1987}.
\bibitem{Vbvg}B. van Geemen and P. van Baal, {\em Proc. K. Ned. Akad. Wet.} B 
              {\bf 89}, 39 (1986); P. van Baal and B. van Geemen, 
              \Journal{\JMP}{27}{455}{1986}.
\bibitem{LePo}D.R. Lebedev and M.I. Polikarpov, \Journal{\NPB}{269}{285}{1986}.
\bibitem{GO83}A. Gonz\'alez-Arroyo and M. Okawa, \Journal{\PRD}{27}{2397}{1983}.
\bibitem{Ambj}J. Ambj\o rn and H. Flyvbjerg, \Journal{\PLB}{79}{241}{1980}.
\bibitem{GJKA}J. Groeneveld, J. Jurkiewicz and C.P. Korthals Altes,\\
             {\em Physica Scripta} {\bf 23}, 1022 (1981).
\bibitem{VB84}P. van Baal, {\em Twisted Boundary Conditions: A Non-Perturbative 
	      Probe for Pure Non-Abelian Gauge Theories}, PhD thesis (Utrecht, 
              July 1984).
\bibitem{Tho3}G. 't~Hooft, {\em Acta Physica Austriaca}, Suppl. {\bf XXII}, 531
              (1980).
\bibitem{DGKS}A. Gonz\'alez-Arroyo and C. Korthals Altes,
              \Journal{\NPB}{311}{433}{1988/89}; D. Daniel, 
              A. Gonz\'alez-Arroyo, C. Korthals Altes and B. S\"oderberg, 
              \Journal{\PLB}{221}{136}{1989}.
\bibitem{DGKA}D. Daniel, A. Gonz\'alez-Arroyo and C. Korthals Altes,\\
              \Journal{\PLB}{251}{559}{1990}.
\bibitem{GPGS}M. Garc\'{\i}a P\'erez, A. Gonz\'alez-Arroyo and B. S\"oderberg,\\
              \Journal{\PLB}{235}{117}{1990}.
\bibitem{StTe}P. Stephenson and M. Teper, \Journal{\NPB}{327}{307}{1989}.
\bibitem{Step}P. Stephenson, \Journal{\NPB}{356}{318}{1991}.
\bibitem{RTNc}The RTN collaboration, M. Garc\'{\i}a P\'erez {\em et al},\\
              \Journal{\PLB}{305}{366}{1993} (hep-lat/9302007);\\
              M. Garc\'{\i}a P\'erez, A. Gonz\'alez-Arroyo and P. 
              Mart\'{\i}nez,\\ 
              \Journal{\NPBPS}{34}{228}{1994} (hep-lat/9312066).
\bibitem{GAMa}A. Gonz\'alez-Arroyo and P. Mart\'{\i}nez, 
              \Journal{\NPB}{459}{337}{1996} (hep-lat/9507001).
\bibitem{Klma}F. R. Klinkhamer and M. Manton, \Journal{\PRD}{30}{2212}{1984}.
\bibitem{Taub}C. Taubes, \Journal{\JDG}{19}{517}{1984}.
\bibitem{DoKr}S. Donaldson and P. Kronheimer, {\em The Geometry of Four 
              Manifolds}\\ (Oxford University Press, 1990).
\bibitem{BrvB}P.J. Braam and P. van Baal, \Journal{\CMP}{122}{267}{1989}.
\bibitem{BrMT}P.J. Braam, A. Maciocia and A. Todorov, {\em Inv. Math.} 
              {\bf 108}, 419 (1992).
\bibitem{Cool}B. Berg, \Journal{\PLB}{104}{475}{1981}; J. Hoek, M. Teper and 
              J. Waterhouse, \Journal{\NPB}{288}{589}{1987}.
\bibitem{GGVS}M. Garc\'{\i}a P\'erez, A. Gonz\'alez-Arroyo, J. Snippe and P.
              van Baal,\\ \Journal{\NPB}{413}{535}{1994} (hep-lat/9309009);\\
              \Journal{\NPBPS}{34}{222}{1994} (hep-lat/9311032).
\bibitem{GoAr}A. Gonz\'alez-Arroyo in {\em Advanced School for Non-perturbative
              Quantum Field Theory}, eds. M. Asorey {\em et al} (World 
              Scientific, Singapore, 1998) p~57 (hep-th/9807108).
\bibitem{Sedl}S. Sedlacek, \Journal{\CMP}{86}{515}{1982}.
\bibitem{Schw}A.S. Schwarz, \Journal{\CMP}{64}{233}{1979}.
\bibitem{BaPs}B. S\"oderberg, {\em Topology in Twisted Lattice Gauge Theories},
              LU~TP~87-2 (Lund preprint, February 1987, unpublished);\\
              I.M. Barbour and S.J. Psycharis, \Journal{\NPB}{334}{302}{1990}.
\bibitem{Twis}M. Garc\'{\i}a P\'erez and A. Gonz\'alez-Arroyo,
              \Journal{\JPA}{26}{2667}{1993} (hep-lat/9206016);
              M. Garc\'{\i}a P\'erez, A. Gonz\'alez-Arroyo and A. Montero,
              \Journal{\NPBPS}{63}{501}{1998} (hep-lat/9709107);\\
              A. Montero, \Journal{\JHEP}{05}{022}{2000} (hep-lat/0004009).
\bibitem{LNdW}B. de Wit, M. L\"uscher and H. Nicolai,
              \Journal{\NPB}{320}{135}{1989}.
\bibitem{Itoy}H. Itoyama and B. Razzaghe-Ashrafi, \Journal{\NPB}{354}{85}{1991}.
\bibitem{Smil}A.V. Smilga, \Journal{\NPB}{266}{45}{1986};
              \Journal{\YF}{43}{215}{1986}.
\bibitem{NSVZ}V. Novikov, M. Shifman, A. Vainshtein and V. Zakharov,\\
              \Journal{\NPB}{229}{407}{1983}.
\bibitem{ShVa}V. Novikov, M. Shifman, A. Vainshtein and V. Zakharov,\\
              \Journal{\NPB}{260}{157}{1985};\\
              M.A. Shifman and A.I. Vainshtein, \Journal{\NPB}{296}{445}{1988}.
\bibitem{Shif}M. Shifman in {\em Confinement, Duality and Nonperturbative 
              Aspects of QCD}, ed. P. van Baal (Plenum, New York, 1998) p~477.
\bibitem{AmVe}D. Amati, K. Konishi, Y. Meurice, G. Rossi and G. Veneziano,\\
              \Journal{\PRP}{162}{169}{1988}.
\bibitem{KrvB}T.C. Kraan and P. van Baal, \Journal{\NPB}{533}{627}{1998}
              (hep-th/9805168); \Journal{\PLB}{435}{389}{1998} 
              (hep-th/9806034).
\bibitem{LLYi}K. Lee and P. Yi, \Journal{\PRD}{56}{3711}{1997} 
              (hep-th/9702107);\\ K. Lee and C. Lu, 
              \Journal{\PRD}{58}{025011}{1998} (hep-th/9802108).
\bibitem{Khoz}N.M. Davies, T.J. Hollowood, V.V. Khoze and M.P. Mattis,\\
              \Journal{\NPB}{559}{123}{1999} (hep-th/9905015).
\bibitem{WiBr}E. Witten, \Journal{\JHEP}{02}{006}{1998} (hep-th/9712028).
\bibitem{KeRS}A. Keurentjes, A. Rosly and A.V. Smilga, 
              \Journal{\PRD}{58}{081701}{1998} (hep-th/9805183).
\bibitem{Keur}A. Keurentjes, \Journal{\JHEP}{05}{001}{1999} (hep-th/9901154);
              \Journal{\JHEP}{05}{014}{1999} (hep-th/9902186); {\em New Vacua 
              for Yang-Mills Theory on a 3-Torus}, PhD thesis (Leiden, June 
              2000)\\ (hep-th/0007196).
\bibitem{KaSm}V.G. Kac and A.V. Smilga, {\em Vacuum Structure in Supersymmetric
              Yang-Mills Theories with any Gauge Group}, hep-th/9902029 v.3.
\bibitem{BoFM}A. Borel, M. Friedman and J.W. Morgan, {\em Almost Commuting
              Elements in Compact Lie Groups}, math/9907007.
\bibitem{WiRe}E. Witten, {\em Supersymmetric Index in Four-Dimensional 
              Gauge Theories}, hep-th/0006010.
\bibitem{Savv}G.K. Savvidy, \Journal{\PLB}{159}{325}{1985}.
\bibitem{Tho4}G. 't~Hooft, \Journal{\CMP}{81}{267}{1981}.
\bibitem{CMP1}P. van Baal, \Journal{\CMP}{94}{397}{1984}.
\bibitem{LePR}D.R. Lebedev, M.I. Polikarpov and A.A. Rosly, 
              \Journal{\NPB}{325}{138}{1989}.
\bibitem{Pert}M. Garc\'{\i}a P\'erez, A. Gonz\'alez-Arroyo and C. Pena, {\em 
              Perturbative Construction of Self-dual Solutions on the Torus}, 
              hep-th/0007113.
\bibitem{Muka}S. Mukai, {\em Nagoya Math. J.} {\bf 81},153 (1981).
\bibitem{ADHM}M. Atiyah, N. Hitchin, V. Drinfeld and Yu. Manin,\\ 
              \Journal{\PLA}{65}{185}{1978}.
\bibitem{CoGo}E. Corrigan and P. Goddard, \Journal{\AP}{154}{253}{1984}.
\bibitem{AtSi}M.F. Atiyah and I.M. Singer, {\em Ann. Math.} {\bf 93}, 
              119 (1971);\\ {\em Proc. Natl. Acad. Sci. USA}, Vol. {\bf 81}, 
              2597 (1984).
\bibitem{Lect}P. van Baal, {\em Complex Structures in Gauge Theories}, Graduate
              Course Lectures, unpublished, (Stony Brook, spring 1986).
\bibitem{Sche}H. Schenk, \Journal{\CMP}{116}{177}{1988}.
\bibitem{Tony}A. Gonz\'alez-Arroyo, \Journal{\NPB}{548}{626}{1999} 
              (hep-th/9811041).
\bibitem{Synt}M. Garc\'{\i}a P\'erez, A. Gonz\'alez-Arroyo, C. Pena and P. van
              Baal, \Journal{\NPB}{564}{159}{2000}.
\bibitem{Buck}P. van Baal, \Journal{\NPBPS}{49}{238}{1996}\\ (hep-th/9512223).
\bibitem{PLB0}P. van Baal, \Journal{\PLB}{448}{26}{1999} (hep-th/9811112).
\bibitem{GAPe}A. Gonz\'alez-Arroyo and C. Pena, \Journal{\JHEP}{09}{013}{1998}\\
              (hep-th/9807172).
\bibitem{GGMV}M. Garc\'{\i}a P\'erez, A. Gonz\'alez-Arroyo, A. Montero and 
              P. van Baal,\\ \Journal{\JHEP}{06}{001}{1999} (hep-lat/9903022).
\bibitem{Vbda}P. van Baal and N. D. Hari Dass, \Journal{\NPB}{385}{185}{1992}.
\bibitem{Hoso}Y. Hosotani, \Journal{\PLB}{147}{44}{1984}.
\bibitem{Vbvd}P. van Baal and B. van den Heuvel, 
              \Journal{\NPB}{417}{215}{1994}\\ (hep-lat/9310005).
\bibitem{Cutk}R.E. Cutkosky, \Journal{\JMP}{25}{939}{1984};\\
              R.E. Cutkosky and K. Wang, \Journal{\PRD}{37}{3024}{1988};\\
              R.E. Cutkosky, {\em Czech. J. Phys.} {\bf 40}, 252 (1990).
\bibitem{Lue2}M. L\"uscher, \Journal{\PLB}{70}{321}{1977}.
\bibitem{Vbcu}P. van Baal and R.E. Cutkosky, {\em Int. J. Mod. Phys.}
              A(Proc.~Suppl.) {\bf 3A}, 323 (1993) (hep-lat/9208027).
\bibitem{Vdhe}B. van den Heuvel, \Journal{\PLB}{368}{124}{1996} 
              (hep-lat/9509019); \Journal{\PLB}{386}{233}{1996} 
              (hep-lat/9604017); \Journal{\NPB}{488}{282}{1997}
              (hep-lat/9608101); {\em Non-perturbative Phenomena in Gauge
              Theory on $S^3$}, PhD thesis (Leiden, September 1996).
\bibitem{Diek}B. Diekmann, {\em Eine Modellraum-Methode zur Berechnung des
              Glueballspektrums im kompaktifizierten Minkowskiraum}, PhD 
              thesis [TK-95-26] (Bonn, October 1995).
\bibitem{Shur}E.V. Shuryak, \Journal{\NPB}{302}{559,574,599,621}{1988};
              T. Sch\"afer and E.V. Shuryak, \Journal{\RMP}{70}{323}{1998}
              (hep-ph/9610451);\\ E. Shuryak in {\em Confinement, Duality and 
              Nonperturbative Aspects of QCD}, ed. P. van Baal (Plenum, New 
              York, 1998) p~307.
\bibitem{ShSh}T. Sch\"afer and E.V. Shuryak, \Journal{\PRL}{75}{1707}{1995}\\
              (hep-ph/9410372).
\bibitem{Tepe}M. Teper, {\em Glueball Masses and other Physical Properties
              of SU(N) Gauge Theories in D=(3+1): A Review of Lattice Results 
              for Theorists}, Oxford preprint OUTP-98-88-P, hep-lat/9812187.
\bibitem{Edin}P. van Baal, \Journal{\NPBPS}{63}{126}{1998}.
\bibitem{Lue4}M. L\"uscher, \Journal{\CMP}{105}{153}{1986}.
\bibitem{Lue5}M. L\"uscher, \Journal{\NPB}{354}{531}{1991};\\
              \Journal{\NPB}{364}{237}{1991}.
\bibitem{La90}P. van Baal, \Journal{\NPBPS}{20}{3}{1991}. 
\bibitem{HuYa}K. Huang and C.N. Yang, \Journal{\PR}{105}{767}{1957};\\
              C.N. Yang, {\em Chinese J. Phys.} {\bf 25},80 (1987).
\bibitem{LuWo}M. L\"uscher and U. Wolff, \Journal{\NPB}{339}{222}{1990}.
\bibitem{GaLa}C.R. Gattringer and C.B. Lang, \Journal{\PLB}{274}{95}{1992};\\
              \Journal{\NPB}{391}{463}{1993} (hep-lat/9206004).
\bibitem{MoWe}I. Montvay and P. Weisz, \Journal{\NPB}{290}{327}{1987};
\bibitem{Fric}Ch. Frick, K. Jansen, J. Jers\'ak, I. Montvay, P. Seuferling and
              G. M\"unster, \Journal{\NPB}{331}{515}{1990}.
\bibitem{LeLu}L. Lellouch and M. L\"uscher, {\em Weak Transition Matrix 
              Elements from Finite Volume Correlation Functions}, 
              hep-lat/0003023.
\bibitem{Lute}M. L\"uscher, \Journal{\NPB}{180}{317}{1981}.
\bibitem{HaLe}J. Gasser and H. Leutwyler, \Journal{\PLB}{188}{477}{1987};\\
              P. Hasenfratz and H. Leutwyler, \Journal{\NPB}{343}{241}{1990};\\
              H.C. Hansen and H. Leutwyler, \Journal{\NPB}{350}{201}{1991}.
\bibitem{LeSm}H. Leutwyler and A.V. Smilga, \Journal{\PRD}{46}{5607}{1992}.
\bibitem{BaCa}T. Banks and A. Casher, \Journal{\NPB}{169}{103}{1980}.
\bibitem{ShVe}E.V. Shuryak and J.J.M. Verbaarschot, 
              \Journal{\NPA}{560}{306}{1993}\\ (hep-th/9212088).
\bibitem{Verb}J. Verbaarschot in {\em Confinement, Duality and Nonperturbative
              Aspects of QCD}, ed. P. van Baal (Plenum, New York, 1998) p~343
              (hep-th/9710114).
\bibitem{WiCh}E. Witten, \Journal{\PLB}{86}{283}{1979}.
\bibitem{Tho7}G. 't~Hooft, \Journal{\NPB}{190}{455}{1981}; {\em Physica 
              Scripta}, Vol. {\bf 25}, 133 (1982); in {\em Confinement, 
              Duality and Nonperturbative Aspects of QCD}, ed. P. van Baal 
              (Plenum, New York, 1998) p~379.
\bibitem{Cole}S. Coleman, ``The Uses of Instantons", in {\em The Whys of 
              Subnuclear Physics}, ed. A. Zichichi (Plenum Press, New York, 
              1979) p~805.
\bibitem{Raja}R. Rajaraman, {\em Solitons and Instantons} (North-Holland, 
              Amsterdam, 1982).
\bibitem{Volo}V.N. Gribov, \Journal{\EPJ}{10}{71}{1999} (hep-ph/9807224);\\
              \Journal{\EPJ}{10}{91}{1999} (hep-ph/9902279).
\bibitem{Tho6}G. 't~Hooft, \Journal{\NPB}{72}{461}{1974};
              \Journal{\CMP}{86}{449}{1982}; \Journal{\CMP}{88}{1}{1983};
              in {\em Progress in Gauge Field Theory}, eds.  G. 't~Hooft
              {\em et al} (Plenum, New York, 1984) p~271.
\bibitem{Make}Yu. Makeenko, {\em Large-N Gauge Theories}, hep-th/0001047.
\bibitem{Poly}A.M. Polyakov, {\em Gauge Fields and Strings} 
              (Harwood, Chur, 1987);\\ 
              \Journal{\NPB}{486}{23}{1997} (hep-th/9607049);\\
              \Journal{\NPBPS}{68}{1}{1998} (hep-th/9711002);\\
              {\em Int. J. Mod. Phys.} A{\bf 14}, 645 (1999) (hep-th/9809057).
\bibitem{Clay}See 
    www.claymath.org/prize\underline{~\,}problems/yang\underline{~\,}mills.htm.
\bibitem{Mand}S. Mandelstam, \Journal{\PRP}{23}{245}{1976}.
\bibitem{Tho5}G.'t~Hooft in {\em High Energy Physics}, ed. A. Zichichi 
              (Editrice Compositori, Bolognia, 1976); 
              \Journal{\NPB}{138}{1}{1978}.
\bibitem{SeWi}N. Seiberg and E. Witten, \Journal{\NPB}{426}{19}{1994}
              [erratum B{\bf 430}, 485 (1994)] (hep-th/9407087).
\end{thebibliography}
\end{document}